 \documentclass[opre]{cls_files/informs4} 
\usepackage{tgtermes}
\usepackage{newtxtext}
\usepackage{newtxmath}
\usepackage{bm}
\usepackage{endnotes}

\OneAndAHalfSpacedXII 

\usepackage[normalem]{ulem}

\usepackage{algorithm}
\usepackage{algpseudocode}
\usepackage{tikz}
\usepackage{pdfpages}

\usepackage[hidelinks]{hyperref}
\usepackage[textsize=tiny]{todonotes}
\PassOptionsToPackage{usenames,dvipsnames}{xcolor} 

\newenvironment{myproof}{ \paragraph{Proof: } } {\hfill$\square$}
\newenvironment{myproof2}[1]{ \paragraph{Proof of #1: } } {\hfill$\square$}

\newtheorem{observation}{Observation}
\usepackage{bbm}
\usepackage{graphicx}
\usepackage{caption}
\usepackage{subcaption}
\usepackage{empheq}
\usepackage{float}
\usepackage{booktabs}
\usepackage{multirow}

\usepackage{adjustbox}
\usepackage{thmtools}

\usepackage{soul} 

\newcommand{\defeq}{\mathrel{\overset{\mathrm{def}}{=}}}

\newcommand{\indSi}{\mathbb{S}_i } 
\newcommand{\indSitilde}{\widetilde{ \mathbb{S} }_i } 
\newcommand{\indFi}{\mathbb{F}_i } 
\newcommand{\indFitilde}{\widetilde{ \mathbb{F} }_i } 
\newcommand{\indPi}{\mathbb{P}_i }

\newcommand{\E}[1]{\mathbb{E} \left[ {#1} \right]}
\newcommand{\p}[1]{\mathbf{Pr} \left[ {#1} \right]}
\newcommand{\pr}[1]{\mathbf{Pr} \left[ {#1} \right]}
\newcommand{\one}[1]{\BFI \left[ {#1} \right]}
\newcommand{\paren}[1]{ \left( {#1} \right) }

\newcommand{\bigOmega}[1]{ \Omega \left( {#1} \right) }
\newcommand{\abs}[1]{ \left | {#1} \right | }
\newcommand{\brackets}[1]{\left[ {#1} \right]}
\newcommand{\braces}[1]{\left\{{#1}\right\}}

\usepackage{color}              
\usepackage[suppress]{color-edits}
\usepackage{color-edits}
\addauthor{Ali}{red}

\newcommand{\bfx}{\mathbf{x}}

\usepackage{natbib}
 \bibpunct[, ]{(}{)}{,}{a}{}{,}%
 \def\bibfont{\small}%
\EquationsNumberedThrough    

\TheoremsNumberedThrough     
\ECRepeatTheorems  %

\MANUSCRIPTNO{MOOR-0001-2024.00}

\begin{document}


\RUNAUTHOR{ Aouad, Ji, and Shaposhnik }

\RUNTITLE{The Pandora's Box Problem with Sequential Inspections}

\TITLE{The Pandora's Box Problem \\ with Sequential Inspections}

\ARTICLEAUTHORS{%
\AUTHOR{Ali Aouad}
\AFF{Sloan School of Management, Massachusetts Institute of Technology, \EMAIL{maouad@mit.edu}} 
\AUTHOR{Jingwei Ji}
\AFF{Management Science and Engineering, Stanford University, \EMAIL{jingwei.ji@stanford.edu}} 
\AUTHOR{Yaron Shaposhnik}
\AFF{Simon Business School, University of Rochester, \EMAIL{yaron.shaposhnik@simon.rochester.edu} }
} 

\ABSTRACT{%
The Pandora's box problem \citep{weitzman1979optimal}  is a core model in economic theory that captures an agent's (Pandora's) search for the best alternative (box).
We study an important generalization of the problem where the agent can either {\em fully open} boxes for a certain fee to reveal their exact values or {\em partially open} them at a reduced cost. This introduces a new tradeoff between information acquisition and cost efficiency.
We establish a hardness result and employ an array of techniques in stochastic optimization to provide a comprehensive analysis of this model. This includes (1) the identification of structural properties of the optimal policy that provide insights about optimal decisions; (2) the derivation of problem relaxations and provably near-optimal solutions; 
(3) the characterization of the optimal policy in special yet non-trivial cases; and (4) an extensive numerical study that compares the performance of various policies, and which provides additional insights about the optimal policy. Throughout, we show that intuitive threshold-based policies that extend the Pandora's box optimal solution can effectively guide search decisions. 
}%




\KEYWORDS{Pandora's box problem, dynamic programming, approximate algorithms} 

\maketitle


\section{Introduction}
\label{sec:introduction}

Sequential search problems refer to a class of optimization problems that have long been studied in the economics, operations research (OR), and computer science (CS) literature.
In his seminal work, \citet{weitzman1979optimal} studies the Pandora's box sequential search  problem,\endnote{We refer to this model as the \textit{standard Pandora's problem.}} in which Pandora is given a collection of boxes containing random prizes drawn from known and independent probability distributions.
She can adaptively pay box-dependent fees to open boxes and observe their realizations until she decides to select the prize held within one of the open boxes, or stop exploring without selecting any box. Her objective is to maximize the expected profit, that is, the expected value of the collected prize minus the expected combined box-opening fees. This fundamental model, which captures an agent's search for the best alternative, found applications in consumer search, project management, job recruitment, and asset-selling problems, to name a few.

In this paper, we consider a generalization of this problem in which Pandora can also exercise a cheaper and possibly less accurate option of {\em partially opening} boxes.
When partially opening a box, Pandora observes its {\em type}, an informative signal based on which she updates her belief about the prize inside the box.
We refer to this variant as \textit{Pandora's Box with Sequential Inspection} (PSI for short).
Similarly to the standard problem, she may only select a prize contained in a box that was {\em fully opened}, that is, a box whose exact prize has been observed.
However, the ability to partially open boxes introduces a new tradeoff. On the one hand, partially opening a box reveals certain  information about the prize value that allows Pandora to potentially detect ``bad outcomes'' at a lower cost, and thus, avoid paying the full opening fee. On the other hand, while the cost of  partial opening is cheaper than the cost of full opening, it may not be negligible, in which case overusing the option to partially open may lead to excessive costs.

To put this model in concrete terms, consider  job recruitment, an application discussed in~\citet{weitzman1979optimal}.  
Recent technological advances transform job recruitment processes in various industries.
For example,  video conferencing is often used for remote (offsite) interviews.  More recently, AI systems have been  introduced to conduct such initial remote interviews without human intervention; the transcriptions are then reviewed by human evaluators. Remote interviews provide convenient ways for companies to learn more about candidates at reduced costs, compared with onsite interviews. Nevertheless,  remote interviews cannot fully replace onsite visits, which are often still necessary to  eventually make job offers.
Moreover, while cheaper than onsite visits, substantive remote interviews -- beyond the preliminary screenings based on CVs -- still require time from human interviewers or evaluators to  assess the candidates' fit for the position. In selecting a candidate for a vacant position, companies must therefore decide how to synthesize the different mediums to identify and select qualified candidates while minimizing costs.
Put into the framework of our model, Pandora represents the company, boxes correspond to candidates whose prior evaluations are based on screenings and automated assessments, and box opening actions correspond to conducting interviews -- offsite or onsite. The goal is to effectively administer the hiring process by deciding which candidates should be interviewed and in which format: cheaper and less accurate offsite interviews, or more expensive and informative onsite visits. 

More generally and beyond the particular application, our work tries to study how to integrate different \textit{sequential information acquisition} methods in the context of stochastic optimization problems. To our knowledge, this is the first work that examines this feature, which is quite common in various practical sequential search problems (a more detailed discussion can be found in Section~\ref{sec:literature} that surveys related work).

\citet{weitzman1979optimal} showed that there exist optimal policies to the standard Pandora's problem that admit an elegant and simple threshold-based structure. A threshold is defined for each box, indicating the order in which boxes should be considered for opening, reducing the problem to an optimal stopping time problem. The decision about when to stop is determined using the same thresholds by comparing the largest prize amongst open boxes to the largest threshold amongst closed boxes; if the incumbent prize is larger, Pandora should stop; otherwise, she should proceed to the next box, and so forth.
However, it has been observed in the literature (see Section~\ref{sec:literature}) 
that seemingly small changes in the model assumptions break the structure of the optimal policy and lead to problems whose optimal solution cannot be analytically characterized.
Our goal is to study this fundamental sequential search process when partial information acquisition is possible. 
Specifically, we strive to answer the following questions: How should partial opening and full opening be prioritized? Could threshold-based policies, in the same vein as those defined in the standard Pandora's box problem, describe the optimal policy in settings with sequential inspections?  Beyond the characterization of optimal policies, we are also interested in developing simple and intuitive policies with provably good performance guarantees.

\subsection{Results and contributions}
We summarize below the main findings and contributions of our work.

\textit{Modeling.}
    We introduce a new model for sequential search by an agent, which incorporates sequential information acquisition into stochastic optimization problems. 
    Such information acquisition process can be represented by a directed acyclic graph describing the information states of each box (see Figure~\ref{fig:states_diagram}).
    Our graph layout is unique in that boxes can be inspected repeatedly and there is more than one way to inspect boxes. 
    To our knowledge, this generalized graph layout has not been previously studied.
    Moreover, it raises a fundamental technical challenge, which has been a long-lasting open question in the literature (we further discuss this in Section~\ref{sec:literature}).
    
    \textit{Hardness.} We prove that computing optimal policies for PSI is NP-hard, contrary to the standard Pandora's box problem. The proof is based on a reduction defined using PSI instances that consist of boxes with binary prizes and where partial opening indicates one of two types.

    \textit{Characterization of an optimal policy.} 
    Despite the hardness result, we identify threshold-based rules that almost fully describe the optimal policies.  Similarly to the Pandora's box problem, one type of threshold is associated with decisions to fully open boxes. To capture the added value of partial inspections, we define two additional types of thresholds associated with decisions to partially open boxes.     
    We show that, in most cases, the optimal actions can be prescribed using the newly defined thresholds. In particular, we characterize the optimal stopping criterion (i.e., when to stop searching and collect the best prize presently available), and identify sufficient conditions under which it is optimal to fully or partially open a box. 
    We identify special non-trivial cases where our characterization fully describes an optimal policy, and show that a simple policy, which is defined using these thresholds, is asymptotically optimal. 
    Numerical experiments show that optimal actions are identified by our theoretical characterization in more than 92\% of all systems states, on average, and this percentage increases in the number of boxes of the problem instance.

    \textit{Simple robust policies.} Next, we focus on the class of \textit{committing policies}.\endnote{Our approach is inspired by the class of committing algorithms studied by~\cite{beyhaghi2019pandora}, with the notable difference that our setting involves sequential inspections.} These semi-adaptive policies specify a priori whether each box should be partially or fully opened the first time it is inspected. Interestingly, each committing policy induces a corresponding multi-armed bandit problem (MAB) in which the optimal indices coincide with our proposed thresholds. 
    By leveraging fundamental results on the adaptivity gap of stochastic submodular optimization problems, we argue 
    that the optimal a committing policy can be efficiently approximated within a ratio of $1-1/e$.

\textit{Numerical experiments.} We conduct extensive numerical experiments to evaluate the performance of various policies on a diverse set of synthetic instances and obtain additional insights about the optimal policy:

\begin{itemize}
    \item Solving the problem to optimality using exact dynamic programming methods is computationally prohibitive, even when the number of boxes is around 10, the prize distributions have a small support, and the number of types is small. These computational challenges warrant the development of simple suboptimal policies. 

    \item    The number of times an optimal policy exercises the partial opening option increases as (i) there are more boxes, (ii) the partial opening thresholds tend to be larger than the corresponding full opening thresholds, and (iii) the distributions of the prizes are similar across boxes. In contrast, when boxes tend to be different, much about the order by which boxes are inspected is determined without partial opening. 
    
    \item The newly introduced thresholds are key to the development of effective policies. Specifically, the class of committing policies, which utilize our thresholds, achieves a robust performance. In particular, the best committing policy is very close to the optimum, both in terms of average and worst-case performance. In line with the asymptotic analysis, we observe that an intuitive threshold-based committing policy\endnote{Here, the policy commits to partial opening vs. full opening simply by comparing the corresponding thresholds for each box. } is near-optimal when the number of boxes is sufficiently large.

\end{itemize}

Our proofs combine several methods used in recent literature, including sample path coupling and interchange arguments \citep{attias2017stochastic}, reductions to submodular maximization problems \citep{beyhaghi2019pandora}, and results based on the information relaxation duality approach \citep{balseiro2019approximations}.
In this context, our main technical contribution lies in analyzing threshold-based rules using coupling arguments. We also reveal an unnoticed connection between approximation methods independently proposed in the literature. Specifically, we show that the cost accounting technique of \cite{beyhaghi2019pandora} amounts to a relaxation of Whittle's integral, a more general upper bound for multi-armed bandit superprocesses.

Overall, our results suggest that generalized Weitzman's thresholds play an important role in characterizing  both optimal and near-optimal inspection and selection decisions when the decision-maker can sequentially inspect boxes. This aspect of selection processes has not been thoroughly studied in previous literature, as we discuss next.

\subsection{Related Work} \label{sec:literature}

Our work relates to several streams of literature in economics, operations research, and computer science. 
We divide them imperfectly into three parts: generalization of the standard problem, Markovian MAB problems, and applications of sequential search.

\textit{The Pandora's box problem and its variants.} 
    The seminal work of \cite{weitzman1979optimal} led to substantial research on more general and realistic settings. 
    For example, \cite{vishwanath1992parallel} consider a generalization with parallel searches, where Pandora can open several boxes simultaneously at a discounted cost. The authors identify distributional assumptions under which it is optimal to open boxes according to Weitzman's thresholds.    
    \cite{olszewski2015more} examine a generalization where the objective function depends on all discovered prizes.\endnote{In the standard Pandora's box problem, the objective function is simply the maximum observed prize.}  They show that if a threshold-based rule is unconditionally optimal, then the problem is reducible to the standard Pandora's box setting.
    \cite{boodaghians2020pandora} study a variant of the Pandora's box problem in which precedence constraints are imposed on the inspections of boxes. They show that when the precedence constraints are tree-shaped, the optimal policy can still be described as a threshold-based rule.
    {  
    \cite{gibbard2022model} extend the Pandora's box problem by assuming that the prize is additive and that a two-step sequential inspection reveals the exact value of each part. Conceptually, such a model studies how information acquisition should be prioritized when the acquisition process can be broken into multiple steps. The model is a special case of the MAB problem \citep{bertsimas1996conservation} that inherits the indexability structure of the optimal policy. 
    In contrast, our model attempts to capture a different tradeoff between two information acquisition methods (a cheaper and less precise versus a more expensive and accurate), a characteristic which we show to be challenging analytically and computationally.
    }
    
    Recently, there has been a growing interest in non-obligatory inspections~\citep{attias2017stochastic, doval2018whether,beyhaghi2019pandora} where Pandora may select any closed box without inspecting it (in which case the prize is random at selection). Building on a clever cost accounting method, \cite{beyhaghi2019pandora} show that this setting is a special case of maximizing a stochastic non-negative submodular function under a matroid constraint, a breakthrough that paved the way to the design of approximation algorithms. Our algorithmic results in Section~\ref{sec:apx} draw from this line of reasoning. Closer to our work, \cite{beyhaghi2019approximately} studies a setting in which Pandora can choose from multiple opening actions for each box. However, unlike our work, the search process in that scenario does not feature the notion of sequential inspections. Therefore, such a model does not capture the key tradeoff which we aim to study  between reducing uncertainty significantly at a higher cost, and reducing it only partially at a lower cost with an option to further reduce the uncertainty later at an overall higher cost. This very tradeoff is present in the applications that motivate our work.   
    Finally, we note the work of \cite{clarkson2020fast} that study a problem where an object is hidden at one of several locations according to some known distribution. The goal is to find it in the minimum expected number of attempts using two modes of inspection, fast or slow. This can be seen as a variant of the Pandora's box problem, where the reward is binary, and boxes are correlated and can be inspected indefinitely.

\textit{Markovian multi-armed bandits.}
     From a technical perspective, the standard Pandora's box problem is a special case of the Markovian MAB \citep{Gittins1979Bandit}, in which an agent sequentially activates (pulls) one out of many bandits (arms). Arms correspond to independent Markov reward processes, which generate a reward and transition to a new state upon activation. The optimal solution admits an index structure, known as the Gittins index\endnote{Gittins indices and thresholds in Weitzman's policy are equivalent in the special case of the Pandora's box problem \citep[see, e.g., an early version of ][]{olszewski2015more}.}, where every state of every bandit is associated with an index, and it is always optimal to activate the arm whose current state's index is maximal. However, the optimality of the Gittins index breaks under small variations of the model assumptions, such as correlated arms \citep{pandey2007multi}, switching costs      \citep{asawa1996multi}, 
     or delayed feedback \citep{eick1988gittins}. We similarly observe that index policies are  suboptimal in our setting (see Example \ref{example:not_indexable} in Section~\ref{subsec:motivating_examples}).

    The problem considered in our paper can be  viewed as a special case of MAB superprocesses \citep{gittins2011multi}, in which there exist multiple ways to pull each arm. That is, after an arm is selected, one has to choose the mode of activation, which affects transitions and rewards. General knowledge about this class of problems is rather limited. 
    Unlike the standard MAB problem, MAB superprocesses need not admit an optimal index policy.
    Much of the existing literature explores special cases where index policies are ``unambiguously'' optimal~\citep{whittle1980multi}, as well as heuristic solution methods and upper bounds~\citep{brown2013optimal,balseiro2019approximations}. 
    The fact that we identify that our problem is NP-Hard suggests that a complete characterization of the optimal policy is not possible, not even through the MAB framework. 
    
\textit{Sequential search and applications.}
    The stylized sequential search process introduced by \cite{weitzman1979optimal} has been widely used in economics, marketing, and operations research.
    This includes empirical work~\citep{de2012testing, ursu2018power}, behavioral research~\citep{gabaix2006costly}, and economic models that use the Pandora’s box as a discrete-choice framework~\citep{miller1984job, robert1993informative, choi2018consumer, derakhshan2018product, chu2020position, immorlica2020information}.
    Our work offers a richer modeling structure to represent decision-making processes where alternatives can be inspected in stages with varying costs, although our current contribution remains theoretical.
    
    We note in passing about the existence of a literature on sequential search problems modeled using a Bayesian framework where there is a single box that can be sampled repeatedly for a fee and the decision maker tries to decide when to stop sampling with the goal of maximizing the expected value of the highest observed sample minus the sampling costs \cite[see, e.g., the recent work of][and the references therein]{baucells2023search}. Although the idea of sequential inspection is similar to our work, the key characteristics of our model, a tradeoff between cheap and costly inspections across multiple alternatives, are not captured by this literature.


\section{Problem Formulation}   \label{sec:formulation}

Pandora (the decision-maker) is given $N \in \mathbb{N}^*§$ boxes containing random prizes. Each box $i \in [N]$ is associated with a priori unknown prize $V_i$ and type $T_i$  drawn from a known joint probability distribution $\mathcal{D}_i$. We assume that $V_i$ is either a continuous or discrete scalar random variable and that $T_i$ is a discrete scalar random variable with support $\Gamma_i$.\endnote{Throughout, we use capital letters to denote random variables and lowercase letters to denote their realizations.}
We assume that the support of $\mathcal{D}_i$ is contained in $\mathbb{R}^+  \times \Gamma_i$, and that the random variables $(V_1,T_1),\ldots,(V_N,T_N)$ are mutually independent (for each box $i$, the random variables $V_i$ and $T_i$  are dependent; otherwise the type would not be informative about the prize).
Pandora can open a box in one of two inspection modes: full opening (F-opening) and partial opening (P-opening).
The realization of type $T_i$ is revealed immediately by P-opening box $i \in [N]$, while the corresponding F-opening action reveals the realization of $(V_i,T_i)$.
These opening actions incur constant costs, respectively denoted by $c_i^F$ and $c_i^P$, which are also known to Pandora.

\begin{figure}[htbp]
  \centering
  \includegraphics[width=10cm]{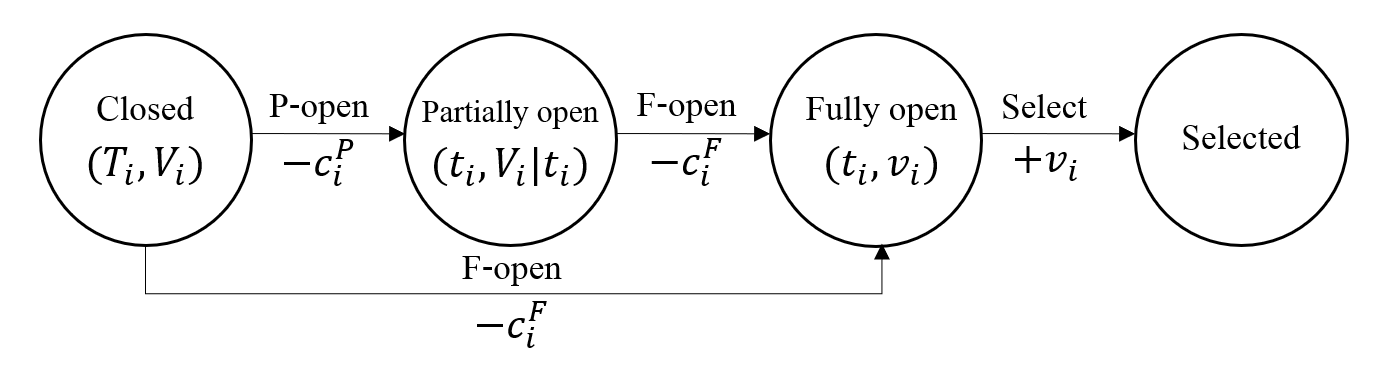}
  \caption{Illustration of the dynamics of a single box.}
  \label{fig:states_diagram}
\end{figure}

Figure~\ref{fig:states_diagram} illustrates the state evolution of each box following various actions.
We call the initial state of a box \textit{closed}; here, the realizations of $V_i$ and $T_i$ are unknown.
 P-opening a closed box changes its state to \textit{partially open}. In this case, Pandora observes the realization of $T_i$, which we denote by $t_i \in \Gamma_i$, and going forward, she updates her belief about $V_i$ to the conditional distribution $\operatorname{Pr}[V_i \in \cdot \mid T_i = t_i]$. We sometimes refer to this conditional distribution as $V_i  \mid  t_i$ for short. We say that a box is \textit{fully open} once F-opening takes place. Here, Pandora observes the realizations of $V_i$ and $T_i$, which we denote by $v_i, t_i$, respectively.
Lastly, the problem terminates when a fully open box is selected.
We note that full opening is a prerequisite for selection in a way similar to the standard Pandora's box problem.
The state of the system is jointly described by the collective individual state of all boxes.

An adaptive policy in our generalized Pandora's box problem is a mapping between system states and feasible actions. There are three types of  feasible actions: (1) P-opening a closed box, (2) F-opening a closed or partially open box, and (3) selecting a fully open box.
Alternatively, this can be seen as a cascade process of (1) deciding whether to select a box (stop) or keep inspecting (continue), (2) choosing a box to inspect, and (3) deciding on a box-opening mode. Our goal is to develop a policy that maximizes the expected profit, defined as the selected prize minus the box-opening costs.


\subsection{Dynamic programming (DP) formulation}
The problem naturally lends itself to a DP formulation. While the state space is exponentially large (and therefore, the DP cannot be approached using standard solution methods), we exploit this formulation to derive structural properties of the optimal policy.

\textit{State space.}
Each state of the DP can be described by a tuple $(\mathcal{C}, \mathcal{P}, y)$, where $\mathcal{C} \subseteq [N]$ is the set of closed boxes,
$\mathcal{P}$ is formed by all pairs $(i,t_i)$, describing each P-opened box $i$ and its corresponding type $t_i$ (revealed by P-opening), and
$y$ is the maximal prize found so far (it is easy to see that all fully open boxes whose prize is smaller than $y$ can be discarded, since at most one prize can be claimed).
We observe that the total number of states is at least $\prod_{i=1}^{N}\left( 1+ \mid \Gamma_i \mid  \right)$, which is exponentially large.

\textit{Actions.}
At each state $(\mathcal{C}, \mathcal{P}, y)$, we can fully open a closed or partially open box,  partially open a closed box, or select a fully open box and stop.
After F-opening a box $i \in \mathcal{C}$ and observing the realization $v_i$ of its prize $V_i$, we transition to state $(\mathcal{C} \setminus \{i\}, \mathcal{P}, \max \{y, v_i\})$. 
Hereinafter, we slightly abuse the notation by using $i \in \mathcal{P} $ in place of $(i,t_i) \in  \mathcal{P} $.
If we choose to F-open a box $i \in \mathcal{P}$, we transition to state $(\mathcal{C}, \mathcal{P} \setminus \{(i,t_i)\}, \max\{y, v_i \})$, where $v_i$ is the realization of $V_i|t_i$ revealed by our action.
Likewise, if we choose to P-open a box $i \in \mathcal{C}$, we transition to state $(\mathcal{C} \setminus \{i\}, \mathcal{P} \cup  \{(i,t_i)\}, y)$, where  $t_i$ is the type realization of $T_i$ revealed by P-opening box $i$.
Finally, when we select the fully open box whose value is equal to $y$, the decision process terminates.

\textit{Bellman's Equation.}
Let $J(\mathcal{C}, \mathcal{P}, y)$ denote the value function in state $(\mathcal{C}, \mathcal{P}, y)$ under an optimal policy.
Then, $J(\mathcal{C}, \mathcal{P}, y)$ satisfies the Bellman's Equation:
\begin{eqnarray}
J(\mathcal{C}, \mathcal{P}, y) &=& \max \left\{
    \begin{array}{ll}
    y    & \text{select (stop)}     \\
    -c_i^F+ \mathbb{E}_{V_i} \left[J(\mathcal{C} \setminus \{i\}, \mathcal{P}, \max\{V_i, y  \}   ) \right]   & \text{F-open box }    i \in \mathcal{C}    \\
    -c_i^F+   \mathbb{E}_{V_i \mid T_i=t_i} \left[J(\mathcal{C}, \mathcal{P} \setminus \{(i,t_i)\}, \max\{V_i, y  \}   ) \right]  &\text{F-open box }  i \in \mathcal{P}    \\
    -c_i^P+ \mathbb{E}_{T_i} \left[J  \left(\mathcal{C} \setminus  \{i\}, \mathcal{P} \cup \{ (i,T_i)\} , y \right)  \right] & \text{P-open box }  i \in \mathcal{C}
      \end{array}
      \right. \label{eq:Bellman_Equation}
      \\
J(\emptyset, \emptyset, y) &=& y.   \nonumber
\end{eqnarray}
In what follows, we use the shorthand $s_0 = ([N],\emptyset,0)$ to designate the initial system state.

\textit{Policies.}
A feasible policy  $\pi$  is a mapping from each state to a corresponding action, which is selected amongst all those that are feasible in the state in question.  For simplicity, we assume that there exists a unique optimal policy that satisfies Bellman's Equation.

\subsection{Illustrative examples} \label{subsec:motivating_examples}

To provide some intuition about why P-opening could be beneficial, consider the following simple examples.
\begin{example} 
There is a large number of i.i.d. boxes; each contains a prize of 110 with probability $\frac{1}{2}$ and zero otherwise. The box-opening costs are $c_i^F = 100$ and $c_i^P = \epsilon $, and P-opening acquires the same information as F-opening (nevertheless, F-opening is still required  to collect the prize). It is easy to see that given only the  F-opening inspection mode, it is not worthwhile to F-open any box and the expected profit of any policy that opens boxes is negative. However, given the option to P-open, one could identify a box that contains the prize prior to F-opening it. In that case, the expected profit is $110-100=10$  minus $\epsilon$ times the expected number of opened boxes, which is equal to two.
\end{example}

This example shows how the option to P-open could transform a problem where opening boxes is not worthwhile into one in which it is. Hence, the optimality gap for the naive policy that ignores P-opening (i.e., Weitzman's policy, which is a natural candidate for Pandora's box-like problems) is unbounded.

\begin{example}  \label{example:not_indexable}
Consider the state $(\{ 1 \}, \emptyset, y)$ where there is one box left with two types: `exceptional' and `average,' which can be revealed by P-opening. 
Box 1 is `exceptional' with probability $\delta = 0.01$ and is `average' with probability $1-\delta$. 
If the box is exceptional,  the prize is known to be $2$ almost surely, but if it is average, the prize is uniformly distributed in $[0,1]$. 
Assuming that the F-opening cost is $c_1^F=\frac{1}{3}$ and the P-opening cost is $c_1^P=\frac{1}{3} \delta $, 
it can be easily seen that when $y=0$, both inspection modes are preferred over stopping (which yields $0$). Moreover, F-opening dominates P-opening because F-opening is inevitable in this case. 
However, when $y=1$, F-opening would be no longer worthwhile compared to stopping, because the cost $c_1^F = \frac{1}{3}$ exceeds the expected gain $ \delta \cdot (2 - y) = \delta$. 
By contrast, P-opening (followed by the F-opening) of an exceptional box is advantageous because the expected cost is $c_1^P + \delta c_1^F = \frac{1}{3}\delta + \frac{1}{3}\delta =\frac{2}{3} \delta$, whereas the expected gain is $\delta$.
\end{example}

Intuitively, in the context of job hiring, the instance captures a candidate who is likely `average' but has a small chance of being `exceptional.' A low-cost online interview can reveal this. If no good candidate has been found ($y = 0$), it is optimal to proceed directly with an onsite interview. However, if a strong candidate has already been identified ($y = 1$), the onsite visit should be reserved for an `exceptional' applicant, identified via a relatively short online screening.

Technically, the example shows that the optimal policy is not indexable. If it were, each action would receive a fixed index, and the best action---corresponding to the highest index---would be independent of $y$, which is not the case in Example~\ref{example:not_indexable}. In MAB terms, the optimal activation of one arm depends on the state of others. This illustrates the added complexity of bandit superprocesses, which do not inherit the indexability of classic MABs (see also Section~\ref{sec:literature}).

\section{Characterizations of Optimal Policies}  \label{sec:characterization}

We begin by studying structural properties of optimal policies, which can be exploited to identify optimal actions using well-designed {\em opening thresholds}. In Section~\ref{sec:stopping}, we show that it is optimal to stop when the largest available prize exceeds all opening thresholds.
In Section~\ref{sec:F_largest}, we establish that whenever the largest opening threshold corresponds to an F-opening action, it is optimal to F-open the corresponding box.
In contrast, we show that the complementary case, in which the largest opening threshold corresponds to a P-opening action, is more complex to analyze, as there exists no threshold-based decision rule that is universally optimal. Nevertheless, in Section~\ref{sec:P_largest}, we identify sufficient conditions for which P-opening the box with the largest threshold is optimal.
In Section~\ref{sec:optimal_binary_prize_iid}, we provide a complete characterization of the optimal policy for a special case of a PSI with binary prizes where partial opening  reveals two types.

\subsection{Opening thresholds and optimal stopping rule} \label{sec:stopping}
We start by defining \textit{opening thresholds}, a type of index associated with each potential action. 
Although there may not exist an optimal index policy (as shown in Section~\ref{subsec:motivating_examples}), we show that such indices explain to a large degree the optimal policy. 
While applicable to general system states, the thresholds can be most easily illustrated by considering the special case where there is a single box that is not fully open. This setting also allows us to express the optimal stopping criterion using our thresholds. 
We let $x^+ = \max\{0,x\}$ for every $x\in \mathbb{R}$.

\begin{definition}[Thresholds]\label{def:testing_threshold}
 For every box $i \in [N]$, we define the opening thresholds:
\begin{itemize}
    \item If box $i$ is closed, the F-threshold $\sigma^{F}_i$ is a solution to
    \begin{equation}  \label{eq:def_F_threshold}
      c_i^F = \mathbb{E}_{V_i}\Big[(V_i-\sigma^{F}_i)^+\Big].
    \end{equation}

    \item If box $i$ is partially opened with type $t_i$, the (conditional) F-threshold $\sigma^{F \mid t_i}_i$ is a solution to
    \begin{equation} \label{eq:def_F_threshold_partial}
      c_i^F = \mathbb{E}_{V_i \mid T_i=t_i}\Big[(V_i - \sigma^{F \mid t_i}_i)^+\Big]. 
    \end{equation}

    \item If box $i$ is closed, the P-threshold $\sigma^{P}_i$ is a solution to
\begin{equation} \label{eq:def_P_threshold}
  c_i^P = \mathbb{E}_{T_i} \Big[\max\{0, -c_i^F + \mathbb{E}_{V_i \mid T_i} \big[ (V_i-\sigma^{P}_i)^+ \big]\}\Big].  
\end{equation}

    \item If box $i$ is closed, the FP-threshold $\sigma^{F/P}_i$ is a solution to
\begin{equation}
c_i^P = \mathbb{E}_{T_i} \Big[\max\{0, c_i^F-\mathbb{E}_{V_i \mid T_i} \big[ (V_i-\sigma^{F/P}_i)^+ \big]\}\Big].  
\end{equation}
Observe that the threshold is well-defined even in the case that $c_i^P \geq c_i^F $, in which case P-opening is not beneficial.
\end{itemize}
\end{definition}

Assuming the prize distributions are non-atomic, it is not difficult to see that the above thresholds $\sigma_i^F$, $\sigma_i^{F \mid t_i}$, $\sigma_i^P$, and $\sigma_i^{F/P}$ exist and are uniquely defined.
Moreover, there is a natural ordering between threshold values:

\begin{observation}[Thresholds order]\quad\label{obs:thresholds_order}
    \begin{enumerate}
    \item For any box $i$, either $ \sigma^{F/P}_i = \sigma^{F}_i = \sigma^{P}_i$,  $ \sigma^{F/P}_i < \sigma^{F}_i \leq \sigma^{P}_i$, or $ \sigma^{P}_i \leq \sigma^{F}_i < \sigma^{F/P}_i$; 
    \item For any box $i$, we have $\underset{t_i \in \Gamma_i}{\min} ~ \sigma_{i}^{F|t_i} \leq \sigma_i^{F/P}$ and $\underset{t_i \in \Gamma_i}{\max} ~\sigma_{i}^{F|t_i}  \geq \sigma_i^P$. 
    \end{enumerate}
\end{observation}

\paragraph{Economic interpretation.}
The set of F-thresholds $\{\sigma_i^{F}\}$ in Equation~\eqref{eq:def_F_threshold} describes the optimal index policy for the standard Pandora's box problem \citep{weitzman1979optimal}. 
The conventional economic interpretation for this opening threshold is that of a fair strike price of a hypothetical option contract. That is, if $c_i^F$  is the cost of a call option to buy prize $V_i$ with strike price $\sigma_i^{F}$, then $\sigma_i^{F}$ is the ``fair'' strike price that equalizes the cost and the expected profit of buying this option. 
 {This intuition carries over to our problem where inspections are sequential using the notion of a compound option or, informally, a ``two-stage call option.'' This analogy is developed in Appendix~\ref{apx:economic_interpretation}.}

\paragraph{Single-box setting and optimal stopping rule.}
We justify our definition of the opening thresholds through their ability to express optimal policies in single closed-box settings, and to prescribe an optimal stopping rule in  general.
Namely, the \textit{single closed-box setting} is the collection of states $({\cal C},{\cal P},y)$  where there is only one closed box that can be inspected, i.e., $\mathcal{C}=\{ i\}$  for some box $i\in [N]$ and $\mathcal{P}=\emptyset$.
The difference between the expected profit from F-opening and from stopping is exactly
\begin{equation} \label{eq:optimal_stopping_F}
    -c_i^F + \mathbb{E}_{}\Big[\max \left \{ V_i, y \right \}\Big] - y  = -c_i^F + \mathbb{E}_{}\Big[ \max \{ V_i - y, 0  \} \Big] = - \mathbb{E}_{}\Big[ \max \{ V_i - \sigma_i^F, 0  \} \Big] +  \mathbb{E}_{}\Big[ \max \{ V_i - y, 0  \} \Big] \ . 
\end{equation}

The F-threshold can be therefore interpreted as a break-even point for the value of $y$ under which F-opening is preferred to stopping, and over which stopping is preferred.
Similarly, in this single-box case, the difference between the expected profit from P-opening (and following the optimal action thereafter) and from stopping is
\begin{eqnarray}
 & &  - c_i^P + \mathbb{E}_{T_i} \Big[\max\{ y , -c_i^F + \mathbb{E}_{V_i \mid T_i} \big[ \max \left\{ V_i , y \right\} \big]\} \Big] - y \nonumber \\
 &=&  - c_i^P + \mathbb{E}_{T_i} \Big[\max\{0, -c_i^F + \mathbb{E}_{V_i \mid T_i} \big[ \max \left\{ V_i - y , 0 \right\} \big]\} \Big] \nonumber \\ 
 &=& - \mathbb{E}_{T_i} \Big[\max\{0, -c_i^F + \mathbb{E}_{V_i \mid T_i} \big[ (V_i-\sigma^{P}_i)^+ \big]\}\Big] + \mathbb{E}_{T_i} \Big[\max\{0, -c_i^F + \mathbb{E}_{V_i \mid T_i} \big[ \max \left\{ V_i - y , 0 \right\} \big]\} \Big].
 \label{eq:optimal_stopping_P}  
\end{eqnarray} 

Similarly, the P-threshold can be interpreted as a break-even point for the value of $y$ under which P-opening is preferred to stopping, and over which stopping is preferred.

Regarding the FP-threshold $\sigma^{F/P}_i$, we note that the difference between the expected profits from the optimal course of actions starting from P-opening box $i$ and that from the optimal course of actions starting from F-opening box $i$ is 
\begin{eqnarray}
 & & \left( - c_i^P + \mathbb{E}_{T_i} \Big[\max\left\{y, -c_i^F + \mathbb{E}_{V_i \mid T_i} \big[ \max \left\{ V_i , y \right\} \big]\right\} \Big]  \right) - \left(-c_i^F + \mathbb{E}_{}\Big[\max \left\{ V_i, y \right\}\Big]\right)    \nonumber\\
 &&=  - c_i^P + \mathbb{E}_{T_i} \Big[\max\{0, c_i^F-\mathbb{E}_{V_i \mid T_i} \big[ (V_i-y)^+ \big]\}\Big] \ . \label{eq:optimal_stopping_FP}
\end{eqnarray}

One can verify that the threshold $\sigma_{i}^{F/P}$ is a break-even point on the value of $y$ that determines whether P-opening or F-opening achieves a higher expected profit. 
Intuitively, in the single-box case, the definition for the set of thresholds $\sigma_i^F$, $\sigma_i^{F \mid t_i}$, $\sigma_i^P$, and $\sigma_i^{F/P}$ exploits the single-crossing property between pairs of action-value functions. In the more general case, the definition can be viewed as using a myopic approximation for the action-value functions that assumes that only one box may be opened.

Based on the above discussion, we construct a natural \textit{threshold-based policy} for the single closed box setting. 
We assign the index $y$ to the action of stopping, the index $\sigma_i^P$ to the action of P-opening, and the index $ \sigma^{F}_{i}$ to the action of F-opening. 
Our policy prescribes which action to take by comparing their corresponding indices.
If the largest threshold is $y$,  then we stop and choose the largest prize at hand. Otherwise, if the largest threshold is $\sigma^F_i$, then we $F$-open box $i$ and choose the largest prize at hand. Finally, if the highest threshold is $\sigma^P_i$, we compare $y$ to $\sigma^{F/P}_i$: if $y\leq \sigma^{F/P}_i$, we F-open, otherwise we P-open box $i$. 
We refer the reader to Figure~\ref{fig:example_1} for a pictorial illustration; the policy is also summarized as Algorithm~\ref{algo:single-box} in Appendix~\ref{appendix:thm_stopping}.

The next theorem states the optimality of our threshold-based rules in the single close box setting. Moreover, in the general case (where there is more than one box), the same threshold-based logic yields an optimal stopping criterion.
For any event $A$, we let $\BFI\{A\}$ denote the indicator function that equals $1$ if $A$ occurs and $0$ otherwise.

We define the largest opening threshold $\sigma_M(\mathcal{C}, \mathcal{P})$:
\begin{equation} \label{eq:sigma_M_C_P}
    \sigma_M(\mathcal{C}, \mathcal{P})  =  \max \left\{  \underset{j \in \mathcal{C} }{\max}  \left\{  \sigma_j^F, \sigma_j^P \right\}, \underset{ j \in \mathcal{P}}{\max }  \left\{ \sigma_j^{F \mid t_j}  \right\}   \right\}. 
\end{equation}
We refer to the box attaining the largest threshold $\sigma_M(\mathcal{C}, \mathcal{P})$ as the {\em leading box}. 

\begin{theorem} \label{thm:myopic_stopping} 
The threshold-based policy is optimal in the single closed box setting $({\cal C},{\cal P},y) = (\{i\}, \emptyset, y)$, and it achieves the following expected profit:
$$
J\left( \{i\}, \emptyset, y \right) = \mathbb{E} \left[ \max \left\{ y,
 \BFI\{ y > \sigma_i^{F/P}  \}  \cdot \min\{V_i, \sigma_i^{F \mid T_i}, \sigma_i^P \} ,
  \BFI\{ y \leq \sigma_i^{F/P}  \} \cdot  \min\{V_i,  \sigma_i^F \}
  \right\} \right].
$$
In general, for every state $(\mathcal{C}, \mathcal{P}, y)$, it is optimal to stop if and only if
\begin{equation}\label{eq:myopic_stopping_condition}
  y \geq \sigma_M(\mathcal{C}, \mathcal{P}) .
\end{equation}

\end{theorem}
The formal proof of the stopping rule is based on a coupling argument, which is presented in Appendix~\ref{appendix:thm_stopping}. 
We argue that the stopping decision regarding a box (i.e., never to open this box in the future) only depends on $y$. Hence, the optimal stopping rule shown for the single box setting holds for the general case.

\subsection{Sufficient conditions for optimal F-opening} 
\label{sec:F_largest}

Having characterized the optimal stopping criterion, we now focus on states $(\mathcal{C}, \mathcal{P}, y)$ in which stopping is not optimal. 
Here, there are two possibilities: $\sigma_M(\mathcal{C}, \mathcal{P})$  corresponds either to an F-threshold or a P-threshold. The next theorem establishes that in the former case, it is optimal to F-open the leading box. This condition holds regardless of whether the leading box is closed or partially-opened.

\begin{theorem} \label{thm:just_strong}
There exists an optimal policy that immediately F-opens the leading box in every state $(\mathcal{C}, \mathcal{P}, y)$ where stopping is suboptimal and $\sigma_M (\mathcal{C}, \mathcal{P})$ is an F-threshold.
\end{theorem}
The proof of this result appears in Appendix~\ref{appendix:thm_just_strong}. 
The crux of it is constructing a policy that F-opens the leading box, and then showing that this policy achieves a larger expected profit than any alternative one using interchange and coupling arguments. Compared to the standard  Pandora's box problem, 
the main difficulty here is that P-opening introduces more courses of action to consider (e.g., P-open the leading box or alternative ones). Hence, our proof carefully distinguishes between different cases, where the leading box is P-opened and those where it is closed, and proceeds by induction. We need to quantify the value of information gathered by P-opening in order to eventually rule out this option. To this end, we upper bound this quantity using a probabilistic coupling, which exploits the dynamic programming tree locally.

We mention in passing that as part of the proof of the theorem we also show that boxes whose F-threshold is higher than their P-threshold should never be P-opened. The opposite direction does not hold, as our illustrative examples suggest.

\subsection{Sufficient conditions for optimal P-opening} \label{sec:P_largest}

We consider the alternative case where $\sigma_M (\mathcal{C}, \mathcal{P})$ is a P-threshold.  
Not surprisingly, in this case, there is no straightforward decision rule that prescribes the optimal action as a function of the opening thresholds. This was illustrated in Section~\ref{subsec:motivating_examples}, even in the simplest case of a single closed-box where the optimal inspection mode could well be an F-opening or P-opening action, depending on the value of $y$ (we provide another example in Appendix~\ref{subsec:example_many_boxes} that shows the intricacy of optimal policies).  
We identify a sufficient condition under which P-opening the leading box is optimal.

\begin{definition} \label{def:well_class}
   Given a state $(\mathcal{C}, \mathcal{P}, y)$ and a closed box $i \in \mathcal{C}$, we define $
  \sigma_{-i}  = \sigma_M (\mathcal{C}  \setminus \{ i \}, \mathcal{P})  
  $. Suppose $\sigma_i^P > \sigma_i^F $. We say that box $i $ is \textit{well-classified} if $y > \sigma_i^{F/P}$ and for every type $t_i \in \Gamma_i$, one of the following conditions holds: (1) $\sigma_i^{F \mid t_i} \geq \sigma_{-i}$ or (2) $y > \sigma_i^{F \mid t_i}$.
\end{definition}
To unpack this notion, we remark that, when dealing with a well-classified box $i$, there is no ambiguity about the action that follows P-opening.
Put simply, upon P-opening, the box can be classified as  ``good'' or ``bad.''
After observing its type $t_i \in \Gamma_i$, our definition delineates two cases. In
case (1), we infer from Theorem~\ref{thm:just_strong} that it is optimal to immediately F-open the same box (good case), whereas in case (2), we infer from Theorem~\ref{thm:myopic_stopping} that box $i$ will never be F-opened (bad case). Importantly, our notions of ``good'' and ``bad'' types are preserved in later stages of the decision process, regardless of which actions are taken.
As a result, for a well-classified box, it is relatively easier to quantify the  value of information afforded by the P-opening. 
\begin{theorem} \label{thm:just_weak_closed}
There exists an optimal policy that P-opens the leading box in every state $(\mathcal{C}, \mathcal{P}, y)$ where stopping is not optimal, $\sigma_M (\mathcal{C}, \mathcal{P})$ is a P-threshold, and the leading box is well-classified.
\end{theorem}
The proof, presented in Appendix~\ref{appendix:thm_just_weak_closed}, is based on induction using interchange arguments. 
Similarly to Theorem~\ref{thm:just_strong}, the analysis requires a probabilistic coupling of different cases.

Unfortunately, combining Theorems~\ref{thm:myopic_stopping}, \ref{thm:just_strong}, and \ref{thm:just_weak_closed} does not  determine an optimal policy in every  state of the DP. Unlike the optimal stopping condition, Theorems~\ref{thm:just_strong}-\ref{thm:just_weak_closed} provide sufficient but not necessary conditions. Nevertheless, in Section~\ref{sec:numerical_jay}, we numerically solve synthetic instances and show that these theorems capture optimal decisions in most system states. Theorem~\ref{thm:just_weak_closed} is relatively more useful in ``later stages'' of the decision process, where $y$ tends to be larger, and condition (2) of Definition~\ref{def:well_class} becomes easier to satisfy. 
Moreover, our theorems speed up the DP computation. When the conditions of Theorems~\ref{thm:myopic_stopping}--\ref{thm:just_weak_closed} hold, there is no need to apply the DP recursion.

That said, we identify a special case in which Theorems~\ref{thm:myopic_stopping}--\ref{thm:just_weak_closed} fully characterize an optimal policy. This happens when  every box $i$ such that $\sigma_i^P > \sigma_i^F$, has only two types (i.e., $ |\Gamma_i| = 2 $), and the FP-threshold $\sigma_i^{F/P}$ is negative (i.e., $\sigma_i^{F/P} < 0$).
To see this, note that whenever stopping is suboptimal, we have two cases: If the largest opening threshold is an F-threshold, it is optimal to F-open the leading box (Theorem~\ref{thm:just_strong}). If the largest threshold is a P-threshold, then the leading box $i$ is well-classified. Indeed, in view of the discussion that follows Definition~\ref{def:testing_threshold}, we know that $ \underset{t_i \in \Gamma_i}{\min} ~ \sigma_{i}^{F|t_i} \leq \sigma_i^{F/P} < 0 \leq y  $ and $\underset{t_i \in \Gamma_i}{\max} ~ \sigma_{i}^{F|t_i} \geq \sigma_i^P > \sigma_{-i}$. Intuitively, this setting captures scenarios in which alternatives can be either very good or very bad. If partial opening is cheap enough, it should always be exercised. 
We summarize this fact in the next corollary. 
\begin{corollary} \label{cor:support_2}
Suppose that every box $i \in [N]$ such that $\sigma_i^P > \sigma_i^F$ satisfies $ |\Gamma_i| = 2 $ and $\sigma_i^{F/P} < 0$. Then, Theorems~\ref{thm:myopic_stopping}, \ref{thm:just_strong} and \ref{thm:just_weak_closed} fully determine an optimal policy. 
\end{corollary}

\subsection{An optimal policy for a problem with binary prizes} \label{sec:optimal_binary_prize_iid}

Next, we study a special case of PSIs whereby boxes are statistically identical and independent, contain binary Bernoulli prizes, and belong to one of two types---``good'' or ``bad.'' We refer to this problem as PSI-B2I (Bernoulli, 2-type, identical and independent).
This setting applies to situations where a decision maker is searching for a satisfactory alternative among options that are initially perceived as equivalent. Partial inspection reveals whether an alternative is promising (``good'') or has low potential (``bad''), while full inspection discloses the exact outcome. 
Several variants of this problem in which the decision maker cannot inspect boxes sequentially have been studied in the literature \citep[see, e.g.][and the references therein]{doval2018whether}.
Our setting is also different from \cite{clarkson2020fast} who assume that there is exactly one prize hidden in the boxes and that opening can be done repeatedly (which does not fit our motivating applications).

In striking contrast to the problem class with binary prizes which is shown to be NP-hard (Section~\ref{sec:np_hardness_mainbody}), the optimal policy to PSI-B2I admits a simple and intuitive structure. 
To formally describe it, we first remark that the equivalence of boxes implies that their opening thresholds are equal; hence, we will refer to them as $\sigma^F$ and $\sigma^P$.
If $\sigma^F \geq \sigma^P$, the optimal policy is fully characterized by Theorems~\ref{thm:myopic_stopping} and~\ref{thm:just_strong}: the optimal policy F-opens closed boxes until a prize is found.

In the difficult case $\sigma^P > \sigma^F$, we denote by $n_C, n_G, n_B$ the number of closed, good and bad boxes, respectively, in the current state. 
Our main result in this section shows that there exists a threshold $N_C$ on the number of closed boxes so that when $n_C > N_C$, an optimal policy exclusively P-opens closed boxes until stopping or reaching a state in which $n_C \leq N_C$, at which point it exclusively F-opens closed boxes. 
We note that the threshold $N_C$ is fixed and independent of the state. 
Moreover, an optimal policy immediately F-opens good-type boxes upon discovery, while postponing bad-type boxes until all closed boxes have been at least partially opened. 
There are exactly $N$ such policies---corresponding to each value of $N_C$---that can be easily enumerated and evaluated for the purpose of optimization.

The structure of the policy above captures the case $\sigma^P \leq \sigma^F$, using the convention $N_c=\infty$.  We also observe that the special case where $\sigma^F < \sigma^P$ and $\sigma^{F/P} \leq 0$ is subsumed by Theorem~\ref{thm:just_weak_closed} (see  Observation~\ref{obs:thresholds_order}), in which case $N_c=0$. Therefore, to ease our analysis, we make  \textbf{without loss of generality} the following simplifying assumption:
\begin{assumption}\label{assumption:b2i}
    The boxes of PSI-B2I satisfy: 
    $0<\sigma^{F/P}<\sigma^F < \sigma^P.$
\end{assumption}

\begin{theorem} \label{thm:iid_optimal}
    For a PSI-B2I instance satisfying Assumption~\ref{assumption:b2i}, there exists an optimal policy $\pi$ that satisfies the following properties: 
    \begin{enumerate}
    

        \item Policy $\pi$ assigns the highest priority to F-opening good-type boxes.

        \item Policy $\pi$ either discards all bad-type boxes, or  assigns the lowest priority to F-opening them.
        
        \item There exists a finite threshold $N_C$ such that policy $\pi$ prioritizes P-opening closed boxes over F-opening them iff the number of closed boxes $n_C > N_C$.

    \end{enumerate}
\end{theorem}

The proof can be found in Appendix~\ref{sec:proof_iid_optimal}. 
{ 
The theorem is presented as an algorithm in Appendix~\ref{sec:algo_psi_b2i}.
}

The structure of the optimal policy suggests that the relation between the opening thresholds is the primary factor that determines whether or not P-opening should be considered. 
If $\sigma^P > \sigma^F$ then P-opening should be considered to identify ``promising'' boxes. This option is exerted whenever there are sufficiently many a priori equivalent alternatives (as captured by the threshold rule $n_C > N_C$).
Intuitively, due to the uncertainty about finding good-type boxes, multiple boxes must be P-opened to obtain with high probability a satisfactory outcome. By contrast, when the total number of boxes is small, the decision-maker will most likely end up opening most of them and therefore P-opening is not beneficial. 

{ 
Observe that this policy satisfies the optimal characterization given in Theorems~\ref{thm:myopic_stopping}-\ref{thm:just_weak_closed}; 
indeed, the proof in Appendix~\ref{sec:proof_iid_optimal} directly builds on these properties.
Interestingly, although PSI-B2I is a special case, much of the structure exhibited by the optimal policy for this instance extends naturally to the broader class of PSI problems.
For example, it is useful to consider a class of policies that commit ex ante to fully or partially opening each closed box before execution; this notion of a committing policy proves useful in the general problem and is discussed in detail in Section~\ref{subsec:apx_policies}.
As another illustration, the instance highlights how an increase in the number of boxes makes partial opening more advantageous, a phenomenon that we further validate numerically in Section~\ref{sec:exploring_p_opening_main}. 
}

\section{Simple Approximately Optimal Policies}
\label{sec:apx}
We establish that computing optimal policies for PSI is NP-hard (Section~\ref{sec:np_hardness_mainbody}). 
We therefore turn our attention to the development and analysis of simple approximate solutions that utilize relaxations that approximate the value function. 
In Section~\ref{subsec:upper_bounds}, we propose two approximation methods: {\em Whittle's integral} and the {\em free-info relaxation}, which adapt tools from related literature. Interestingly, we show that Whittle's integral provides a tighter upper bound, 
and establish the asymptotic optimality of the index policy based on the opening thresholds. 
In general, we exploit these problem relaxations  in Section~\ref{subsec:apx_policies} to devise approximation algorithms, in particular achieving a $(1-1/e)$-approximation ratio, through a reduction to the maximization of stochastic submodular functions~\citep{asadpour2015maximizing,singla2018price,beyhaghi2019pandora}.

\subsection{Hardness of the problem}  \label{sec:np_hardness_mainbody}

\begin{theorem} \label{thm:hardness}
Computing optimal policies for PSI is NP-hard.
\end{theorem}

The proof strategy involves a reduction from a family of subset sum problems. Specifically, we construct a hard instance in which every box $i$ has two types $|\Gamma_i| = 2$:  ``good" and ``bad." 
The good type is sufficiently valuable that an optimal policy F-opens the box upon discovering it, and bad type boxes are discarded.
Hence, determining the optimal policy is essentially equivalent to deciding how to partition the $N$ boxes, i.e., deciding the opening mode for each box. 
We carefully design this PSI instance so that the partition exactly corresponds to the solution of the subset sum problem. The proof is intricate and appears in Appendix~\ref{sec:proof_thm_hardness}. 
In particular, formalizing the preceding argument requires controlling the encoding error due to the bit precision of the input.

It is interesting to note that the hard instances we constructed for the proof of Theorem~\ref{thm:hardness} are very close to the one subsumed by Corollary~\ref{cor:support_2}, for which we were able to provide a polynomial-time algorithm. Specifically, our family of hard instances satisfies that,  every box $i \in [N]$ has $\sigma_i^P > \sigma_i^F$ and  $ |\Gamma_i| = 2 $. The only difference, however, is that $\sigma_i^{F/P} > 0$, contrary to Corollary~\ref{cor:support_2}.  
This suggests that Corollary~\ref{cor:support_2} is ``tight'' in some sense.

Finally, our reduction from subset partition has some similarities with the NP-hardness proof of \cite{fu2022pandora} in a related computational setting; see Remark~\ref{remark:connections_to_fu} in Appendix~\ref{sec:proof_thm_hardness} where we discuss similarities and important differences in both reduction proofs.

\subsection{Relaxations and upper bounds} \label{subsec:upper_bounds}
A common method to solve high-dimensional DPs consists in approximating the value function \citep[see, e.g.,  Chapter 9 in][]{powell2007approximate}.
For example, effective policies are often obtained by solving the approximate Bellman's equation, where the true (unknown) value function is replaced by its approximate counterpart.
In what follows, we construct two natural upper bounds on $J({\cal C},{\cal P},y)$ and compare them analytically.

\textit{Whittle's integral.}
This is a known upper bound on the value function of MAB superprocesses \citep[see, e.g.,][]{brown2013optimal}. Consider a problem with a single non-fully-open box $i$ where the current maximum prize is $y$. 
Let $\omega_i(y)$ be the probability that the current maximum prize $y$ is eventually selected by an optimal policy. 
In other words, we can think of $\omega_i(y)$ as the probability that the effort of opening box $i$ is in vain, provided that only box $i$ is left.  
Given the characterization of the optimal policy for the single-box case in Theorem~\ref{thm:myopic_stopping}, it is not hard to see that, if box $i \in \mathcal{C}$,
$$
 \omega_i\left( y \right)= \BFI\left\{ y > \sigma_i^{F/P}  \right\} \operatorname{Pr} \left[ \min \left\{ V_i, \sigma_i^{F \mid T_i}, \sigma_i^P \right\} \leq y \right] + \BFI\left\{ y \leq \sigma_i^{F/P}  \right\} \operatorname{Pr} \left[  \min \left\{ V_i,  \sigma_i^F \right\} \leq y \right] \ .
$$
Otherwise, if box $i \in \mathcal{P}$,
$$
\omega_i\left( y \right)=  \operatorname{Pr} \left[ \min\{V_i, \sigma_i^{F \mid t_i} \} \leq y \mid T_i=t_i \right] \ .
$$ 
Consequently, Whittle's integral corresponds to  the following quantity:
\begin{equation} \label{eq:whittlesintegral}
    J^W( \mathcal{C}, \mathcal{P}, y ) = \sigma_M (\mathcal{C}, \mathcal{P}) - \int_{y}^{\sigma_M(\mathcal{C}, \mathcal{P})} \prod_{i\in {\cal C}\cup {\cal P}} \omega_i( u) du \ .
\end{equation}
We will establish in Lemma~\ref{thm:upper_bounds} that this function provides an upper bound on the $J( \mathcal{C}, \mathcal{P}, y)$. Conversely, when $y = \sigma_M(\mathcal{C}, \mathcal{P})$, we straightforwardly obtain  $J( \mathcal{C}, \mathcal{P}, y) \geq y \geq \sigma_M(\mathcal{C}, \mathcal{P}) \geq  J^W( \mathcal{C}, \mathcal{P}, y )$, implying that the upper bound~\eqref{eq:whittlesintegral} can be tight in certain states.

\paragraph{Free-info relaxation.} We now consider an alternative upper-bounding method that exploits the interpretation of indices as fair strike prices of two-stage call options. To this end, we introduce indicator random variables that describe the actions chosen by any given policy $\pi \in \Pi$ from state $({\cal C},{\cal P},y)$.\endnote{Recall that, in the state description, $y$ is the reward of a best F-opened box. Without loss of optimality, we restrict attention to policies that discard other $F$-opened boxes.} Let $\mathbb{S}_i^{\pi}, \widetilde{\mathbb{S}}_i^{\pi}$ be the binary random variables that indicate whether box $i$ is eventually selected without a partial opening and after a partial opening, respectively. (Note that we necessarily have $\mathbb{S}_i^{\pi} + \widetilde{\mathbb{S}}_i^{\pi} \leq 1$). Similarly, let $\mathbb{F}_i^{\pi},\widetilde{\mathbb{F}}_i^{\pi}$ be the binary random variables that indicate whether box $i$ is F-opened without any prior P-opening or after a P-opening, respectively. Finally, let $\mathbb{P}_i^{\pi}$ be the binary random variable that indicates whether box $i$ is P-opened.  While these decision variables depend on the policy $\pi$, for ease of notation, we sometimes suppress the superscript $\pi$. 
With this notation at hand, the expected value of policy $\pi$ in state $({\cal C},{\cal P},y)$ can be written as 
\begin{equation}
    J^{\pi} ({\cal C},{\cal P},y) =  
    \E{ \left. {\mathbb{S}}_0^{\pi} \cdot y + \sum_{i\in {\cal C}} \left(  \mathbb{S}_i^{\pi} V_i - \mathbb{F}_i^{\pi}  c_i^F - \mathbb{P}_i^{\pi} c_i^P \right) + \sum_{i\in {\cal C} \cup {\cal P} } \left(  \widetilde{\mathbb{S}}_i^{\pi} V_i -  \widetilde{\mathbb{F}}_i^{\pi}  c_i^F \right) \right| ({\cal C},{\cal P},y) }  \ ,
\end{equation}
where, by a slight abuse of notation, the conditional expectation indicates that policy $\pi$ is executed starting from state $({\cal C},{\cal P},y)$, and thus, we only accrue the  costs of future opening decisions and the rewards of future selections decisions. 
Hence, the Pandora's box problem with sequential inspections can be formulated as the following mathematical program:
\begin{equation}  \label{eq:obj}
   J ({\cal C},{\cal P},y) = \max_{\pi \in \Pi} ~ J^{\pi} ({\cal C},{\cal P},y)\ . 
\end{equation}

The free-info relaxation considers an alternative setting, in which there is no cost of P-opening and F-opening, but instead, the rewards of the boxes are capped according to the threshold values. The notion of a feasible policy does not change in this new formulation, only the cost accounting is modified. Specifically, for every box $i \in \mathcal{C}$, we define the capped value of F-opening to be $K_i = \min\{V_i, \sigma^{F}_i\}$ and the capped value of P-opening to be $\widetilde{K}_i = \min \{ V_i, \sigma^{F \mid T_i}_i, \sigma^{P}_i\} $. Similarly, for every box $i\in {\cal P}$, we define the capped value of type $t_i$ as $\widetilde{K}_i^{t_i} = \min\{ \sigma_i^{F \mid t_i} , V_i \}$.  Consequently, in the free-info formulation, the expected reward of policy $\pi$ in state $({\cal C},{\cal P},y)$ is
\begin{eqnarray}
  J^{\pi,K} (\mathcal{C}, \mathcal{P}, y)  
    &=&   \mathbb{E}\left [ \left. {\mathbb{S}}_0^{\pi} \cdot y + \sum_{i\in {\cal C}}  \left(\mathbb{S}_i^{\pi} K_i + \widetilde{\mathbb{S}}_i^{\pi} \widetilde{K}_i \right)+ \sum_{i\in {\cal P} }  \widetilde{\mathbb{S}}_i^{\pi}  \widetilde{K}^{t_i}_i  \right| ({\cal C},{\cal P},y) \right] \label{eq:free} \ . 
\end{eqnarray}
Recall that we interpret the caps as fair strike prices of call options on the box values, whose expected payoff offsets the opening costs. The formulation~\eqref{eq:free} can be viewed as an alternative cost accounting where Pandora opens boxes free of charge (there are no negative terms), and in exchange, there is a short position on a call option that exactly compensates the opening costs. However, Pandora has the advantage of being subjected to the short contract only if the box is ultimately selected; otherwise, the short contract is void. Under this optimistic assumption, Pandora essentially collects one of the capped rewards, $K_i$, $\widetilde{K}_i$, or $\widetilde{K}^{t_i}_i$, depending on the chosen opening mode. 

Going one step further, the free-info relaxation in state $(\mathcal{C}, \mathcal{P}, y)$ corresponds to the following quantity:
\begin{eqnarray}\label{freeinfo:ub}
J^K (\mathcal{C}, \mathcal{P}, y) = \mathbb{E}\left[  \max \left\{ y,  \underset{i \in \mathcal{C}}{ \max }  \left\{ \max \left\{K_i, \widetilde{K}_{i}  \right\} \right\} ,   \underset{i \in \mathcal{P}}{\max}  \left\{\widetilde{K}_i^{t_i}  \right\} \right\} \right] \ .
\end{eqnarray}
Similarly to Whittle's integral, we eliminate the dependence on policy $\pi$ by choosing the box with the highest capped reward. Therefore, $J^K (\mathcal{C}, \mathcal{P}, y) $ is an upper bound on~\eqref{eq:free}, which might not be attainable through a specific policy $\pi$. 

We are now ready to state our main result with respect to Whittle's integral and the free-info relaxation, showing that both methods provide valid upper bounds on the value function.
\begin{lemma}\label{thm:upper_bounds}
For every state $(\mathcal{C}, \mathcal{P}, y)$, we have
\begin{equation}
   J(\mathcal{C}, \mathcal{P}, y)  \leq J^W (\mathcal{C}, \mathcal{P}, y)  \leq  J^K (\mathcal{C}, \mathcal{P}, y) \ .
\end{equation}
\end{lemma}
The proof appears in Appendix~\ref{appendix:thm:upper_bounds}. 
These upper bounds have been independently derived in the literature for related optimization settings. To our knowledge, this result is the first to formally connect these methods and show that Whittle's integral provides a tighter upper bound. 
The proof relies on results by~\cite{brown2013optimal} who exhumed an unnoticed technical claim in~\cite{whittle1980multi} to derive Whittle's integral (for the cost-discounted case), and \cite{beyhaghi2019pandora} who proposed the free-info relaxation; a technical difference in our setting is that information can be acquired sequentially over multiple stages.

We note in passing that Lemma~\ref{thm:upper_bounds} can be used to establish the asymptotic optimality of the index policy that prioritizes the highest threshold inspection. For brevity, we relegate this result to Appendix~\ref{appendix:ex_infinite_iid_boxes}.

\subsection{Near-optimal algorithms} \label{subsec:apx_policies}

 In this section, we leverage the upper bounds of Section~\ref{subsec:upper_bounds} to construct simple near-optimal policies. Specifically, we develop an algorithm with a worst-case approximation ratio of $(1-1/e)$ using a reduction to the stochastic submodular optimization problem. 
 
 Our approach is an application of the framework by \cite{singla2018price} and \cite{beyhaghi2019pandora}. In this context, the main idea is to focus on policies for which the free-info upper bound $J^{\pi,K}$ of equation~\eqref{eq:free} is exactly equal to the value function $J^{\pi}$ of equation~\eqref{eq:obj}. It turns out there is a simple recipe for constructing such policies: (i) commit before opening any box on the selection mode of each box (i.e., whether we start by F-opening or P-opening it), and (ii) strictly follow the order of decreasing opening thresholds. That is, we partition the set of boxes into two subsets, $[N]= F\cup P$, such that boxes $i\in F$ are initially $F$-opened, whereas boxes $i\in P$ are $P$-opened, if inspected. Next, in each state, we select the action with the highest associated threshold. The resulting class of policies $\pi^{F,P}$ is termed {\em committing} policies. The reader is referred to the Algorithm~\ref{algo:committing} for a complete description.

\begin{algorithm}[H]
\footnotesize
  \caption{Committing policy $\pi^{F,P}$.} \label{algo:committing}
  \begin{algorithmic}[1]
  \State Initialize state $(\mathcal{C}, \mathcal{P}, y) \gets ([N], \emptyset, 0)$;
\While { $y < \max \left\{  
  \underset{j \in \mathcal{C} \cap F }{\max}  \left\{  \sigma_j^F \right\}, 
  \underset{j \in \mathcal{C} \cap P }{\max}  \left\{  \sigma_j^P \right\}, 
  \underset{ j \in \mathcal{P}}{\max }  \left\{ \sigma_j^{F \mid t_j}  \right\}  
  \right\}  $  }
  \State Let $ \sigma_i = \max \left\{  
  \underset{j \in \mathcal{C} \cap F }{\max}  \left\{  \sigma_j^F \right\}, 
  \underset{j \in \mathcal{C} \cap P }{\max}  \left\{  \sigma_j^P \right\}, 
  \underset{ j \in \mathcal{P}}{\max }  \left\{ \sigma_j^{F \mid t_j}  \right\}  
  \right\}$; 
\If{$i \in F$}  
    \State F-open box $i$; $~ y \gets \max \{ y, v_i\}$; $~\mathcal{C} \gets \mathcal{C} \setminus \{ i \}$;
\ElsIf{ $i \in P$}
    \If {$i \in \mathcal{C}$}
        \State P-open box $i$; $\mathcal{C} \gets \mathcal{C} \setminus \{ i \};  ~ \mathcal{P} \gets \mathcal{P} \cup \{ (i, t_i)\}$; 
    \ElsIf{ $ i \in \mathcal{P}$}
        \State F-open box $i$; $\mathcal{P} \gets \mathcal{P} \setminus \{(i, t_i)\}; ~ y \gets \max \{ y, v_i\}$;
    \EndIf
\EndIf

\EndWhile
\State Select $y$ and stop;
  \end{algorithmic}
\end{algorithm}

The next lemma shows that the expected profit of any committing policy indeed coincides with its free-info upper bound, which can be expressed in a simple form.

\begin{lemma}  \label{lemma:committing}
Given a partition $(F,P)$ of the set of boxes $[N]$,
the committing policy $\pi^{F,P}$ generates an expected profit of
\begin{equation} \label{eq:lemma:committing}
    J^{\pi^{F,P}}(s_0) = J^{\pi^{F,P},K}(s_0)
=
\mathbb{E}\left[ \max \left\{ \underset{i \in F}{\max} \{  K_i\}   , \underset{i \in P}{\max}  \{ \widetilde{K}_i \}  \right\}\right].
\end{equation}
\end{lemma}
 Intuitively, committing policies are not exposed to any cost accounting error, in comparing the opening costs to the corresponding short positions on call options. Recall that this free-info cost accounting is optimistic because Pandora is subjected to the short contract only if the box is ultimately selected. In the proof of Lemma~\ref{lemma:committing}, we identify necessary and sufficient conditions such that Pandora always selects any F-opened box (or F-opens any P-opened box) whenever the call option (or two-stage call option) is in the money. This way, Pandora meets the contractual payment of the option, which is equal to the opening costs in expectation. In particular, we show these necessary and sufficient conditions are met by committing policies.
 
Now, this introduction of such committing policies raises the following question: how should we design the partition $(F,P)$? Can we construct committing policies that achieve strong (constant-factor) approximation ratios? Based on the expected profit characterization of  Lemma~\ref{lemma:committing}, we can compare committing policy $\pi^{F,P}$'s payoff $\mathbb{E}[ \max \{ \underset{i \in F}{\max} \{  K_i\}   , \underset{i \in P}{\max}  \{ \widetilde{K}_i \}  \}]$ to the free-info upper-bound $J^K (s_0) = \mathbb{E}[ \underset{i \in [N]}{\max}   \max\{K_i, \widetilde{K}_{i}  \} ]$ (see Equation~\eqref{freeinfo:ub}).
The former ex-ante chooses either $K_i$ or  $\widetilde{K}_{i}$, before observing their realizations, depending on whether $i\in F$ or $i\in P$. By contrast, the latter chooses the maximum reward $\max \left\{K_i, \widetilde{K}_{i}  \right\} $ out of the two opening modes after seeing their realizations. This informal comparison leads to the design of simple approximation algorithms.

First, note that the following upper bound on the free-info setting always holds:
\begin{eqnarray*}
    J^K \left(s_0\right) &=& \mathbb{E}\left[   \max_{i\in [N]} ~ \max\left\{K_i, \widetilde{K}_{i}  \right\}  \right] \\
    &\leq& \mathbb{E}\left[  \max_{i\in [N]} \left({\mathbb I}[i\in F] \cdot K_i + {\mathbb I}[i\in P] \cdot\widetilde{K}_i\right) + \max_{i\in [N]} \left({\mathbb I}[i\in P] \cdot K_i + {\mathbb I}[i\in F] \cdot \widetilde{K}_i\right) \right] \\
    &= & J^{\pi^{F,P}} \left(s_0\right) + J^{\pi^{P,F}} \left(s_0\right) \ ,
\end{eqnarray*}
where the inequality holds since $\max\{a,b\} \leq \max\{c,d\} +\max\{e,f\} $ for every scalars $a\leq \max\{c,e\}$ and $b\leq \max\{d,f\}$, and the last equality proceeds from Lemma~\ref{lemma:committing}. Importantly, the above inequality implies that, for any partition of $(F,P)$ of $[N]$, either the committing policy $\pi^{F,P}$ or the one $\pi^{P,F}$ that flips the opening modes achieves a factor-$\frac{1}{2}$ approximation of the optimum.

More broadly, the performance of a well-designed committing policy is related to the adaptivity gap in stochastic submodular maximization; the question can be reformulated as optimizing a certain submodular function over random entries and the goal is to bound the gap between non-adaptive policies, that select those entries before their realizations are revealed, and adaptive ones. By leveraging fundamental results in this area, we obtain the following theorem.
\begin{theorem}\label{thm:1-1/e}
There exists a committing policy that achieves at least  $1 - 1/e\approx 0.63$ of the optimum 
and can be approached via a polynomial-time approximation scheme.
\end{theorem}
In Appendix~\ref{appendix:thm:1-1/e}, we devise a reduction from Pandora's box problem to the problem of maximizing a stochastic submodular function under additional constraints.
In this context, \cite{asadpour2015maximizing} showed, among other results, that the adaptivity gap (i.e., the performance gap between non-adaptive and adaptive policies) is lower bounded by $1-1/e$, and that an approximate non-adaptive policy can be efficiently constructed using a continuous-greedy algorithm.

\section{Numerical Study}
\label{sec:numerical}

We conduct numerical experiments to further explore the characteristics of optimal policies and assess the effectiveness of natural candidate policies.
Section~\ref{sec:numerical_instances}  describes the process for generating a diverse set of problem instances. 
Section~\ref{sec:numerical_jay} quantifies the extent to which our theoretical characterization captures optimal decisions. 
Section~\ref{sec:numerical_index} compares the performance of various policies inspired by theoretical analysis. 
Finally, in Section~\ref{sec:exploring_p_opening_main}, we examine factors that increase the tendency of the optimal policies to P-open.

\subsection{Problem instances}  \label{sec:numerical_instances}

Our design of numerical experiments is based on the observation that the interaction between opening thresholds significantly impacts the behavior of optimal policies. We therefore create instances that consist of boxes with diverse values of opening thresholds.

\begin{figure}[htbp]
  \centering
  \includegraphics[width=7cm]{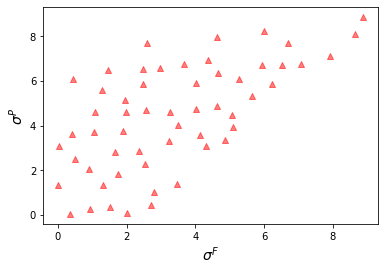}
  \caption{A visualization of how the opening thresholds of the prototypical boxes are distributed. }
  \label{fig:scatter_boxes0}
\end{figure}

We begin by generating a collection of ``prototypical'' boxes. 
To create a prototypical box $i$, we sample 5 values for the support of $V_i$ from the uniform distribution $\mathcal{U}(0,10)$; every type $T_i$ can realize up to at most three types (that is, $ \mid \Gamma_i \mid  \leq 3 $); for each type we uniformly sample the conditional distribution $V_i \mid T_i = t$ over the support; we sample the F-opening and P-opening costs from the uniform distributions $\mathcal{U}(0,5)$ and $\mathcal{U}(0,3)$, respectively. We then generate a candidate set of prototypical boxes out of which we select a small representative set (Figure~\ref{fig:scatter_boxes0}).

Each problem instance consists of randomized copies of the prototypical boxes, in varying compositions with $N=2$ to $N=16$ boxes. Due to running time limitations, there are 1000 instances whose size is 2 to 9 (``small instances'' which can be solved to optimality using DP), and there are 300 instances whose size is 10 to 16 (``large instances'' which can only be solved using heuristics).
To avoid trivial cases, we only generate boxes where $c^F_i \geq c^P_i$ (which is why in Figure~\ref{fig:scatter_boxes0} most prototypical boxes have greater P-thresholds).

\subsection{Quantifying the coverage of analytical results} \label{sec:numerical_jay}

The analysis of Section~\ref{sec:characterization} revealed various insights about the nature of the optimal policy. Since the analysis includes sufficient but not always necessary conditions, we are interested in understanding to what extent the sufficient conditions of Theorems 1--3 describe an optimal policy. 

To this end, we conduct two types of experiments. First, we examine optimal solutions to the ``small instances'' and compute the percentage of states where the sufficient conditions of Theorems 1--3 hold (i.e., where our theorems prescribe optimal actions). 
We implement the DP recursion in \eqref{eq:Bellman_Equation} and compute {\em coverage rates}, defined as the percentages of system states where the optimal action can be determined by our sufficient conditions.

The results are summarized in Table~\ref{tab:util_suff_thms}. The overall coverage rate is presented in the 5th column, and is divided into specific conditions in the 2nd to 4th columns. 
Looking at the 5th column of the table, we see that as the problem increases in size, the sufficient conditions of Theorems 1--3 capture optimal actions in all but a relatively few states. 
The last two columns present the recall of Theorems 2 and 3. These reflect the percentage of states in which the sufficient conditions of the theorems hold out of all states where F-opening (column 6) and P-opening (column 7) are optimal. For example, for instances of size 5 (4th row in the table), Theorem 2 captures 97\% of states where F-opening is optimal, and Theorem 3 captures 77\% of all states where P-opening is optimal.
We observe that identifying P-opening is more challenging than identifying when F-opening is optimal, but overall, the experiment suggests that our characterization explains to a large degree the behavior of the optimal policy. Note that the recall of Theorem 1 is 100\% as the theorem perfectly describes when stopping is optimal.

We conduct a second experiment where we compare two DP implementations that recursively compute the optimal policy. The first policy ``Na\"ive DP'' is implemented using the DP recursion in Equation~\eqref{eq:Bellman_Equation}, which does not leverage any structure of the optimal policy. The second implementation, on the other hand, utilizes the sufficient conditions of Theorems 1--3 which determine optimal actions and do not require calling the DP recursion when the conditions are satisfied.  Table~\ref{tab:comparison_DP_JAY} compares the two approaches in terms of the total number of states created, the total number of visits to system states, and the runtime. We see that the characterization of the optimal policy not only provides insights about the optimal policy, but also bears computational benefits, allowing to solve problems at significantly shorter times. That  said, the ``curse of dimensionality'' cannot be avoided, as can be observed by the rate at which the size of the state space grows.

\subsection{Policy benchmark and computational study}
\label{sec:numerical_index}
We evaluate and compare various policies that arise in our analysis: 
\begin{itemize}
    \item $\pi^\textrm{OPT}$ -- the optimal policy, computed by solving the DP naively.
    \item $\pi^{F^*,P^*}$ -- the best committing policy, obtained through complete enumeration. 
    \item  $\pi^{\textrm{index}}$ -- an index policy defined using the opening thresholds associated with every box and action.
    In every state, the policy acts according to the largest index, where the thresholds $\sigma^F_i$, $\sigma^{F \mid t_i}_i$, and $\sigma^P_i$ serve as indices for the F- and P-opening actions, while $y$ is the index for selecting the best opened box.
    This policy is asymptotically optimal (Corollary~\ref{lemma:infinite_iid_boxes}) and
    satisfies the sufficient optimality conditions. 
    \item $\pi^{\textrm{W}}$ -- a one-step lookahead policy \citep{bertsekas2019reinforcement}  based on approximate value functions 
    using Whittle's integral.\endnote{
Another natural policy to consider is the one-step lookahead policy that uses $J^K$ to approximate the  value function. However, given that $J^{W}$ provides a tighter upper bound (Lemma~\ref{thm:upper_bounds}), we only test $\pi^W$ in the experiments.}
    \item $\pi^{STP}$ -- a  policy that determines its next action based on the maximal myopic improvement (using a single ``test,'' hence Single Test Policy) in comparison to stopping. Computing the improvement from each non-open box becomes tractable by ignoring other non-open boxes, and follows the optimal policy for the single box problem (see Figure~\ref{fig:example_1}).
    
\end{itemize}

To evaluate performance, we measure the value function and runtime of each problem instance under different policies, starting from the initial state. As the expected profit can vary significantly among different instances, we normalize the expected profit of each policy relative to the optimal value obtained by policy $\pi^\textrm{OPT}$. 
For large instances where the optimal policy cannot be computed, we use the best-performing policy to normalize the expected profit.
Thus, the performance of each policy is reported as a number between 0 and 1, where larger values indicate better performance. For each policy and a set of instances, we report on performance using the average  across instances and the worst instance tested.

\textit{Average performance. } All heuristics perform quite well in terms of their average performance, often exceeding 90\% of $\textrm{OPT}$. 
However, $\pi^{\textrm{index}}$ and $\pi^{F^*,P^*}$ are superior, with average performances that are close to 100\%.

\textit{Worst-case performance.} We observe that $\pi^{\textrm{W}}$ can perform poorly, at times nearing zero profit. 
This suggests that Whittle's integral approach may not be a suitable approach to deriving tight bounds for our problem, and perhaps to other Pandora's box problems as well.
We also see that $\pi^{\textrm{index}}$ is superior to $\pi^{STP}$ and that its relative performance improves as the number of boxes increases.
We observe that $\pi^{\textrm{index}}$ approaches optimality as $N$ increases, which is aligned with its asymptotic optimality.
Outperforming all others is policy $\pi^{F^*,P^*}$, which achieves the best worst-case performance, with a loss of at most 7\% for instances of size 2. This suggests that index policies based on our opening thresholds are effective for solving the problem. 
We note, however, that this policy is not optimal, and finding the optimal committing policy is computationally challenging \citep[an approximation algorithm is provided in][]{asadpour2015maximizing}.

\textit{Runtime. } Table~\ref{fig:optimality_boxplot} in Appendix~\ref{apx:tables} summarizes the runtimes. 
As expected, the time for solving instances to optimality grows exponentially fast, making it challenging to solve instances with more than 9 boxes. 
We also see that applying policy $\pi^{\textrm{W}}$ becomes more challenging as the number of boxes increases, due to the calculation of Whittle's integral.
Similarly, the exhaustive enumeration carried out by policy $\pi^{F^*,P^*}$ could result in long runtimes.
We note that the runtime also depends on the size of the support of the distribution (which we assumed to be 5) and the number of box types (which we assumed to be at most 3).
Note that these runtimes include the entire calculation of the value function and that simply applying these policies is very fast.

\textit{Optimality. }
Table~\ref{tab:comparison_of_J} shows the percentage of instances for which different policies attain the optimal expected objective value. 
In contrast to the near-optimal performance displayed in Table~\ref{tab:performance_with_DP}, policy $\pi^{\textrm{W}}$ is almost never optimal. Additionally, even policies $\pi^{F,P}$ and $\pi^{\textrm{index}}$, which perform better, in many cases do not achieve the optimal objective value. We notice that overall, the performance of the latter two index policies improves with the number of boxes, which aligns with the asymptotic optimality result (Corollary~\ref{lemma:infinite_iid_boxes}).

\subsection{When is P-opening worthwhile?}
\label{sec:exploring_p_opening_main}

We next conduct experiments that shed light on the structure of optimal policies and further characterize when P-opening is worthwhile.

To this end, we define the notion of \textit{P-ratio} as the average percentage of P-openings conducted by the optimal policy of all opening actions. A higher P-ratio implies that the optimal policy applies P-opening more frequently. We compute the P-ratio for a set of problem instances and plot it against various problem characteristics. For brevity, we describe the main results below and postpone the complete details of the experiment and the figures to Appendix~\ref{sec:numerical-apx}. 

One of our main findings is that greater homogeneity of boxes in terms of their opening thresholds (i.e., boxes $i$ and $j$ are ``homogeneous'' when the tuples $(\sigma^F_i,\sigma^P_i)$ and $(\sigma^F_j,\sigma^P_j)$ are close) is correlated with  more frequent partial opening. This aligns with intuition, since when boxes are very different, their relative value can be determined without partial inspection; hence, F-opening can simply be conducted as a necessity prior to selection. 

Additionally, in line with our theoretical results, we observe an increased tendency to partially open when there is a higher proportion of boxes whose P-thresholds exceed their corresponding F-thresholds.
This finding underscores the role of opening thresholds in capturing the value of corresponding inspection modes.

Finally, in the case of identical boxes (homogeneity is maximal), the tendency to partially open increases in the number of boxes. This suggests that when the pool of alternatives is larger, partial opening becomes more significant---assuming partial opening is worthwhile to begin with.


\section{Conclusions and Future Directions}
\label{sec:conclusion}

The work studies a generalized Pandora's box problem that captures a tradeoff related to acquiring accurate and costly versus inaccurate but cheaper information about the value of competing alternatives. 
We proved that the problem is NP-hard, but nevertheless, using a synthesis of techniques, we identified structural properties of the optimal policy and provided insights into the key drivers of optimal information acquisition decisions. 
We showed that intuitive threshold-based policies that extend the Pandora's box solution describe optimal decisions and form the basis of approximate solutions. Extensive numerical experiments show that this characterization almost entirely explains the optimal policy, and that the approximate policies perform extremely well.

As our work focused on the simplest yet sufficiently rich context where the aforementioned tradeoff occurs, it would be interesting to consider more general settings. For example, a richer information model that allows multiple types of repeated inspections, inspection costs that decrease after inspections take place, and the option to select partially opened boxes, all could be interesting directions for future work.


\ACKNOWLEDGMENT{We would like to express our sincere gratitude to the editorial team for their  insightful comments, which significantly improved the quality of the paper.}








\theendnotes

\bibliographystyle{apalike} 
\SingleSpacedXI
\bibliography{ref_pandora}
\OneAndAHalfSpacedXI

\section*{Tables}

\begin{table}[htbp]
  \centering
  \caption{Average weighted coverage rates of Theorems \ref{thm:myopic_stopping}--\ref{thm:just_weak_closed} and recall rates of Theorems \ref{thm:just_strong} and~\ref{thm:just_weak_closed}. 
   Theorem~\ref{thm:myopic_stopping} describes the optimal stopping criterion. 
   Theorems~\ref{thm:just_strong} and~\ref{thm:just_weak_closed} provide  sufficient condition for F-opening and P-opening, respectively. }
    \begin{tabular}{ccccccccccccc}
    \toprule
    \multirow{2}[4]{*}{N} &    & \multicolumn{7}{c}{Coverage rate} &    & \multicolumn{3}{c}{Recall } \\
\cmidrule{2-9}\cmidrule{11-13}       &    & Thm. \ref{thm:myopic_stopping} &    & Thm. \ref{thm:just_strong} &    & Thm. \ref{thm:just_weak_closed} &    & Overall &    & Thm. \ref{thm:just_strong} &    &  Thm. \ref{thm:just_weak_closed} \\
\cmidrule{1-1}\cmidrule{3-3}\cmidrule{5-5}\cmidrule{7-7}\cmidrule{9-9}\cmidrule{11-11}\cmidrule{13-13}    2  &    & 0.48 &    & 0.38 &    & 0.06 &    & 0.92 &    & 0.90 &    & 0.53 \\
    3  &    & 0.47 &    & 0.38 &    & 0.09 &    & 0.94 &    & 0.95 &    & 0.66 \\
    4  &    & 0.46 &    & 0.39 &    & 0.10 &    & 0.95 &    & 0.97 &    & 0.72 \\
    5  &    & 0.46 &    & 0.39 &    & 0.11 &    & 0.95 &    & 0.97 &    & 0.77 \\
    6  &    & 0.45 &    & 0.40 &    & 0.11 &    & 0.96 &    & 0.97 &    & 0.81 \\
    7  &    & 0.45 &    & 0.40 &    & 0.11 &    & 0.96 &    & 0.97 &    & 0.83 \\
    8  &    & 0.44 &    & 0.41 &    & 0.11 &    & 0.96 &    & 0.96 &    & 0.86 \\
    9  &    & 0.44 &    & 0.41 &    & 0.11 &    & 0.96 &    & 0.96 &    & 0.87 \\
    \bottomrule
    \end{tabular}%
  \label{tab:util_suff_thms}%
\end{table}%

\begin{table}[htbp]
\scriptsize
  \centering
  \caption{A comparison between a DP implementation that utilizes the sufficient optimality conditions and one that does not, in terms of the average number of states created and revisited, and average runtimes (in seconds).}
    \begin{tabular}{ccccrcccc}
    \toprule
    \multirow{2}[4]{*}{$N$} & \multicolumn{3}{c}{Na\"ive DP} &    & \multicolumn{3}{c}{DP that utilizes Theorems 1--3} & \multirow{2}[4]{*}{runtime ratio} \\
\cmidrule{2-4}\cmidrule{6-8}       & \# states created &\# states revisited & runtime (s) &    & \# states created &\# states revisited & runtime (s) &  \\
    \midrule
    2  & 38.4 & 145.0 & 0.04 &    & 24.8 & 32.37 & 0.03 & 0.731 \\
    3  & 209.0 & 1384.0 & 0.01 &    & 115.3 & 195.09 & 0.00 & 0.225 \\
    4  & 1101.4 & 10513.9 & 0.10 &    & 575.1 & 1204.03 & 0.02 & 0.180 \\
    5  & 5567.0 & 69796.4 & 0.64 &    & 2798.7 & 6703.55 & 0.10 & 0.156 \\
    6  & 28158.6 & 440669.7 & 3.99 &    & 12943.3 & 35306.54 & 0.51 & 0.128 \\
    7  & 137652.0 & 2570720.3 & 23.59 &    & 65826.3 & 208193.38 & 3.00 & 0.121 \\
    8  & 680310.1 & 14825094.6 & 143.80 &    & 298534.1 & 1067723.38 & 15.93 & 0.106 \\
    9  & 3238243.1 & 80398698.6 & 810.50 &    & 1368893.5 & 5450797.57 & 82.81 & 0.094 \\
    \bottomrule
    \end{tabular}%
  \label{tab:comparison_DP_JAY}%
\end{table}%

\begin{table}[htbp]
\scriptsize
  \centering
  \caption{Normalized performance of different policies. }
    \begin{tabular}{cccccccccccccccccccc}
     \toprule
    \multirow{2}[2]{*}{$N$} & \multicolumn{3}{c}{$\pi^\textrm{OPT}$} &    & \multicolumn{3}{c}{$\pi^{\textrm{index}}$} &    & \multicolumn{3}{c}{$\pi^{\textrm{W}}$} &    & \multicolumn{3}{c}{$\pi^{F^*,P^*}$} &    & \multicolumn{3}{c}{$\pi^{STP}$} \\
\cmidrule{2-4}\cmidrule{6-8}\cmidrule{10-12}\cmidrule{14-16}\cmidrule{18-20}       & mean  & std & worst &    & mean  & std & worst &    & mean  & std & worst &    & mean  & std & worst &    & mean  & std & worst \\
\midrule
    2  & 1.000 & 0.000 & 1.000 &    & 0.987 & 0.036 & 0.605 &    & 0.922 & 0.123 & 0.033 &    & 0.999 & 0.005 & 0.933 &    & 0.963 & 0.057 & 0.709 \\
    3  & 1.000 & 0.000 & 1.000 &    & 0.995 & 0.017 & 0.804 &    & 0.901 & 0.101 & 0.298 &    & 1.000 & 0.002 & 0.960 &    & 0.947 & 0.062 & 0.601 \\
    4  & 1.000 & 0.000 & 1.000 &    & 0.998 & 0.007 & 0.882 &    & 0.893 & 0.088 & 0.403 &    & 1.000 & 0.001 & 0.976 &    & 0.943 & 0.063 & 0.611 \\
    5  & 1.000 & 0.000 & 1.000 &    & 0.999 & 0.003 & 0.958 &    & 0.890 & 0.082 & 0.373 &    & 1.000 & 0.000 & 0.993 &    & 0.938 & 0.062 & 0.652 \\
    6  & 1.000 & 0.000 & 1.000 &    & 0.999 & 0.002 & 0.966 &    & 0.890 & 0.085 & 0.438 &    & 1.000 & 0.000 & 0.993 &    & 0.940 & 0.059 & 0.691 \\
    7  & 1.000 & 0.000 & 1.000 &    & 0.999 & 0.002 & 0.984 &    & 0.888 & 0.084 & 0.503 &    & 1.000 & 0.000 & 0.996 &    & 0.933 & 0.063 & 0.575 \\
    8  & 1.000 & 0.000 & 1.000 &    & 0.999 & 0.002 & 0.989 &    & 0.879 & 0.088 & 0.494 &    & 1.000 & 0.000 & 0.996 &    & 0.937 & 0.061 & 0.564 \\
    9  & 1.000 & 0.000 & 1.000 &    & 0.999 & 0.001 & 0.991 &    & 0.877 & 0.090 & 0.522 &    & 1.000 & 0.000 & 0.997 &    & 0.935 & 0.061 & 0.653 \\
\cmidrule{1-1}      
    10 & -  & -  & -  &    & 0.999 & 0.001 & 0.992 &    & 0.876 & 0.099 & 0.515 &    & 1.000 & 0.000 & 1.000 &    & 0.941 & 0.061 & 0.709 \\
    11 & -  & -  & -  &    & 0.999 & 0.001 & 0.993 &    & 0.871 & 0.110 & 0.463 &    & 1.000 & 0.000 & 1.000 &    & 0.939 & 0.063 & 0.686 \\
    12 & -  & -  & -  &    & 0.999 & 0.001 & 0.994 &    & 0.870 & 0.104 & 0.451 &    & 1.000 & 0.000 & 1.000 &    & 0.940 & 0.061 & 0.693 \\
    13 & -  & -  & -  &    & 0.999 & 0.001 & 0.992 &    & 0.869 & 0.099 & 0.568 &    & 1.000 & 0.000 & 1.000 &    & 0.945 & 0.055 & 0.710 \\
    14 & -  & -  & -  &    & 0.999 & 0.001 & 0.996 &    & 0.877 & 0.102 & 0.464 &    & 1.000 & 0.000 & 1.000 &    & 0.940 & 0.058 & 0.691 \\
    15 & -  & -  & -  &    & 0.999 & 0.001 & 0.996 &    & 0.847 & 0.136 & 0.512 &    & 1.000 & 0.000 & 1.000 &    & 0.957 & 0.064 & 0.779 \\
    16 & -  & -  & -  &    & 0.999 & 0.001 & 0.997 &    & 0.884 & 0.096 & 0.633 &    & 1.000 & 0.000 & 1.000 &    & 0.944 & 0.055 & 0.780 \\
    \bottomrule
    \end{tabular}%
  \label{tab:performance_with_DP}%
\end{table}%

\begin{table}[htbp]
     \scriptsize
  \centering
  \caption{Percentage of instances in which the policies attained the optimal objective value.}
   \begin{adjustbox}{width=0.4\textwidth}
    \begin{tabular}{ccccc}
    \toprule
    $N$ & 
    $\pi^\textrm{OPT}$ &
    $\pi^{\textrm{index}}$ & 
    $\pi^{\textrm{W}}$ &    
    $\pi^{F^*,P^*}$ \\
    \midrule
    2  & 1.000 & 0.543 & 0.238  &  0.835  \\
    3  & 1.000 & 0.614 & 0.065  &  0.850  \\
    4  & 1.000 & 0.622 & 0.023  &  0.853  \\
    5  & 1.000 & 0.661 & 0.002  & 0.892  \\
    6  & 1.000 & 0.686 & 0.001  &  0.904  \\
    7  & 1.000 & 0.701 & 0.000  &  0.927  \\
    8  & 1.000 & 0.715 & 0.001  &  0.944  \\
    9  & 1.000 & 0.733 & 0.000  &  0.943  \\
    \bottomrule
    \end{tabular}%
      \end{adjustbox}
  \label{tab:comparison_of_J}%
\end{table}%

\newpage 

\section*{Figures}

\begin{figure}[htbp]
\centering
\begin{subfigure}[b]{.5\textwidth}
  \centering
  \includegraphics[width=1\linewidth]{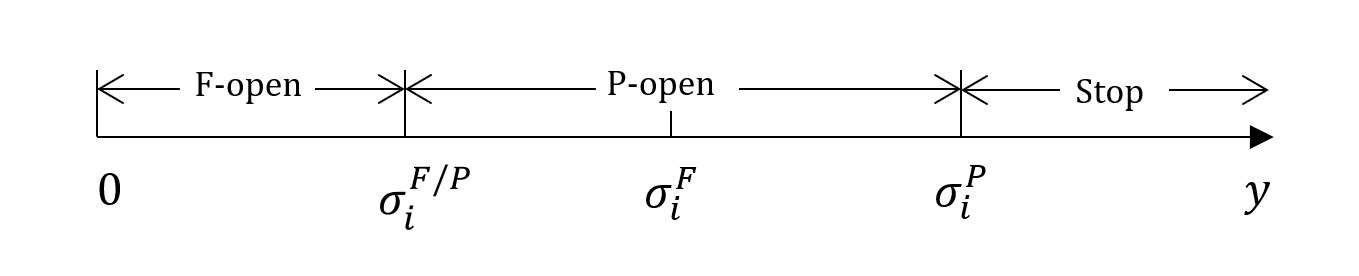}
\end{subfigure}%
\begin{subfigure}[b]{.5\textwidth}
  \centering
  \includegraphics[width=1\linewidth]{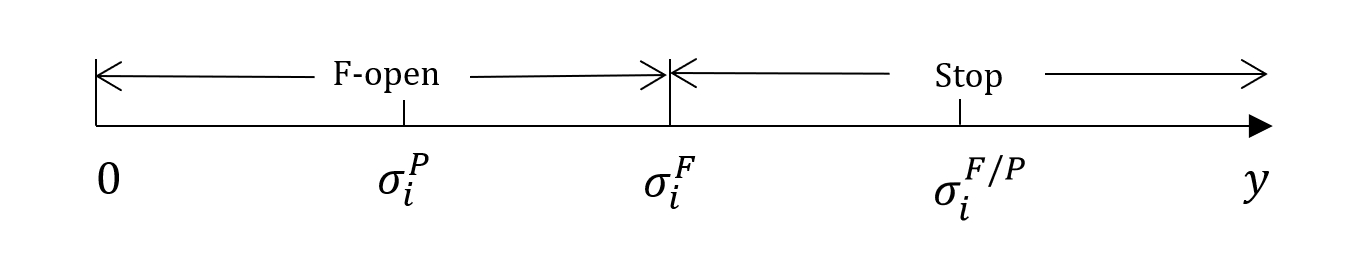}
\end{subfigure}
\caption{Illustration of the optimal policy for the single closed box case. 
Here, we present the case where $\boldsymbol{\sigma_i^{F/P} \geq 0, \sigma_i^P \geq 0}$. }
\label{fig:example_1}
\end{figure}

\ECSwitch

\begin{APPENDICES}

\noindent {\Huge\bfseries Electronic Companion}

\vspace{1em}
\section{Example~\ref{example:many_boxes}}
\label{subsec:example_many_boxes}
We construct an additional example which provides further evidence about the intricacy of the optimal policy in states where the leading threshold corresponds to a P-threshold (see Figure~\ref{fig:example_many_boxes} for a pictorial illustration). 

\begin{example}
\label{example:many_boxes}

We consider the following three boxes: 
\begin{itemize}
    \item Box 1: The opening costs are $c_1^F=1$ and $c_1^P = 0.5$. 
There are two types $\Gamma_1 = \{A,B\}$ and $\p{T_1=A} = 0.2$. 
The joint distribution of $(V_1,T_1)$ is given by: $\p{V_1 = 1 \mid T_1 = A } = 0.1,~ \p{V_1 = 10 \mid T_1 = A } = 0.9$ and $\p{V_1 = 1 \mid T_1 = B } = 0.05, ~\p{V_1 = 10 \mid T_1 = B } = 0.95$.
    \item Box 2: The opening costs are $c_2^F=0.99$ and $c_2^P = 500$. 
There are two types $\Gamma_2 = \{A,B\}$ and $\p{T_2=A} = 0.2$. 
The joint distribution of $(V_2,T_2)$ is given by: $\p{V_2 = 1 \mid T_2 = A } = 0.1, ~\p{V_2 = 10 \mid T_2 = A } = 0.9$ and $\p{V_2 = 1 \mid T_2 = B } = 0.05,~ \p{V_1 = 10 \mid T_1 = B } = 0.95$.
    \item Box 3: The opening costs are $c_3^F=1$ and $c_3^P = 0.5$. 
There are two types $\Gamma_3 = \{A,B\}$ and $\p{T_3=A} = 0.2$. 
The joint distribution of $(V_3,T_3)$ is given by: $\p{V_3 = 1 \mid T_3 = A } = 0.1,~ \p{V_3 = 10 \mid T_3 = A } = 0.9$ and $\p{V_3 = 1 \mid T_3 = B } = 0.05, ~\p{V_3 = 10 \mid T_3 = B } = 0.95$.
\end{itemize}

The thresholds of the boxes are computed numerically and are equal to: $\sigma_1^F = \sigma_3^F \approx 5.45, \sigma_1^P = \sigma_3^P \approx 6.11, \sigma_2^F \approx 5.50, \sigma_2^P < 0$.
The thresholds are visualized in Figure~\ref{fig:example_many_boxes}. 

Depending on the initial state, the unique optimal policy proceeds as follows:
\begin{itemize}
    \item In the initial state of the problem where the only box is box 1, i.e., in state $(\{1\}, \emptyset, 0)$, the optimal policy F-opens box 1.
    \item In the initial state of the problem where there are two boxes, box 1 and box 2, i.e., in state $(\{1,2\}, \emptyset, 0)$, the optimal policy F-opens box 2.
    \item In the initial state of the problem where there are three boxes: box 1, box 2, and box 3, i.e., in state $(\{1,2,3\}, \emptyset, 0)$, the optimal policy P-opens box 1.
\end{itemize}
\end{example}

\begin{figure}[htbp]
  \centering
  \includegraphics[width=9cm]{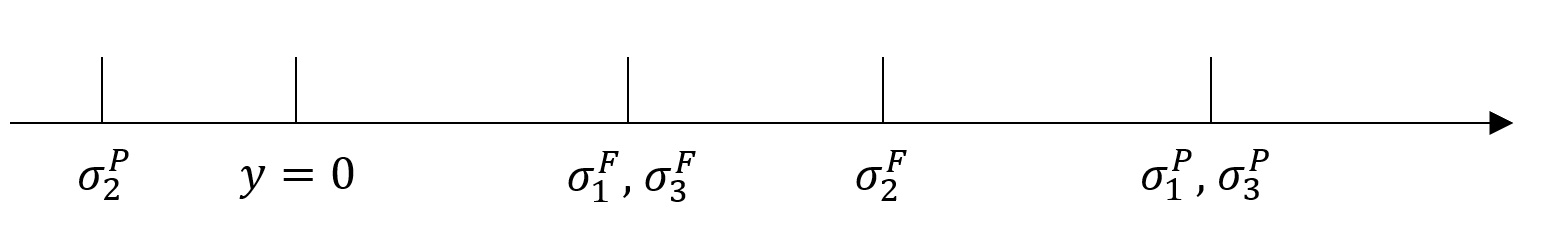}
  \caption{Illustration of thresholds in Example~\ref{example:many_boxes}}
  \label{fig:example_many_boxes}
\end{figure}
As evidenced by Example~\ref{example:many_boxes}, the unique optimal policy is not indexable.
More importantly,  and unlike Example~\ref{example:not_indexable}, it shows that the opening priorities may change when boxes are added, even when the value of $y$ remains unchanged:
in state $(\{1\},\emptyset, 0)$, the action of F-opening box~1 is optimal; after adding box 2 (i.e., in state $(\{1, 2\},\emptyset, 0)$), F-opening box 2 becomes optimal, even if the P-threshold of box~1 remains the largest; however, after adding box 3 (which is statistically identical to box 1), the optimal action is to P-open box 1.
In other words, this example shows three states in which $\sigma_1^P$ is the largest threshold but three distinct actions are in fact optimal. 
\\\\\\

\section{Economic interpretation of opening thresholds}
\label{apx:economic_interpretation}

We can use notion of compound option or, informally, a ``two-stage call option,'' to provide an intuitive interpretation for the opening thresholds. This financial instrument refers to a more complex financial contract that allows the buyer to purchase an option at some future date. 
That is, for the price $c_i^P$ the decision-maker can buy the initial (overlying) option, which allows her at a future date (when more information is available) to purchase the second (underlying) option for a price of $c_i^F$. 
The additional information that is available at the time of purchasing the second option is represented by the realization $t_i$.
The conditional F-threshold represents the fair strike price for purchasing $V_i$, given the information that is available at the future date, while the P-threshold represents the strike price for buying the initial option.
This type of option is referred to as Call on Call. 
On the other hand, the FP-threshold corresponds to a compound option which is referred to as Call on Put. It allows the buyer to short an option at a future time. 
That is, in the first stage, the buyer pays $c_i^P$ to observe the type $t_i$, and then decides whether or not to enter the second stage.
In the second stage, she has the right to choose whether to short a call option with prize $V_i$ and a strike price $\sigma_i^{F/P}$. 
Clearly, if the strike price $\sigma_i^{F/P}$ is overpriced, i.e., greater than $\sigma_i^{F \mid t_i}$, a risk-neutral player will sell the second-stage option and otherwise not. 
Therefore, we can interpret $\sigma_i^{F/P}$ as the fair strike price to equalize the expected costs and expected profits of this two-stage option.

The above interpretation motivates a policy in which the decision maker chooses the highest option strike price---that is, according to the box and opening mode with highest threshold. We show in Section~\ref{sec:stopping} that this index policy is indeed optimal for the single-box case, and we further expand its sufficient optimality conditions in the next sections.

\section{Proofs for Section~\ref{sec:characterization}}
\label{appendix:thm_stopping}

Algorithm~\ref{algo:single-box} below summarizes the optimal policy for the single closed-box setting, which is described in Section~\ref{sec:stopping}.

\begin{algorithm}[H]
\footnotesize
  \caption{The optimal policy for the single closed-box setting.} \label{algo:single-box}
  \begin{algorithmic}[1]
  \State Initialize state $(\mathcal{C}, \mathcal{P}, y) \gets (\{ i \}, \emptyset, y)$;
\If{ $y > \max\{ \sigma_i^F, \sigma_i^P \}$}  
    \State Stop and select $y$; 
\ElsIf{ $\sigma_i^F > \max\{ y,\sigma_i^P\}$   }
    \State F-open box $i$, and select the best prize at hand;
\ElsIf{ $\sigma_i^P > \max\{ y,\sigma_i^F\}$}
    \If{ $ y > \sigma_i^{F/P}$}
    \State P-open box $i$;
    \State Then further F-open box $i$ if and only if the realized type $t_i$ is such that $y < \sigma_i^{F \mid t_i}$ or else stop;
    \Else{ }
    \State F-open box $i$, and select the best prize at hand;
    \EndIf
\EndIf
\end{algorithmic}
\end{algorithm}

In what follows we first prove the optimality of the threshold-based policy for the single closed-box case (Section~\ref{apx:single_box}), 
then prove the expression for the expected profit of the optimal policy in the single closed-box case (Section~\ref{apx:single_box_expression}), and finally, prove the optimal stopping condition for the general case (Section~\ref{apx:stopping}).

\subsection{The optimality of the threshold-based policy for the single closed-box case}\label{apx:single_box}

In state $(\mathcal{C}, \mathcal{P}, y)=(\{i\}, \emptyset, y)$, there are three possible actions that we can take: stop, F-open box $i$ or P-open box $i$.
The optimal stopping rule is argued in Section~\ref{sec:stopping}. We proceed to compare F-opening and P-opening. 
If we F-open the box, then the expected value function is 
\begin{equation}\label{eq:J_A}
    J^F(\{i\}, \emptyset, y) = -c_i^F + \mathbb{E}_{V_i}\big[\max\{y, V_i \}\big] \ . 
\end{equation}
If we P-open the box, we then need to choose between further F-opening and stopping. Therefore, the value function of this course of action is 
\begin{equation}\label{eq:J_B}
  J^P(\{i\}, \emptyset, y) = -c_i^P + \mathbb{E}_{T_i}\bigg[ \max \Big\{y, -c_i^F + \mathbb{E}_{V_i \mid T_i}\big[\max\{y, V_i \}  \big]  \Big\} \bigg] \ . 
\end{equation}

Let $\Lambda_i(y)$ be the set of  ``promising'' types of box $i$ where F-opening is warranted, that is
$$
\Lambda_i(y) := \left\{ t_i \in \Gamma_i  \mid  \sigma_i^{F \mid t_i} > y \right\}= \left \{ t_i \in \Gamma_i  \mid  -c_i^F + \mathbb{E}[(V_i-y)^+ \mid T_i=t_i]>0  \right \} \ .
$$
The difference between F-opening \eqref{eq:J_A} and P-opening and \eqref{eq:J_B} can be written as
\begin{eqnarray}
& &    J^{F}(\{i\}, \emptyset, y) - J^{P}(\{i\}, \emptyset, y)   \nonumber \\ 
 &=& -c_i^F  + \mathbb{E}\big[ \max\{y, V_i \} \big] - \Bigg( -c_i^P + \sum_{T_i \in \Lambda_i(y)} \operatorname{Pr}\big[T_i=t_i\big] \Big( -c_i^F + \mathbb{E}\big[\max\{y, V_i \}  \mid T_i=t_i  \big]  \Big) + \sum_{T_i \notin  \Lambda_i(y)} \operatorname{Pr}\big[T_i=t_i\big] y           \Bigg)  \nonumber \\
   &=&    \sum_{T_i \in \Lambda_i(y)} \operatorname{Pr} \big[T_i=t_i\big]  \Big( -c_i^F+\mathbb{E}\big[ \max\{y, V_i \}  \mid T_i=t_i \big]\Big) + \sum_{T_i \notin \Lambda_i(y)}\operatorname{Pr} \big[T_i=t_i\big]  \Big( -c_i^F+\mathbb{E}\big[ \max\{y, V_i \}  \mid T_i=t_i \big] \Big) \nonumber  \\
   &+& c_i^P + \sum_{T_i \in \Lambda_i(y)} \operatorname{Pr}\big[T_i=t_i\big] \Big( c_i^F - \mathbb{E}\big[\max\{y, V_i \}  \mid T_i=t_i  \big]  \Big)  - \sum_{T_i \notin  \Lambda_i(y)} \operatorname{Pr}\big[T_i=t_i\big] y        \nonumber   \\
   &=& c_i^P + \sum_{T_i \notin  \Lambda_i(y)} \operatorname{Pr}\big[T_i=t_i\big]  \left( \underbrace{-c_i^F+\mathbb{E}\big[  (V_i-y)^+  \mid T_i=t_i \big] }_{\leq 0} \right).   \label{eq:lemma:monotonicity_myopic_gain_SW}
\end{eqnarray}
Using simple arithmetics, one can show that $J^{F}(\{i\}, \emptyset, y) - J^{P}(\{i\}, \emptyset, y)$ is strictly decreasing in $y$. Moreover, the solution to $J^{F}(\{i\}, \emptyset, y) - J^{P}(\{i\}, \emptyset, y) = 0$ is exactly $y=\sigma_i^{F/P}$. 
The optimality of the threshold-based policy for the single-box case then follows from the optimal stopping and the condition about the dominance between F-opening and P-opening, which is determined by the relation between $y$ and $\sigma_i^{F/P}$.

\subsection{The expected profit of the optimal policy in the single closed-box case}\label{apx:single_box_expression}

In what follows, we calculate the expected profit of this optimal policy in closed form. 
\begin{enumerate}
    \item First, if the policy does not stop and F-opens box $i$, the expected profit is
\begin{eqnarray*}
   & & -c_i^F + \operatorname{Pr} \left[ V_i > \sigma_i^F \right] \mathbb{E} \left[ V_i \mid V_i > \sigma_i^F \right]  + \operatorname{Pr} \left[ V_i \leq \sigma_i^F \right]  \mathbb{E} \left[ \max\{y, V_i \}  \mid V_i \leq \sigma_i^F \right]  \\
   &=& \operatorname{Pr} \left[ V_i > \sigma_i^F \right] \sigma_i^F + \operatorname{Pr} \left[ V_i \leq \sigma_i^F \right]  \mathbb{E} \left[ \max\{y, V_i \}  \mid V_i \leq \sigma_i^F \right]    \\
   &=&  \mathbb{E} \left[ \max \left\{y, \min \left\{V_i,  \sigma_i^F \right\}    \right\} \right],
\end{eqnarray*}
where the first equation follows from the definition of F-threshold: $c_i^F = \mathbb{E} \left[ \left( V_i - \sigma_i^F \right)^+   \right] = \operatorname{Pr} \left[ V_i > \sigma_i^F \right] \left( \mathbb{E} \left[ V_i \mid V_i > \sigma_i^F \right] - \sigma_i^F \right)$. Note the terms can be rearranged so that 
$$
-c_i^F + \operatorname{Pr} \left[ V_i > \sigma_i^F \right] \mathbb{E} \left[ V_i \mid V_i > \sigma_i^F \right] = \operatorname{Pr} \left[ V_i > \sigma_i^F \right] \sigma_i^F.
$$
    \item By the same reasoning, following from the definition of the P-threshold $\sigma_i^P$,
\begin{eqnarray*}
  c_i^P &=& \mathbb{E} \Big[\max\{0, -c_i^F + \mathbb{E} \big[ (V_i-\sigma^{P}_i)^+ \mid T_i \big]\}\Big] \\
   &=& \sum_{t_i: \sigma_i^{F \mid t_i} > \sigma_i^P } \operatorname{Pr} \left[ T_i=t_i \right] \Big( -c_i^F + \operatorname{Pr} \left[ V_i > \sigma_i^P \mid T_i=t_i \right]  \left( \mathbb{E}\left[ V_i \mid  V_i > \sigma_i^P, T_i=t_i \right] - \sigma_i^P \right) \Big).
\end{eqnarray*}
Rearranging terms on the right side to the left of the latter equation yields that
\begin{eqnarray}
   & & -c_i^P + \sum_{t_i: \sigma_i^{F \mid t_i} > \sigma_i^P } \operatorname{Pr} \left[ T_i=t_i \right] \Big( -c_i^F + \operatorname{Pr} \left[ V_i > \sigma_i^P \mid T_i=t_i \right] \mathbb{E}\left[ V_i \mid  V_i > \sigma_i^P, T_i=t_i \right] \Big) \nonumber \\
   &=& \sum_{t_i: \sigma_i^{F \mid t_i} > \sigma_i^P } \operatorname{Pr} \left[ T_i=t_i \right]  \operatorname{Pr} \left[ V_i > \sigma_i^P \mid T_i=t_i \right] \sigma_i^P \ .  \label{eq:single_box_P_open}
\end{eqnarray}

Therefore, if the policy does not stop and P-opens box $i$ first, the expected profit is
\begin{eqnarray*}
   & &  -c_i^P + \sum_{t_i: \sigma_i^{F \mid t_i} > \sigma_i^P } \operatorname{Pr} \left[ T_i=t_i \right] \Big ( -c_i^F + \operatorname{Pr} \left[ V_i > \sigma_i^P \mid T_i=t_i \right] \mathbb{E}\left[ V_i \mid  V_i > \sigma_i^P, T_i=t_i \right]  \\
   & & \quad \quad \quad \quad \quad \quad \quad \quad \quad \quad \quad \quad   +   \operatorname{Pr} \left[ V_i \leq \sigma_i^P  \mid T_i=t_i\right]  \mathbb{E} \left[ \max\{ y, V_i \}   \mid V_i \leq \sigma_i^P, \sigma_i^{F \mid t_i} > \sigma_i^P  \right] \Big)     \\
   & & + \sum_{t_i: \sigma_i^{F \mid t_i} \leq \sigma_i^P } \operatorname{Pr} \left[ T_i=t_i \right] \left(  \BFI \left\{  y > \sigma_i^{F \mid t_i} \right\} \cdot y +    \BFI \left\{  y \leq \sigma_i^{F \mid t_i} \right \} \left( -c_i^F + \mathbb{E} \left[ \max\{y, V_i\} \mid y \leq \sigma_i^{F \mid t_i},  \sigma_i^{F \mid t_i} \leq \sigma_i^P  \right] \right)    \right)      \\
   &=& \sum_{t_i: \sigma_i^{F \mid t_i} > \sigma_i^P } \operatorname{Pr} \left[ T_i=t_i \right] \Big( \operatorname{Pr} \left[ V_i > \sigma_i^P \mid T_i=t_i \right] \sigma_i^P +  \operatorname{Pr} \left[ V_i \leq \sigma_i^P  \mid T_i=t_i\right]  \mathbb{E} \left[ \max\{ y, V_i \}   \mid V_i \leq \sigma_i^P, \sigma_i^{F \mid t_i} > \sigma_i^P  \right]   \Big) \\
   & & + \sum_{t_i: \sigma_i^{F \mid t_i} \leq \sigma_i^P } \operatorname{Pr} \left[ T_i=t_i \right]  \left(\BFI \left\{  y > \sigma_i^{F \mid t_i} \right\} \cdot y +  \BFI \left\{  y \leq \sigma_i^{F \mid t_i} \right \}   \mathbb{E} \left[ \max \{ y, \min \{ V_i, \sigma_i^{F \mid t_i} \} \} \mid y \leq \sigma_i^{F \mid t_i},  \sigma_i^{F \mid t_i} \leq \sigma_i^P  \right]   \right)    \\
   &=& \mathbb{E} \left[ \max \left\{ y, \min\{V_i, \sigma_i^{F \mid T_i}, \sigma_i^P \}
  \right\} \right].
\end{eqnarray*}
Here, the first equality follows by introducing \eqref{eq:single_box_P_open} and using the definition of F-threshold again. Indeed, for a type $t_i$ such that $y \leq \sigma_i^{F \mid t_i} \leq \sigma_i^P$, $c_i^F = \p{V_i > \sigma_i^{F \mid t_i} } \E{ V_i - \sigma_i^{F \mid t_i} \mid V_i > \sigma_i^{F \mid t_i}, y \leq \sigma_i^{F \mid t_i} \leq \sigma_i^P}$. Therefore, we have 
\begin{small}
\begin{eqnarray*}
    & & -c_i^F + \mathbb{E} \left[ \max\{y, V_i\} \mid y \leq \sigma_i^{F \mid t_i},  \sigma_i^{F \mid t_i} \leq \sigma_i^P  \right]  \\
    &=& - \p{V_i > \sigma_i^{F \mid t_i} } \E{ V_i - \sigma_i^{F \mid t_i} \mid V_i > \sigma_i^{F \mid t_i}, y \leq \sigma_i^{F \mid t_i} \leq \sigma_i^P} \\
    & & + \p{V_i > \sigma_i^{F \mid t_i} } \mathbb{E} \left[ V_i \mid V_i > \sigma_i^{F \mid t_i} ,  y \leq \sigma_i^{F \mid t_i},  \sigma_i^{F \mid t_i} \leq \sigma_i^P  \right] 
    + \p{V_i \leq \sigma_i^{F \mid t_i} } \mathbb{E} \left[ \max\{y, V_i\} \mid V_i \leq \sigma_i^{F \mid t_i} ,  y \leq \sigma_i^{F \mid t_i},  \sigma_i^{F \mid t_i} \leq \sigma_i^P  \right]  \\
    &=& \p{V_i > \sigma_i^{F \mid t_i} } \E{ \sigma_i^{F \mid t_i} \mid  V_i > \sigma_i^{F \mid t_i}, y \leq \sigma_i^{F \mid t_i} \leq \sigma_i^P  }  
    + \p{V_i \leq \sigma_i^{F \mid t_i} } \mathbb{E} \left[ \max\{y, V_i\}  \mid V_i \leq \sigma_i^{F \mid t_i} ,  y \leq \sigma_i^{F \mid t_i},  \sigma_i^{F \mid t_i} \leq \sigma_i^P  \right]  \\ 
    &=& \mathbb{E} \left[ \max \{ y, \min \{ V_i, \sigma_i^{F \mid t_i} \} \} \mid y \leq \sigma_i^{F \mid t_i},  \sigma_i^{F \mid t_i} \leq \sigma_i^P  \right] \ . 
\end{eqnarray*}
\end{small}

\end{enumerate}

\subsection{Optimal stopping}\label{apx:stopping}

We proceed to prove the optimal stopping rule in the general case. 
The overall proof strategy is to show that the stopping decision regarding a box (i.e., never open this box in the future) is independent of the existence of other boxes. 
Then, the optimal stopping rule shown in Section~\ref{sec:stopping} for the single box setting can be applied to the general case.

We call a closed box $i$  ``F-expired'' if $y \geq \max \sigma_i^F$. 
We call a partially open box $i$ ``F-expired'' if $y \geq \sigma_i^{F \mid t_i}$. 
We call a closed box $i$ P-expired if $y \geq \sigma_i^P$. 
In what follows, we show that there exists an optimal policy that never F-opens an F-expired box and P-opens a P-expired box.

\paragraph{Case 1: F-expired box.}
First, we show that an optimal policy never F-opens an F-expired box. 
The proof proceeds by contradiction. 
Suppose that in state $(\mathcal{C}, \mathcal{P}, y)$ the optimal policy $\pi$ F-opens an F-expired box $i$. 
We define the information set $I_i$ of box $i$ to be $I_i \triangleq \emptyset$ if box $i$ is closed and $I_i \triangleq \{T_i=t_i \}$ if box $i$ is partially opened. 
By the Bellman equation, we have
$$
J^\pi(\mathcal{C}, \mathcal{P}, y) = -c_i^F + \mathbb{E}[J^\pi (\mathcal{C}', \mathcal{P}', \max\{y, V_i \} ) \mid I_i ],
$$
where $\mathcal{C}'=\mathcal{C}  \setminus \{i\}, P'=P$ if $i \in \mathcal{C}$ and $\mathcal{C}'=\mathcal{C}, \mathcal{P}'=\mathcal{P}\setminus \{i\}$ if $i \in \mathcal{P}$.

We construct an alternative policy $\pi'$ that does not open box $i$; instead, it ``simulates'' a value for $V_i$, discards box $i$ for the remainder of the inspection process, and imitates the policy $\pi$ after that, with the only difference that $\pi'$ selects $y$ whenever policy $\pi$ selects box $i$. 
With this definition at hand, it is not difficult to show that the expected profit under policy $\pi'$ is given by
$$
J^{\pi'}(\mathcal{C}, \mathcal{P}, y) = \mathbb{E}\left[J^\pi (\mathcal{C}', \mathcal{P}', y  ) \mid I_i \right].
$$
The remainder of the proof proceeds by coupling the realizations of $V_i$ under the two policies to compare their expected profits. For any sample path $\rho$, we denote the realization of $V_i$ by $V_i^\rho$. Then, the difference in the value functions of the two policies is
\begin{eqnarray*}
 J_\rho^{\pi}(\mathcal{C}, \mathcal{P}, y) - J_\rho^{\pi'}(\mathcal{C}, \mathcal{P},y ) &=&  \left\{\begin{array}{ll}{ -c_i^F  } & { \text{if box i is not selected}} \\ {- c_i^F + \max \{y, V_i^\rho \} - y  }& { \text{ if box i is selected }}   \end{array}\right. \\
   &\leq& \left\{\begin{array}{ll}{ -c_i^F +  \max \{y, V_i^\rho \} - y  } & { \text{if box i is not selected}} \\ {- c_i^F + \max \{y, V_i^\rho \} - y }& { \text{ if box i is selected }}   \end{array}\right..
\end{eqnarray*}
Therefore, by taking expecations on both sides, we obtain 
$$
J^{\pi}(\mathcal{C}, \mathcal{P}, y ) - J^{\pi'}(\mathcal{C}, \mathcal{P}, y ) = \mathbb{E}_\rho \left[ J_\rho^{\pi}(\mathcal{C}, \mathcal{P}, y) - J_\rho^{\pi'}(\mathcal{C}, \mathcal{P},y ) \right] \leq \mathbb{E}\left[- c_i^F + \max \{y, V_i\} - y  \mid I_i\right] \ .
$$
Hence, other boxes do not affect the stopping decision regarding box $i$. 
Recalling the optimal stopping policy in the single box setting, one can easily verify that $\mathbb{E}\left[- c_i^F + \max \{y, V_i\} - y  \mid I_i\right] \leq 0 $ if box $i$ is F-expired.  
We conclude that there exists an optimal policy that does not F-open an F-expired box.

\paragraph{Case 2: P-expired box.}
Next, we prove that there exists an optimal policy that never P-opens a P-expired box. 
Suppose that in state $(\mathcal{C}, \mathcal{P}, y)$, the optimal policy $\pi$ P-opens a box $i$ with $y \geq \sigma_i^P $. 
We denote its value function by $J^\pi(\mathcal{C}, \mathcal{P}, y)$. Hence, we have 
$$
J^\pi(\mathcal{C}, \mathcal{P}, y) = -c_i^P + \mathbb{E}[J^\pi (\mathcal{C} \setminus \{i\}, \mathcal{P} \cup \{(i,T_i)\}, y ) ].
$$

Similar to Case 1, we construct an alternative policy $\pi'$ that ``simulates'' a type for $T_i$, discards box $i$ for the remainder of the inspection process, and imitates the policy $\pi$ after that, with the only difference that $\pi'$ selects $y$ whenever policy $\pi$ selects the box $i$. 
For any sample path $\rho$, we denote the realization of $T_i$ and $V_i$ by $t_i^\rho$ and $V_i^\rho$, respectively.
Then the difference between the expected profits of the two policies is
\begin{eqnarray}
& &  J_\rho^{\pi}(\mathcal{C}, \mathcal{P}, y) - J_\rho^{\pi'}(\mathcal{C}, \mathcal{P}, y) \nonumber \\ 
&=&
 \left\{\begin{array}{ll}
 { -c_i^P  } & { \text{if box i is not F-opened by }\pi} \\
 {-c_i^P - c_i^F   }& { \text{ if box i is F-opened, but not selected by } \pi }  \\
  {-c_i^P - c_i^F  + \max\{ V_i^\rho,y \}-y }              & { \text{ if box i is F-opened, and selected by } \pi  }
 \end{array}\right. \nonumber\\
   &\leq&
   \left\{\begin{array}{ll}
 { -c_i^P  } & { \text{if box i is not F-opened by }\pi} \\
 {-c_i^P - c_i^F  + \max\{ V_i^\rho,y \}-y    }& { \text{ if box i is F-opened, but not selected by } \pi }  \\
  {-c_i^P - c_i^F  + \max\{ V_i^\rho,y \}-y }              & { \text{ if box i is F-opened, and selected by } \pi  }
 \end{array}\right..   \label{eq:expired_box_sample_path_1}
\end{eqnarray}
We denote by $q_i^\pi(t_i^\rho) \in [0,1]$ the probability that policy $\pi$ F-opens box $i$ when the revealed type is $t_i^\rho$, and we define the set of  ``promising'' types  $\Lambda_i(y)=\{t_i \in \Gamma_i \mid y < \sigma_i^{F \mid t_i} \}$ which warrant F-opening. 

We next write an explicit expression for the expectation of the difference, splitting to cases based on \eqref{eq:expired_box_sample_path_1} and obtain the following:
\begin{eqnarray}
  J^{\pi}(\mathcal{C}, \mathcal{P},y ) - J^{\pi'}(\mathcal{C}, \mathcal{P},y )  &=&\mathbb{E}_\rho \left[J_\rho^{\pi}(\mathcal{C}, \mathcal{P}, y) - J_\rho^{\pi'}(\mathcal{C}, \mathcal{P}, y)    \right] \nonumber \\
   &\leq& -c_i^P + \sum_{t_i^\rho \in \Lambda_i(y)}\operatorname{Pr}[T_i=t_i^\rho]q_i^\pi(t_i^\rho) \left( \underbrace{-c_i^F + \mathbb{E}[(V_i-y)^+ \mid T_i=t_i^\rho]}_{>0}   \right)  \nonumber  \\
   & & + \left(\sum_{t_i^\rho \notin \Lambda_i(y)}\operatorname{Pr}[T_i=t_i^\rho]+\sum_{t_i^\rho \in \Lambda_i(y)}\operatorname{Pr}[T_i=t_i^\rho]\left(1-q_i^\pi(t_i^\rho)\right) \right) \cdot 0    \label{eq:expired_box_sample_path_2}  \\
   &\leq& -c_i^P + \sum_{t_i^\rho \in \Lambda_i(y)}\operatorname{Pr}[T_i=t_i^\rho] \left( \underbrace{-c_i^F + \mathbb{E}[(V_i-y)^+ \mid T_i=t_i^\rho]}_{>0}   \right) \label{eq:expired_box_sample_path_3} \ . 
\end{eqnarray}
Inequality \eqref{eq:expired_box_sample_path_2} is based on \eqref{eq:expired_box_sample_path_1}, multiplied by the conditional probability of type $T_i$.

We showed that there exists an optimal policy that stops when all boxes F-expire and P-expire. On the other hand, if there is even a single box that did not F-expire or P-expire, opening such box is better than stopping. The latter follows from the the single-box case. Therefore, the stopping condition is both sufficient and necessary.

\subsection{Proof of Theorem~\ref{thm:just_strong}}
\label{appendix:thm_just_strong}

The proof consists of two parts, dealing with cases where the box that attains the largest F-opening threshold is either partially open or closed. We prove these cases as separate lemmas.

\begin{lemma} \label{lemma:just_strong_partial}
For every state $(\mathcal{C}, \mathcal{P}, y)$ where it is suboptimal to stop,  if the F-threshold of a partially opened box $k$ is the largest opening threshold among boxes in that state, then there exists an optimal policy that F-opens box $k$ at that state.
\end{lemma}

\begin{myproof}
The lemma applies to cases in which there is at least one partially opened box, implying $\mid\mathcal{P}\mid \geq 1$. We prove the lemma by induction on the maximal potential number of box openings from a given state, which is captured by the expression:  $2\mid\mathcal{C}\mid + \mid\mathcal{P}\mid$ (each partially opened box can be opened at most once, while closed boxes can be opened at most twice, first being P-opened and then F-opened). We denote by $k$ the index of the partially open box whose F-threshold is the largest opening threshold in state $(\mathcal{C}, \mathcal{P}, y)$. With a slight abuse of notation, we denote by $\sigma_k$  the conditional F-threshold of box $k$:
$$
\sigma_k 
\triangleq 
\sigma_k^{F \mid t_k} = \sigma_M\left(\mathcal{C},\mathcal{P} \right).
$$
Since stopping is assumed to be sub-optimal, the optimal stopping condition implies that $y< \sigma_k$.

As base cases, consider states where $2  \mid \mathcal{C} \mid + \mid \mathcal{P} \mid  \in \{1,2\}$. Since $\mid\mathcal{P}\mid \geq 1$, this implies that $\mid\mathcal{C}\mid=0$. We observe that whenever $\mid\mathcal{C}\mid=0$ the problem reduces to the standard Pandora's box problem, from which the optimality of the lemma follows.

For the induction step, suppose that the lemma holds in states where $2  \mid \mathcal{C} \mid + \mid \mathcal{P} \mid  =  3,\ldots,n $, where $n\geq 3$. We will show that the property still holds when $2  \mid \mathcal{C} \mid + \mid \mathcal{P} \mid  =  n + 1$.
Assume by contradiction that at a state $(\mathcal{C}, \mathcal{P}, y)$ where $2  \mid \mathcal{C} \mid + \mid \mathcal{P} \mid  =  n + 1$ any  optimal policy does not F-open box $k$. We will construct an alternative policy $\pi'$ that does F-open box $k$ at that state and which performs at least as well as policy $\pi$.
Since stopping is suboptimal in state $(\mathcal{C}, \mathcal{P}, y)$, there are two possibilities for the next action under policy $\pi$; it either F-opens or P-opens another box.

\paragraph{Case~1: Policy $\pi$ F-opens a box $i\neq k$.} 

Suppose first that policy $\pi$  F-opens a box $i \in \mathcal{C}$. Then the system transitions to a state $(\mathcal{C}', \mathcal{P}', y')$ where $2  \mid \mathcal{C} ' \mid  + \mid \mathcal{P}' \mid  = n-1$. 
From Theorem~\ref{thm:myopic_stopping}, in the next state it is optimal to stop if $V_i > \sigma_k$, since $\sigma_k$ remains the largest threshold. Therefore, if $V_i > \sigma_k$, policy $\pi$ stops, otherwise, using the induction hypothesis, we know that policy $\pi$ will F-open box $k$ and stop afterward if $V_k > \sigma_k$. 
Defining $p_s = \operatorname{Pr} \Big[V_s > \sigma_k \Big]$ for each box $s \in \{i,k\}$, we can write the expected profit under policy $\pi$ as
\begin{eqnarray*}
  J^{\pi}(\mathcal{C}, \mathcal{P}, y) &=& - c_i^F + p_i \cdot \mathbb{E}\left[V_i \mid V_i > \sigma_k\right]  \\
   &+&   (1-p_i) \cdot \Big( -c_k^F +   p_k \cdot \mathbb{E}\left[V_k \mid V_k > \sigma_k\right] \\ 
   && +  (1-p_k) \cdot \mathbb{E} \left[   J^{\pi}(\mathcal{C}', \mathcal{P}', \max \left \{y, V_k, V_i \right \}  ) \mid   V_k \leq \sigma_k, V_i \leq \sigma_k \right] \Big),
\end{eqnarray*}
where $\mathcal{C}' = \mathcal{C} \setminus \{ i\} $ and $\mathcal{P}' = \mathcal{P} \setminus \{ k \}$.

We construct an alternative policy $\pi'$ that first F-opens box $k$ and selects box $k$ if $V_k > \sigma_k$; otherwise, policy $\pi'$ F-opens box $i$, and proceeds to  selecting box $i$ if $V_i > \sigma_k$. In the event that $V_i \leq \sigma_k$ (and $V_k \leq \sigma_k$), policy $\pi'$ imitates  policy $\pi$ thereafter. Therefore, the expected profit of policy $\pi'$ is
\begin{eqnarray*}
  J^{\pi'}(\mathcal{C}, \mathcal{P}, y) &=& - c_k^F + p_k  \cdot \mathbb{E}\left[V_k \mid V_k > \sigma_k\right]  \\
   &+&   (1-p_k) \cdot \Big( -c_i^F +   p_i \cdot \mathbb{E}\left[V_i \mid V_i > \sigma_k\right] \\ 
   && + (1-p_i) \cdot \mathbb{E} \left[    J^{\pi}(\mathcal{C}', \mathcal{P}', \max\{y, V_k, V_i \}  ) \mid  V_k \leq \sigma_k, V_i \leq \sigma_k \right] \Big).
\end{eqnarray*}
Using simple arithmetics, we can write the difference between $ J^{\pi}(\mathcal{C}, \mathcal{P}, y) $ and $ J^{\pi'}(\mathcal{C}, \mathcal{P}, y) $ as follows: 
\begin{eqnarray}
  J^{\pi'}(\mathcal{C}, \mathcal{P}, y) -J^{\pi}(\mathcal{C}, \mathcal{P}, y) &=& - c_k^F + p_k \cdot \mathbb{E}\left[V_k \mid V_k > \sigma_k\right] + (1-p_k) \cdot \left(   -c_i^F +   p_i \cdot \mathbb{E}\left[V_i \mid V_i > \sigma_k\right] \right) \nonumber \\
   & &  +  c_i^F  - p_i \cdot \mathbb{E}\left[V_i \mid V_i > \sigma_k\right] - (1-p_i) \cdot \left(   -c_k^F +   p_k \cdot \mathbb{E}\left[V_k \mid V_k > \sigma_k\right] \right)  \nonumber \\
   &=& p_k p_i \cdot \left( \mathbb{E}\left[V_k \mid V_k > \sigma_k\right] - \mathbb{E}\left[V_i \mid V_i > \sigma_k\right] \right)  + p_k c_i^F - p_i c_k^F \nonumber \\
   &=& p_k p_i \cdot \left( \mathbb{E}\left[V_k -\sigma_k \mid V_k > \sigma_k\right] - \mathbb{E}\left[V_i -\sigma_k  \mid V_i > \sigma_k\right] \right)  + p_k c_i^F - p_i c_k^F \nonumber \\
   &=& p_i \cdot \mathbb{E}\left[(V_k -\sigma_k )^+\right] - p_k \cdot \mathbb{E}\left[(V_i -\sigma_k )^+\right] + p_k c_i^F - p_i c_k^F   \nonumber \\
   &=& p_i c_k^F  - p_k \cdot \mathbb{E}\left[(V_i -\sigma_k )^+\right] + p_k c_i^F - p_i c_k^F \nonumber \\
   &=& - p_k \cdot \mathbb{E}\left[(V_i -\sigma_k )^+\right] + p_k \cdot \mathbb{E} \left[(V_i -\sigma_i^F )^+\right] \nonumber .
\end{eqnarray}
The latter is nonnegative whenever $\sigma_k \geq \sigma_i^F$, which is the case considered in the lemma. Therefore,  policy $\pi'$ achieves at least the same expected profit as policy $\pi$, which concludes the proof of this case. 

We briefly note that the case in which policy $\pi$ F-opens a box $i \in \mathcal{P}$ can be proved similarly and, for brevity, is omitted.

\paragraph{Case~2: Policy $\pi$ starts by P-opening a box $i \in \mathcal{C}$.}
Suppose policy $\pi$ P-opens a box $i \in \mathcal{C}$. Then the system transitions to a state $(\mathcal{C}', \mathcal{P}', y)$ for which $2  \mid \mathcal{C} ' \mid  + \mid \mathcal{P}' \mid  = n$. By our induction hypothesis, after this P-opening action, if the revealed type $t_i$ is such that $\sigma_i^{F \mid t_i} > \sigma_k$, policy $\pi$ will immediately F-open box $i$. Consequently,  by the induction hypothesis, after box $i$ is F-opened, if it is suboptimal to stop, policy $\pi$ will then F-open box $k$.
Otherwise, if the revealed type $t_i$ is such that $\sigma_i^{F \mid t_i} \leq \sigma_k$, then, by the induction hypothesis, policy  $\pi$ will F-open box $k$. Policy $\pi$ is illustrated Figure~\ref{diag:just_strong_interchange_1}.
The expected profit of policy $\pi$ is
          \begin{eqnarray}
            J^{\pi}(\mathcal{C}, \mathcal{P}, y) &=& -c_{i}^{p}+ \operatorname{Pr} \left[ \sigma_{i}^{F \mid  T_{i}}>\sigma_{k}\right] \bigg( -c_{i}^{F} + \operatorname{Pr} \left[  V_{i} > \sigma_{k} \middle\vert \sigma_{i}^{F \mid  T_{i}}>\sigma_{k} \right] \mathbb{E} \left[  V_{i} \mid V_{i}>\sigma_{k}, \sigma_{i}^{F  \mid  T_{i}} > \sigma_{k}\right]  \nonumber \\
             &+&    \operatorname{Pr}\left[   V_{i} \leq \sigma_{k} ~\middle \vert~  \sigma_{i}^{F \mid  T_{i}} > \sigma_{k} \right] \left( -c_{k}^{F}+ \operatorname{Pr} \left[V_{k}>\sigma_{k}\right] \cdot \mathbb{E}\left[V_{k} \mid V_{k}>\sigma_{k}\right]
             + \operatorname{Pr}\left[V_{k} \leq  \sigma_{k}\right] \cdot J_1  \right) \bigg) \nonumber  \\
             &+&  \operatorname{Pr} \left[\sigma_{i}^{F \mid  T_i} \leq  \sigma_{k}\right] \Big( -c_{k}^{F}+\operatorname{Pr}\left[V_{k}>\sigma_{k}\right]  \mathbb{E}\left[V_{k} \mid V_{k}>\sigma_{k}\right]+ \operatorname{Pr}\left[V_{k} \leq \sigma_{k}\right] \cdot J_2 \Big),  \nonumber
          \end{eqnarray}
          where $ J_1 = \mathbb{E} \left[ J^{\pi} \left( \mathcal{C} \setminus \{ i \}, \mathcal{P} \setminus \{ k \}, \max\{ y, V_i, V_k\} \right) ~\middle \vert~ \sigma_{i}^{F \mid  T_{i}}>\sigma_{k}, V_i \leq \sigma_k, V_k \leq \sigma_k \right] $, and  \\
          $ J_2 = \mathbb{E} \left[ J^{\pi}\left( \mathcal{C} \setminus \{ i \}, \mathcal{P} \setminus \{ k \} \cup \{i \}, \max\{ y,  V_k\} \right) ~\middle \vert~ \sigma_{i}^{F \mid  T_{i}} \leq \sigma_{k}, V_k \leq \sigma_k \right] $.

\begin{figure}[htbp]
\centering
\begin{subfigure}{.5\textwidth}
  \centering
  \includegraphics[width=0.95\linewidth]{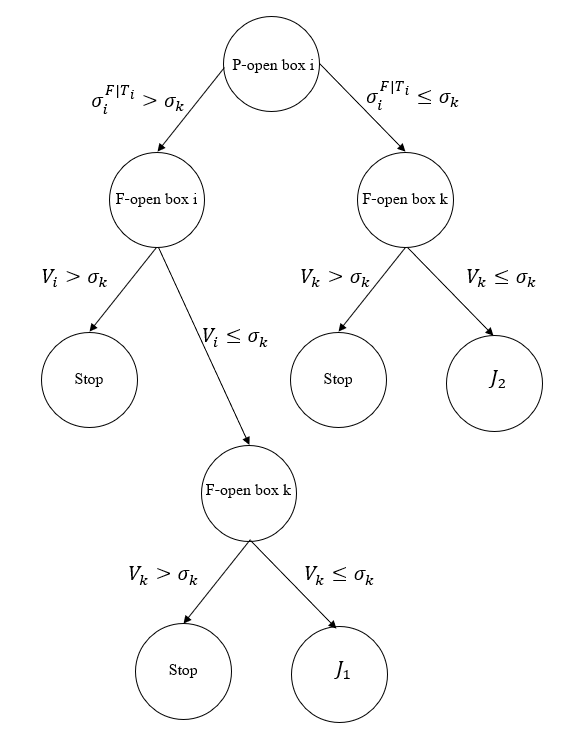}
    \caption{Policy $\pi$}
   \label{diag:just_strong_interchange_1}
\end{subfigure}%
\begin{subfigure}{.5\textwidth}
  \centering
  \includegraphics[width=0.8\linewidth]{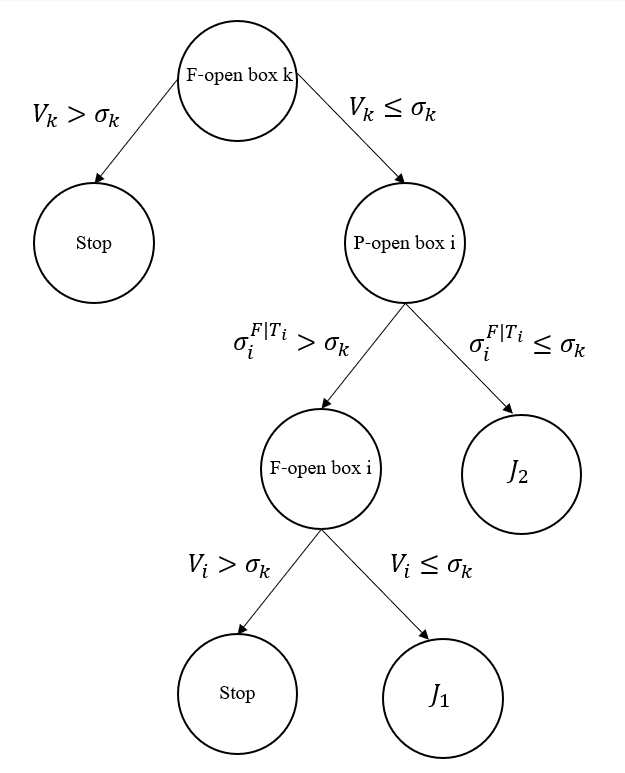}
  \caption{Policy $\pi'$}
  \label{diag:just_strong_interchange_2}
\end{subfigure}
\caption{ A graphical illustration of policy $\pi$ and policy $\pi'$ for Case 2, in the Proof of Lemma~\ref{lemma:just_strong_partial}.   }
\label{fig:just_strong_interchange_policy}
\end{figure}

Similarly to Case 1, we construct an alternative policy $\pi'$ that first F-opens box $k$, as shown in Figure~\ref{diag:just_strong_interchange_2}. If $V_k \leq \sigma_k$, policy $\pi'$  proceeds to P-opening box $i$, and it then F-opens box $i$ if the type $T_i$ is such that $\sigma_i^{F \mid T_i} > \sigma_k$.
Otherwise, $\pi'$ imitates policy $\pi$ in all subsequent states. 
Therefore, the expected profit of policy $\pi'$ can be written as follows: 
\begin{eqnarray}
& & J^{\pi'}(\mathcal{C}, \mathcal{P}, y)  \nonumber \\
 &=&  -c_{k}^{F} + \operatorname{Pr}\left[V_{k} > \sigma_{k}\right] \mathbb{E}\left[V_{k} \mid V_{k} > \sigma_{k}\right] \nonumber \\
 &+& \operatorname{Pr}\left[V_{k} \leq  \sigma_{k}\right] \bigg(  - c_i^P + \operatorname{Pr} \left[\sigma_{i}^{F \mid  T_{i}}>\sigma_{k}\right]  \Big(  -c_i^F + \operatorname{Pr}\left[ V_{i}>\sigma_{k} ~\middle \vert~ \sigma_{i}^{F \mid  T_{i}}>\sigma_{k} \right] \mathbb{E} \left[ V_i ~\middle \vert~ V_i > \sigma_k,  \sigma_{i}^{F \mid  T_{i}}>\sigma_{k}  \right] \nonumber \\
 &+& \operatorname{Pr}\left[V_{i} \leq \sigma_{k} ~\middle \vert~ \sigma_{i}^{F \mid  T_{i}}>\sigma_{k} \right] \cdot J_1 \Big) + \operatorname{Pr}\left[ \sigma_{i}^{F \mid  T_{i}} \leq \sigma_{k} \right] \cdot  J_2 \bigg).   \nonumber
\end{eqnarray}

Let $p_k = \operatorname{Pr}\left[V_{k}>\sigma_{k}\right]$, $q_i = \operatorname{Pr} \left[\sigma_{i}^{F \mid  T_{i}}>\sigma_{k}\right]$, and $p_i = \operatorname{Pr} \left[V_i > \sigma_k \mid \sigma_{i}^{F \mid  T_{i}}>\sigma_{k}\right]$. The difference between $ J^{\pi}(\mathcal{C}, \mathcal{P}, y) $ and $ J^{\pi'}(\mathcal{C}, \mathcal{P}, y) $ can be written as 
\begin{eqnarray}
& & J^{\pi'}(\mathcal{C}, \mathcal{P}, y) - J^{\pi}(\mathcal{C}, \mathcal{P}, y) \nonumber  \\
&=& \left( -c_k^F + q_i(1-p_i) c_k^F + (1-q_i) c_k^F \right) + \left( -(1-p_k)c_i^P + c_i^P \right)  + \left( -(1-p_k)q_ic_i^F + q_i c_i^F \right)  \nonumber \\
 & & +  p_k q_i p_i \cdot \mathbb{E}\left[V_{k} \mid V_{k}>\sigma_{k}\right]  -   p_k q_i p_i \cdot \mathbb{E} \left[  V_i ~ \middle \vert~    V_i > \sigma_k,  \sigma_{i}^{F \mid  T_{i}}>\sigma_{k}  \right]      \nonumber  \\
 &=& -q_i p_i c_k^F + p_k c_i^P + p_k q_i c_i^F   \nonumber  \\
 & & +  p_k q_i p_i \cdot \left( \mathbb{E}\left[V_{k} \mid V_{k}>\sigma_{k}\right]-\sigma_k \right)  -   p_k q_i p_i  \cdot \left( \mathbb{E} \left[ V_i ~\middle \vert~ V_i > \sigma_k,  \sigma_{i}^{F \mid  T_{i}}>\sigma_{k}  \right] -\sigma_k \right) \nonumber  \\
 &=& -q_i p_i c_k^F + p_k c_i^P + p_k q_i c_i^F   \nonumber  \\
 & & +   q_i p_i  \cdot \mathbb{E}\left[ \left( V_{k}  -\sigma_k  \right)^+\right]  -   p_k q_i \cdot \mathbb{E} \left[ \left( V_i - \sigma_k \right)^+ ~\middle \vert~ \sigma_{i}^{F \mid  T_{i}}>\sigma_{k}  \right]   \nonumber \\
 &=&  p_k \cdot \left(  c_i^P + q_i c_i^F    -   q_i \cdot \mathbb{E} \left[ \left( V_i - \sigma_k \right)^+ ~\middle \vert~ \sigma_{i}^{F \mid  T_{i}}>\sigma_{k}  \right] \right) \nonumber \\
 &=&  p_k \cdot \left(  c_i^P -  \operatorname{Pr} \left[\sigma_{i}^{F \mid  T_{i}}>\sigma_{k}\right] \left( - c_i^F    +    \mathbb{E} \left[ \left( V_i - \sigma_k \right)^+ ~\middle \vert~ \sigma_{i}^{F \mid  T_{i}}>\sigma_{k}  \right] \right) \right)  \nonumber   \\
 &=&  p_k \cdot \left(  \mathbb{E}\left[ \max \left\{0, -c_i^F + \mathbb{E}\left[ (V_i - \sigma_i^P )^+\right] \right\} \right] -  \mathbb{E}\left[ \max \left \{0, -c_i^F + \mathbb{E}\left[ (V_i - \sigma_k )^+\right] \right \} \right] \right)  \nonumber .
\end{eqnarray}
The latter is nonnegative whenever $\sigma_k \geq \sigma_i^P$, which is the case considered in the lemma. 
This shows that policy $\pi'$ achieves at least the same expected profit as policy $\pi$, thereby concluding the proof of Case 2 and the lemma.
\end{myproof}

We next turn to the second part of the theorem. To this end, we first make a notable observation. 

\begin{restatable}[]{lemma}{Pdominated} \label{lemma:P_dominated}
For a closed box $i \in \mathcal{C}$, P-opening is dominated if $\sigma_i^F > \sigma_i^P$.
\end{restatable}

Given this claim, it is straightforward to establish the second part of Theorem~\ref{thm:just_strong}. 
\begin{lemma} \label{lemma:just_strong_closed}
For every state $(\mathcal{C}, \mathcal{P}, y)$ such that it is suboptimal to stop,  if the F-threshold of a closed box $i$ with $\sigma_i^F > \sigma_i^P$ is the largest  opening threshold, then there exists an optimal policy that F-opens box $i$ immediately. 
\end{lemma}

\begin{myproof2}{Lemma \ref{lemma:just_strong_closed}}
According to Lemma~\ref{lemma:P_dominated}, P-opening box $i$ is a dominated action. 
Therefore, we can view box $i$ as a partially opened box that can only be F-opened. From Lemma~\ref{lemma:just_strong_partial} and the fact that $\sigma_i^F$ is the leading threshold, it follows that an optimal policy exists that F-opens box $i$ immediately.  
\end{myproof2}

\begin{myproof2}{Lemma \ref{lemma:P_dominated}}

Let $m$ be the index of the box with the largest opening threshold excluding box $i$, and let $\sigma_m$ be its opening threshold. i.e.,
$$
\sigma_m =  \max \left\{  \underset{j \in \mathcal{C} \setminus \{ i\}}{\max}  \left\{  \sigma_j^F, \sigma_j^P \right\}, \underset{ j \in \mathcal{P}}{\max }  \left\{ \sigma_j^{F \mid t_j}  \right\}   \right\}.
$$

If $y > \sigma_m$, then boxes other than box $i$ should not be opened (such boxes F-expire and P-expire, see the proof of Theorem~\ref{thm:myopic_stopping}), and the problem reduces to the single box case for which Theorem~\ref{thm:just_strong} is a direct consequence of Theorem~\ref{thm:myopic_stopping}. As such, in what follows, we consider the case where $y \leq \sigma_m$.

Suppose an optimal policy $\pi$ P-opens the box $i$ in state $(\mathcal{C}, \mathcal{P}, y)$. 
We will construct an alternative policy $\pi'$ that F-opens box $i$ instead at that state, and show that it achieves an expected profit  at least as high as that of policy $\pi$. 
According to Lemma \ref{lemma:just_strong_partial}, after P-opening box $i$, policy $\pi$ proceeds to F-opening box $i$ if $\sigma_i^{F \mid T_i} > \sigma_m $, and it will then stop if $V_i > \sigma_m$. 
  We denote by $G_i$ the set of types associated with box $i$ in which policy $\pi$ is guaranteed to F-open box $i$ immediately upon P-opening, i.e.,  $T_i \in G_i \Leftrightarrow \sigma_i^{F \mid T_i} > \sigma_m $.

  Therefore, the expected profit of policy $\pi$ is
  \begin{eqnarray}
    J^{\pi}(\mathcal{C}, \mathcal{P}, y) &=&  -c_i^P + \operatorname{Pr} \left[ T_i \in G_i  \right] \bigg( -c_i^F + \operatorname{Pr} \left[ V_i > \sigma_m \mid T_i \in G_i  \right] \mathbb{E}\left[ V_i \mid V_i > \sigma_m, T_i \in G_i   \right]    \nonumber  \\
     & & +  \operatorname{Pr} \left[ V_i \leq \sigma_m \mid T_i \in G_i  \right] \cdot J_1   \bigg)   +  \operatorname{Pr} \left[ T_i \notin G_i   \right] \cdot  J_2, 
    \label{eq:lemma:P_dominated_J}
  \end{eqnarray}
  where $J_1 = \mathbb{E} \left[  J^{\pi}(\mathcal{C} \setminus \{ i \} , \mathcal{P}, \max \{y, V_i  \} )  \mid V_i \leq \sigma_m , T_i \in G_i \right]$ and $J_2 = \mathbb{E}\left[ J^{\pi}(\mathcal{C} \setminus \{ i \} , \mathcal{P} \cup \{i, t_i \}, y )  \mid T_i \notin G_i  \right]$.

\begin{figure}
  \centering
  \includegraphics[width=10cm]{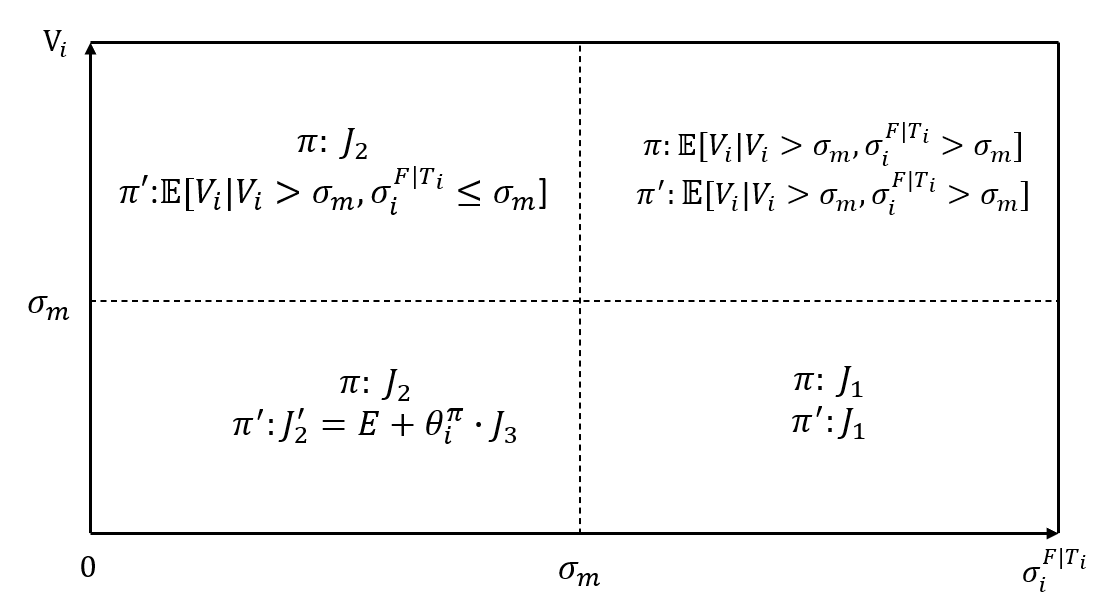}
  \caption{Illustration of payoffs of two policies in different scenarios. The y-axis denotes the realization of the prize, and the x-axis denotes the realization of the conditional threshold.}
  \label{fig:policy_align_just_strong_closed}
\end{figure}

\paragraph{Constructing an alternative policy $\pi'$.}
Let $\pi'$ be the policy that F-opens box $i$ in state $(\mathcal{C}, \mathcal{P}, y)$, and selects box $i$ if $V_i > \sigma_m$; otherwise, it further imitates policy $\pi$ with the only difference that, whenever policy $\pi$ F-opens box $i$, policy $\pi'$ who already observed the value of box $i$, does nothing and no longer incurs the cost $c_i^F$. Therefore, the expected profit of policy $\pi'$ is given by
  \begin{eqnarray}
    J^{\pi'}(\mathcal{C}, \mathcal{P}, y) &=&  -c_i^F + \operatorname{Pr} \left[ T_i \in G_i  \right]  \operatorname{Pr} \left[ V_i > \sigma_m \mid T_i \in G_i  \right] \mathbb{E}\left[ V_i \mid V_i > \sigma_m, T_i \in G_i   \right]  \nonumber  \\
    &+&    \operatorname{Pr} \left[ T_i \notin G_i  \right]  \operatorname{Pr} \left[ V_i > \sigma_m \mid T_i \notin G_i  \right] \mathbb{E}\left[ V_i \mid V_i > \sigma_m, T_i \notin G_i   \right]              \nonumber    \\
    &+& \operatorname{Pr} \left[ T_i \in G_i  \right]  \operatorname{Pr} \left[ V_i \leq \sigma_m \mid T_i \in G_i  \right] \cdot  J_1            \nonumber      \\
    &+& \operatorname{Pr} \left[ T_i \notin G_i  \right]  \operatorname{Pr} \left[ V_i \leq \sigma_m \mid T_i \notin G_i  \right]\cdot J_2', \label{eq:lemma:P_dominated_J'}  
  \end{eqnarray}
  where $J_2' =  \mathbb{E}\left[ J^{\pi'}(\mathcal{C} \setminus \{ i \} , \mathcal{P} , \max\{ y, V_i\} )  \mid V_i \leq \sigma_m ,  T_i \notin G_i  \right]$, which is the expected profit of policy $\pi'$ starting with the imitation of policy $\pi$, after observing that  $T_i \notin G_i$ and $V_i \leq \sigma_m $.

  In what follows, we argue that $J^{\pi'}(\mathcal{C}, \mathcal{P}, y) \geq J^{\pi}(\mathcal{C}, \mathcal{P}, y)$. 
  The argument is established by first decomposing $J_2$ and $J_2'$ in a way that facilitates the comparison between $J^{\pi}$ and $J^{\pi'}$.

\paragraph{Decomposition of $J_2$ and $J_2'$.}
   In order to compare the expected profits $J^{\pi}$ and $J^{\pi'}$, we decompose the terms $J_2$ and $J_2'$ appearing in equations~\eqref{eq:lemma:P_dominated_J} and~\eqref{eq:lemma:P_dominated_J'}, respectively. For this purpose, we can think of a policy as a decision tree consisting of nodes and edges. Each node represents a state-action pair and each edge captures a transition.  We call the nodes where the policy does not stop and opens a box \textit{interior} nodes, and the nodes where the policy stops are called \textit{terminal} nodes. Let $\text{box}(\tau)$ be the box that the policy  selects at a terminal node $\tau$. Let $c(\eta)$ be the opening cost paid at an interior node $\eta$.  Then, we have
  \begin{eqnarray}
    J_2 &=& \sum_{\text{terminal node } \tau   } \operatorname{Pr}\left[ \pi \text{ reaches } \tau \mid T_i \notin G_i \right] \cdot \mathbb{E} \left[ V_{\text{box}(\tau)} \mid \text{info regarding } \text{box}(\tau) \right]   \nonumber \\
     &-& \sum_{\text{interior node } \eta   }  \operatorname{Pr}\left[ \pi \text{ reaches } \eta \mid T_i \notin G_i \right] \cdot c(\eta) \nonumber .
  \end{eqnarray}
  For simplicity, we assume that there is only one node in the decision tree of policy $\pi$ that  F-opens box $i$ when $T_i \notin G_i $. By a slight abuse of language, we refer to this state-action pair as node $i$. 
  If there are multiple such nodes, the same reasoning applies. 
  Let $tree(i)$ be the subtree rooted at the node $i$. Then, on all paths outside $tree(i)$, policy $\pi$ does not F-open box $i$.
  According to Lemma~\ref{lemma:just_strong_partial}, if a conditional threshold greater than $\sigma_m$ is revealed by a P-opening inspection, the corresponding box is immediately F-opened in an optimal policy.
  Therefore, when node $i$ is reached, box $i$ will be F-opened, and it will be selected if $V_i > \sigma_m$. 
  We let $\theta_i^\pi = \operatorname{Pr}\left[ \pi \text{ reaches node } i \mid T_i \notin G_i \right]$. Based on the preceding discussion, $J_2$ can be expressed as
   \begin{eqnarray}
    J_2 &=& E + \theta_i^\pi  \cdot \Big( -c_i^F +   \operatorname{Pr} \left[ V_i > \sigma_m  \mid T_i \notin G_i  \right]   \mathbb{E}\left[ V_i \mid V_i > \sigma_m, T_i \notin G_i   \right]     + \operatorname{Pr} \left[ V_i \leq \sigma_m  \mid T_i \notin G_i  \right]  \cdot J_3  \Big),  \nonumber
  \end{eqnarray}
where we denote, for brevity, 
\begin{eqnarray}
 E &\triangleq& \sum_{ \substack{ \text{terminal node } \tau \\ \text{ outside } tree(i) }    } \operatorname{Pr}\left[ \pi \text{ reaches } \tau \mid T_i \notin G_i \right] \cdot \mathbb{E} \left[ V_{\text{box}(\tau)} \mid \text{info regarding } \text{box}(\tau) \right]  \nonumber \\
     &-& \sum_{\substack{ \text{interior node } \eta \\ \text{ outside } tree(i) }    }  \operatorname{Pr}\left[ \pi \text{ reaches } \eta \mid T_i \notin G_i \right] \cdot c(\eta)  .       \nonumber
\end{eqnarray}
We now turn our attention to the decomposition of $J_2'$. Let $J_3$ be the expected profit conditional on $T_i \notin G_i$ over all the sample paths that reach node $i$ in the decision tree associated with policy $\pi$.
After F-opening box $i$, policy $\pi'$ generates the same total profit as policy $\pi$ if node $i$ is not reached. 
If node $i$ is reached, policy $\pi'$ does not need to pay an additional cost for inspection. Since policy $\pi'$ has already observed that $T_i \notin G_i$ and $ V_i \leq \sigma_m $, we obtain $J_2' = E +  \theta_i^\pi  \cdot J_3$. In the next claim, we provide an upper bound on $E$ using the above decision-tree decomposition of costs. The proof appears at the end of this section.

\begin{figure}
  \centering
  \includegraphics[width=12cm]{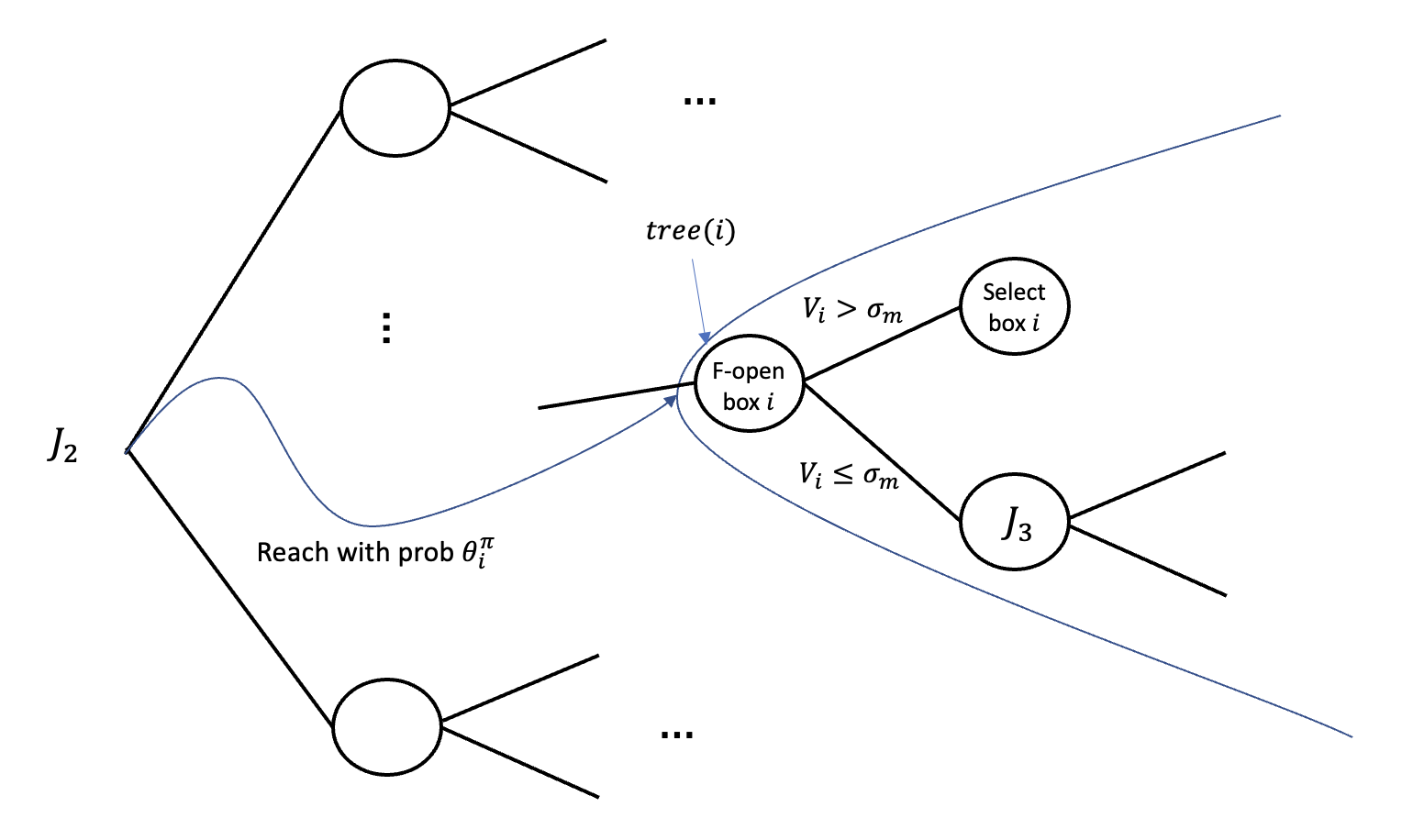}
  \caption{Illustration of $J_2$ via decision-tree-based coupling.}
  \label{fig:tree}
\end{figure}

\begin{claim}\label{lemma:E_bound} $E \leq  \left( 1- \theta_i^{\pi} \right) \cdot \sigma_m$ .    
\end{claim}

\paragraph{Bounding the difference between $J^{\pi'}$ and $J^{\pi}$.} Following the preceding discussion, we have 
\begin{eqnarray}
    & & J^{\pi'}(\mathcal{C}, \mathcal{P}, y) -  J^{\pi}(\mathcal{C}, \mathcal{P}, y) \nonumber \\
     &=& c_i^P + \operatorname{Pr} \left[ T_i \notin G_i  \right] \bigg( - \left(1- \theta_i^{\pi} \right) c_i^F  +  \left(1- \theta_i^{\pi} \right) \operatorname{Pr} \left[ V_i > \sigma_m \mid T_i \notin G_i  \right]  \mathbb{E}\left[ V_i \mid V_i > \sigma_m, T_i \notin G_i   \right]  \nonumber  \\
     & & -  \operatorname{Pr} \left[ V_i > \sigma_m \mid T_i \notin G_i  \right] \cdot E    \bigg)  \nonumber \\
     &\geq&    c_i^P + \operatorname{Pr} \left[ T_i \notin G_i  \right] \bigg( - \left(1- \theta_i^{\pi} \right) c_i^F  +  \left(1- \theta_i^{\pi} \right) \operatorname{Pr} \left[ V_i > \sigma_m \mid T_i \notin G_i  \right]  \mathbb{E}\left[ V_i \mid V_i > \sigma_m, T_i \notin G_i   \right]   \nonumber \\
     & &  -  \operatorname{Pr} \left[ V_i > \sigma_m \mid T_i \notin G_i  \right] \cdot \left(1- \theta_i^{\pi} \right) \sigma_m   \bigg)   \label{eq:thm_just_strong_closed_diff_1}
\end{eqnarray}
\begin{eqnarray}
     &=&  c_i^P +  \left(1- \theta_i^{\pi} \right)  \operatorname{Pr} \left[ T_i \notin G_i  \right] \bigg( - c_i^F  +   \operatorname{Pr} \left[ V_i > \sigma_m \mid T_i \notin G_i  \right] \left( \mathbb{E}\left[ V_i \mid V_i > \sigma_m, T_i \notin G_i   \right] - \sigma_m  \right) \bigg)   \nonumber \\
     &=&   c_i^P +  \left(1- \theta_i^{\pi} \right)  \operatorname{Pr} \left[ T_i \notin G_i  \right] \bigg( \underbrace{ - c_i^F  +  \mathbb{E} \left[ (V_i - \sigma_m )^+ \mid T_i \notin G_i \right]}_{\leq 0 \text{ since } \sigma_m \geq \sigma_i^{F \mid T_i}  \text{ for } T_i \notin G_i  } \bigg) \nonumber \\
     &\geq& c_i^P +  \operatorname{Pr} \left[ T_i \notin G_i  \right] \bigg(  - c_i^F  +  \mathbb{E} \left[ (V_i - \sigma_m )^+ \mid T_i \notin G_i \right]  \bigg) \nonumber \\
     &\geq& 0 \ .  \label{eq:to_explain} 
\end{eqnarray}
Equation~\eqref{eq:thm_just_strong_closed_diff_1} follows from Lemma~\ref{lemma:E_bound}.
The last inequality proceeds in the same reasoning as \eqref{eq:lemma:monotonicity_myopic_gain_SW} in the single-box setting, by noting that $ \left\{  T_i \notin G_i \right\} \Leftrightarrow  \left\{  T_i \notin \Lambda_i ( \sigma_m) \right\} $ and $\sigma_i^{F/P} > \sigma_m$.

To complete the proof of Theorem~\ref{thm:just_strong}, it remains to establish Claim~\ref{lemma:E_bound}.

\begin{myproof2}{Claim \ref{lemma:E_bound}}

We decompose the expected cumulative costs $E$ into three types of nodes (inspections or selections) in the decision-tree: (i) F-opening a box at an interior node, then selecting its reward when it is greater than $\sigma_m$; (ii) $P$-opening a box at an interior node, then $F$-opening it and selecting the reward when the conditional threshold and reward are greater than $\sigma_m$, respectively; and (iii) selecting a box at a terminal node with reward smaller or equal to $\sigma_m$. These three types partition all the nodes of the decision tree. Indeed, (i) when $\sigma_m$ is greater than all remaining thresholds, by Theorem \ref{thm:myopic_stopping}, it is optimal to select a box if its prize is found to be greater than $\sigma_m$; (ii) when a conditional threshold greater than $\sigma_m$ is found, the optimal policy immediately F-opens it by Lemma \ref{lemma:just_strong_partial}; and (iii) in all other cases, the prize value is bounded by $\sigma_m$.
Consequently, we obtain the following expression:
\begin{eqnarray}
  E &=& \sum_{\substack{ \text{interior node } \eta \\ \text{ outside } tree(i)   \\ \text{action} (\eta)= \text{F-open}  \\ \sigma_{\text{box}(\eta)} \leq \sigma_m  }    }  \operatorname{Pr} \left[ \pi \text{ reaches } \eta \mid T_i \notin G_i \right] \Bigg( - c_{\text{box}(\eta)}^{F} +  \operatorname{Pr} \left[ V_{\text{box}(\eta) } > \sigma_m \right] \mathbb{E} \left[ V_{\text{box}(\eta)} \mid V_{\text{box}(\eta)} > \sigma_m \right]  \Bigg)  \nonumber    \\
  &+& \sum_{\substack{ \text{interior node } \eta \\ \text{ outside } tree(i)   \\ \text{action} (\eta)= \text{P-open} }}  \operatorname{Pr} \left[ \pi \text{ reaches } \eta \mid T_i \notin G_i \right]  \Bigg(  - c_{\text{box}(\eta)}^P + \operatorname{Pr} \left[ \sigma_{\text{box}(\eta)}^{F \mid t_{\text{box}(\eta)}} > \sigma_m   \right] \nonumber  \\
  & & \cdot \bigg(  -c_{\text{box}(\eta)}^F + 
  \operatorname{Pr} \left[  V_{\text{box}(\eta)} > \sigma_m ~\middle \vert~ \sigma_{\text{box}(\eta)}^{F \mid t_{\text{box}(\eta)}} > \sigma_m  \right]  
  \mathbb{E} \left[  V_{\text{box}(\eta)} ~\middle \vert~ V_{\text{box}(\eta)} > \sigma_m ,  \sigma_{\text{box}(\eta)}^{F \mid t_{\text{box}(\eta)}} > \sigma_m  \right] \bigg) \Bigg) \nonumber    \\
  &+& \sum_{\substack{ \text{terminal node } \tau \\ \text{ outside } tree(i)   \\ V_{\text{box}(\tau)} \leq \sigma_m } }  \operatorname{Pr} \left[ \pi \text{ reaches } \tau \mid T_i \notin G_i \right]   \mathbb{E} \left[ V_{\text{box}(\tau)} \mid V_{\text{box}(\tau)} \leq \sigma_m \right]  \nonumber \\
  &\leq& \sum_{\substack{ \text{interior node } \eta \\ \text{ outside } tree(i)   \\ \text{action} (\eta)= \text{F-open} \\ \sigma_{\text{box}(\eta)} \leq \sigma_m  }}  \operatorname{Pr} \left[ \pi \text{ reaches } \eta \mid T_i \notin G_i \right] \Big(  \operatorname{Pr} \left[ V_{\text{box}(\eta) } > \sigma_m \right] \cdot \sigma_m \Big) \nonumber    \\
  &+& \sum_{\substack{ \text{interior node } \eta \\ \text{ outside } tree(i)   \\ \text{action} (\eta)= \text{P-open} }}  \operatorname{Pr} \left[ \pi \text{ reaches } \eta \mid T_i \notin G_i \right]  \Bigg(  \operatorname{Pr} \left[ \sigma_{\text{box}(\eta)}^{F \mid t_{\text{box}(\eta)}} > \sigma_m   \right] \operatorname{Pr} \left[  V_{\text{box}(\eta)} > \sigma_m ~\middle \vert~ \sigma_{\text{box}(\eta)}^{F \mid t_{\text{box}(\eta)}} > \sigma_m  \right] \cdot \sigma_m   \nonumber   \Bigg) \nonumber \\
  &+& \sum_{\substack{ \text{terminal node } \tau \\ \text{ outside } tree(i)   \\ V_{\text{box}(\tau)} \leq \sigma_m }}  \operatorname{Pr} \left[ \pi \text{ reaches } \tau \mid T_i \notin G_i \right]  \cdot \sigma_m  \nonumber \\
   &=& \Bigg( \sum_{ \substack{ \text{terminal node } \tau \\ \text{ outside } tree(i) }    }  \operatorname{Pr}\left[ \pi \text{ reaches } \tau \mid T_i \notin G_i \right] \Bigg) \cdot \sigma_m   \nonumber \\
  &=& \left( 1- \theta_i^{\pi} \right) \cdot \sigma_m.  \nonumber
\end{eqnarray}

  The above inequality is due to the following two facts:
   \begin{itemize}
     \item {\em Fact~1:} For any closed box $j$, by the definition of the F-threshold and monotonicity of the function  $ \mathbb{E} \left [ ( V_j  -  x )^+  \right]$ in $x$, we obtain $c_j^F \geq  \widetilde{ c_j^F}  \triangleq \mathbb{E} \left [ ( V_j  -  \sigma_m )^+  \right] $, which follows because $\sigma_m \geq \sigma_j^F$. Thus,
         \begin{equation}\label{eq:cost_accounting_ineq_1}
             \operatorname{Pr} \left[  V_j > \sigma_m \right] \sigma_m = - \widetilde{ c_j^F} +  \operatorname{Pr} \left[  V_j > \sigma_m \right]  \mathbb{E}\left[   V_j \mid V_j > \sigma_m \right]  \geq -  c_j^F +  \operatorname{Pr} \left[  V_j > \sigma_m \right]  \mathbb{E}\left[   V_j \mid V_j > \sigma_m \right].\nonumber
         \end{equation}
     \item  {\em Fact~2:} For any closed box $j$, by definition of the P-threshold and the monotonicity of $ \mathbb{E} \left[ \max \left\{ 0, - c_j^F + \mathbb{E}[ ( V_j - x )^+ ]  \right \} \right ]$ in $x$, we obtain $c_j^P \geq  \widetilde{ c_j^P}  \triangleq \mathbb{E} \left[ \max \left\{ 0, - c_j^F + \mathbb{E}[ ( V_j - \sigma_m )^+ ]  \right \} \right ]  $ since $\sigma_m \geq \sigma_j^P$. Thus,
  \begin{eqnarray}
     & & - c_j^P + \operatorname{Pr} \left[ \sigma_j^{F \mid T_j} > \sigma_m   \right] \left(  -c_j^F + \operatorname{Pr} \left[  V_j > \sigma_m ~\middle\vert~ \sigma_j^{F \mid T_j} > \sigma_m  \right] \mathbb{E} \left[  V_j ~\middle\vert~ V_j > \sigma_m ,  \sigma_j^{F \mid T_j} > \sigma_m \right]    \right) \nonumber \\
     &\leq &  - \widetilde{ c_j^P} + \operatorname{Pr} \left[ \sigma_j^{F \mid T_j} > \sigma_m   \right] \left(  -c_j^F + \operatorname{Pr} \left[  V_j > \sigma_m ~\middle\vert~ \sigma_j^{F \mid T_j} > \sigma_m  \right] \mathbb{E} \left[  V_j ~\middle\vert~ V_j > \sigma_m ,  \sigma_j^{F \mid T_j} > \sigma_m \right]    \right)  \nonumber  \\
     &=& \operatorname{Pr} \left[ \sigma_j^{F \mid T_j} > \sigma_m   \right] \operatorname{Pr} \left[  V_j > \sigma_m ~\middle\vert~ \sigma_j^{F \mid T_j} > \sigma_m  \right] \sigma_m . \label{eq:cost_accounting_ineq_2}\nonumber
  \end{eqnarray}
 \end{itemize}
 
\end{myproof2}

\end{myproof2}

\subsection{Proof of Theorem~\ref{thm:just_weak_closed}}
\label{appendix:thm_just_weak_closed}

We start by establishing an auxiliary lemma that states an optimal policy should not F-open a well-classified box. We then proceed to prove Theorem~\ref{thm:just_weak_closed}.

\begin{lemma}\label{lemma:well_classified_F_dominated}
For every state $(\mathcal{C}, \mathcal{P}, y)$, F-opening a well-classified box $i \in \mathcal{C}$ is suboptimal. 
\end{lemma}

\begin{myproof2}{Lemma \ref{lemma:well_classified_F_dominated}}

 We prove this by contradiction. 
 Suppose that in the current state 
 $(\mathcal{C}, \mathcal{P}, y)$ an optimal policy $\pi$ F-opens box $i$. In such case, the expected profit can be written as
  \begin{eqnarray}
    J^{\pi} (\mathcal{C}, \mathcal{P}, y) &=& - c_i^F + \mathbb{E}\left[  J^{\pi} (\mathcal{C} \setminus \{i \}, \mathcal{P}, \max\left\{ y, V_i \right\} )    \right]  \nonumber \\
     &=& -c_i^F +  \sum_{t_i: \sigma_i^{F \mid t_i} > \sigma_{-i}} \operatorname{Pr}\left[T_i=t_i \right]  \cdot  \mathbb{E}\left[ J^{\pi}  (\mathcal{C} \setminus \{i \}, \mathcal{P}, \max\left\{ y, V_i \right\} ) \mid T_i=t_i\right]   \nonumber \\
     &+&  \sum_{t_i: \sigma_i^{F \mid t_i} \leq \sigma_{-i} } \operatorname{Pr}\left[T_i=t_i \right]  \cdot \mathbb{E}\left[ J^{\pi}  (\mathcal{C} \setminus \{i \}, \mathcal{P}, \max\left\{ y, V_i \right\} ) \mid T_i=t_i\right] .    \nonumber
  \end{eqnarray}

We define an alternative policy $\pi'$ and show that it attains a higher expected profit than policy $\pi$. 
At state $ (\mathcal{C}, \mathcal{P}, y)$, policy $\pi'$ P-opens box $i$.  
If the observed type $t_i$ satisfies $ \sigma_i^{F \mid t_i} > \sigma_{-i}$, then policy $\pi'$ F-opens box $i$ and imitates policy $\pi$ thereafter. 
Otherwise, if $ \sigma_i^{F \mid t_i} \leq \sigma_{-i}$, then instead of F-opening the partially open box $i$, policy $\pi'$ generates an independent sample denoted by $\hat{v_i}$ from the prize distribution of $V_i$ conditional on $T_i=t_i$. 
Policy $\pi'$ treats $\hat{v_i}$ as the realized prize of box $i$ and imitates policy $\pi$ afterward, with the only difference that, whenever policy $\pi$ selects box $i$, policy $\pi'$ selects the box whose value is $y$. 
Therefore, the expected profit of policy $\pi'$ can be expressed as follows:
  \begin{eqnarray}
    J^{\pi'} (\mathcal{C}, \mathcal{P}, y) &=&  - c_i^P + \sum_{t_i: \sigma_i^{F \mid t_i} > \sigma_{-i}} \operatorname{Pr}\left[T_i=t_i \right] \left( -c_i^F + \mathbb{E} \left[ J^{\pi} (\mathcal{C} \setminus \{i \}, \mathcal{P}, \max\left\{ y, V_i \right\} )  \mid T_i=t_i \right] \right) \nonumber  \\
     &+&  \sum_{t_i: \sigma_i^{F \mid t_i} \leq \sigma_{-i} } \operatorname{Pr}\left[T_i=t_i \right] \cdot J^{\pi'} \left(\mathcal{C} \setminus \{i \}, \mathcal{P} \cup \{(i,t_i) \}  , y \right) . \nonumber
  \end{eqnarray}
  It follows that the difference in the expected profits between the two policies satisfies
  \begin{eqnarray}  
  & & J^{\pi'} (\mathcal{C}, \mathcal{P}, y) - J^{\pi} (\mathcal{C}, \mathcal{P}, y)   \nonumber \\
    &=&  - c_i^P +  \sum_{t_i: \sigma_i^{F \mid t_i} \leq \sigma_{-i} } \operatorname{Pr}\left[T_i=t_i \right] \left( c_i^F + J^{\pi'} \left(\mathcal{C} \setminus \{i \}, \mathcal{P} \cup \{(i,t_i) \}  , y \right) - \mathbb{E}\left[ J^{\pi}  (\mathcal{C} \setminus \{i \}, \mathcal{P}, \max\left\{ y, V_i \right\} ) \mid T_i=t_i\right]  \right)   \nonumber   \\
    &=& - c_i^P +  \sum_{t_i: \sigma_i^{F \mid t_i} \leq y } \operatorname{Pr}\left[T_i=t_i \right] \left( c_i^F + J^{\pi'} \left(\mathcal{C} \setminus \{i \}, \mathcal{P} \cup \{(i,t_i) \}   , y \right) - \mathbb{E}\left[ J^{\pi}  (\mathcal{C} \setminus \{i \}, \mathcal{P}, \max\left\{ y, V_i \right\} ) \mid T_i=t_i\right]  \right)  \nonumber \\ 
    &\geq& - c_i^P +  \sum_{t_i: \sigma_i^{F \mid t_i} \leq y } \operatorname{Pr}\left[T_i=t_i \right] \left( c_i^F - \mathbb{E} \left[ \left( V_i - y \right)^+  ~\middle \vert~ T_i=t_i \right] \right)   \label{eq:lemma_sufficient_F_dominated_2}. 
  \end{eqnarray}
The second equation follows from the assumption that box $i$ is well-classified.
The inequality holds since the difference in the total expected profits garnered by policies $\pi'$ and $\pi$ is at most $(V_i - y)^+$ on any sample path.
We note that equation~\eqref{eq:lemma_sufficient_F_dominated_2} is exactly the negative of 
equation~\eqref{eq:lemma:monotonicity_myopic_gain_SW}. 
Therefore, by Theorem~\ref{thm:myopic_stopping}, we have $ J^{\pi'} (\mathcal{C}, \mathcal{P}, y) - J^{\pi} (\mathcal{C}, \mathcal{P}, y)  > 0$, provided that $y > \sigma_i^{F/P} $ (which holds since box $i$ is well-classified box).
\end{myproof2}

\begin{myproof2}{Theorem~\ref{thm:just_weak_closed}}
Similarly to the proof of Lemma~\ref{lemma:just_strong_partial},  we prove the theorem by induction over the integer $2  \mid \mathcal{C} \mid + \mid \mathcal{P} \mid  $, which corresponds to the maximal possible number of openings from state $({\cal C},{\cal P},y)$ onwards. 
  \begin{itemize}
    \item {\em Base case: $2  \mid \mathcal{C} \mid + \mid \mathcal{P} \mid  = 2$.} There is only one closed box. The well-classified condition is clearly satisfied, since the problem reduces to the single box case. 
    The desired claim immediately follows from Theorem~\ref{thm:myopic_stopping}.
\begin{figure}[htbp]
\centering
\begin{subfigure}{.5\textwidth}
  \centering
  \includegraphics[width=0.9\linewidth]{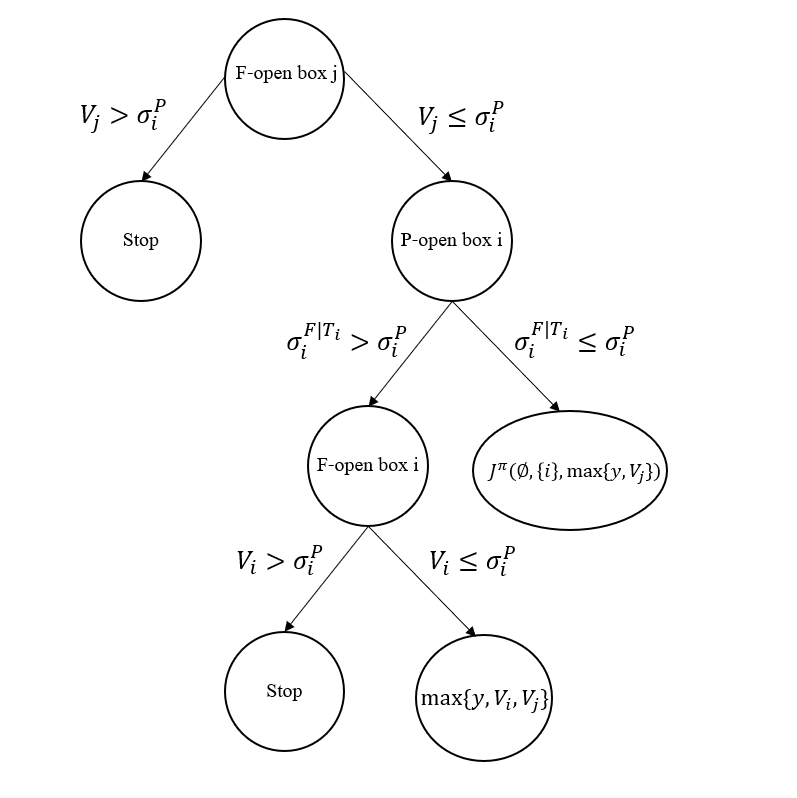}
    \caption{Policy $\pi$}
   \label{fig:just_weak_interchange_base_policy_1}
\end{subfigure}%
\begin{subfigure}{.5\textwidth}
  \centering
  \includegraphics[width=0.9\linewidth]{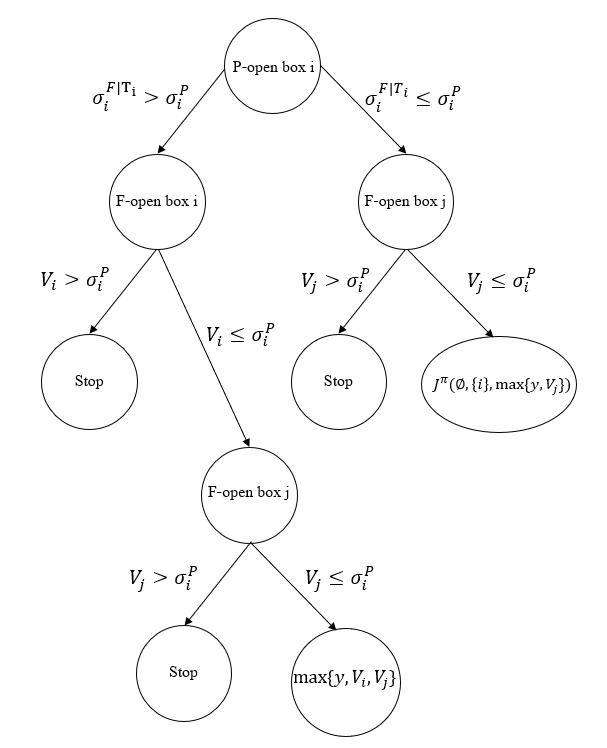}
  \caption{Policy $\pi'$}
  \label{fig:just_weak_interchange_base_policy_2}
\end{subfigure}
\caption{ A graphical illustration of policy $\pi$ and policy $\pi'$, for the base case $2  \mid \mathcal{C} \mid + \mid \mathcal{P} \mid  = 3$ in the proof of Theorem~\ref{thm:just_weak_closed}.   }
\label{fig:just_weak_interchange_base_policy}
\end{figure}
    \item {\em Base case: $2  \mid \mathcal{C} \mid + \mid \mathcal{P} \mid  = 3$.} Let $i$ be the single box contained in ${\cal C}$ and let $j$ be the single box contained in ${\cal P}$. 
    By Lemma~\ref{lemma:well_classified_F_dominated}, F-opening box $i$ is suboptimal. 
    As such, suppose that an optimal policy $\pi$ does not P-open box $i$, and instead it F-opens box $j$, and then the system transitions to a state in which $2  \mid \mathcal{C} \mid + \mid \mathcal{P} \mid  = 2$.
    By Theorem~\ref{thm:myopic_stopping}, policy $\pi$ selects box $j$ if $V_j > \sigma_i^P$. Otherwise, when $V_j \leq \sigma_i^P$, we observe that box $i$ is the only one remaining and it is still well-classified, therefore Theorem~\ref{thm:myopic_stopping} implies that policy $\pi$ proceeds to P-open this box.
    Applying Theorem~\ref{thm:just_strong}, policy $\pi$ then F-opens box $i$ if the revealed type $t_i$ is such that $\sigma_i^{F \mid t_i} > \sigma_i^P$, and further selects box $i$ if $V_i > \sigma_i^P$ (this follows from Theorem~\ref{thm:myopic_stopping}). The policy is illustrated in Figure~\ref{fig:just_weak_interchange_base_policy_1}.
    We let $q_i = \operatorname{Pr} \left[ \sigma_i^{F \mid T_i} > \sigma_i^P  \right]$, $p_i = \operatorname{Pr} \left[ V_i > \sigma_i^P   \middle \vert \sigma_i^{F \mid t_i} > \sigma_i^P  \right]$, and $p_j =  \operatorname{Pr} \left[ V_j > \sigma_i^P  \right]$.     
    Therefore, the expected profit evaluated in state $(\{ i\}, \{ j \}, y)$ is
        \begin{eqnarray}
          & &J^{\pi}((\{ i\}, \{ j \}, y)) \nonumber\\
          &=& -c_j^F +  p_j \cdot \mathbb{E}\left[V_j \mid V_j > \sigma_i^P  \right]  \nonumber \\
          &+& (1-p_j) \cdot  \Bigg( -c_i^P + q_i \Big( -c_i^F + p_i \cdot \mathbb{E} \left[ V_i \mid V_i > \sigma_i^P \right] + (1-p_i) \cdot \mathbb{E} \left[ \max \{y, V_i, V_j \} \mid V_i \leq \sigma_i^P, V_j \leq \sigma_i^P\right] \Big)  \nonumber \\
           &+&   (1-q_i) \cdot \mathbb{E}\left[ J^{\pi}((\emptyset, \{ (i,t_i) \}, \max\{y, V_j \}))  ~\middle\vert~ V_j \leq \sigma_i^P, \sigma_i^{F \mid t_i} \leq \sigma_i^P   \right] \Bigg).   \nonumber
        \end{eqnarray}

        We construct an alternative policy $\pi'$ that P-opens box $i$ in state $(\{ i\}, \{ j \}, y)$, as specified in Figure~\ref{fig:just_weak_interchange_base_policy_2}.
        The expected profit generated by policy $\pi'$ is
        \begin{eqnarray}
          J^{\pi'}((\{ i\}, \{ j \}, y)) &=&  -c_i^P + q_i \cdot \Bigg(  -c_i^F + p_i \cdot  \mathbb{E} \left[ V_i \mid V_i > \sigma_i^P \right]  \nonumber \\
          &+&  (1-p_i) \cdot  \Big( - c_j^F + p_j \cdot \mathbb{E}\left[V_j \mid V_j > \sigma_i^P  \right]  + (1-p_j) \cdot \mathbb{E} \left[ \max \{y, V_i, V_j \} \mid V_i \leq \sigma_i^P, V_j \leq \sigma_i^P\right]   \Big)   \Bigg)         \nonumber \\
          &+&  (1-q_i)  \cdot  \Bigg( -c_j^F + p_j \cdot \mathbb{E} \left[ V_j \mid V_j > \sigma_i^P \right]     \nonumber \\
          &+&  (1-p_j) \cdot \mathbb{E}\left[ J^{\pi}((\emptyset, \{ (i,t_i) \}, \max\{y, V_j \}))  ~\middle\vert~ V_j \leq \sigma_i^P, \sigma_i^{F \mid t_i} \leq \sigma_i^P   \right]  \Bigg). \nonumber
        \end{eqnarray}
        The difference between the expected profit obtained by the two policies is
        \begin{eqnarray}
        & & J^{\pi'}((\{ i\}, \{ j \}, y)) -  J^{\pi}((\{ i\}, \{ j \}, y))\nonumber \\
            &=&  \left(-q_i(1-p_i) - (1-q_i) +1 \right) \cdot c_j^F + \left( -1 +(1-p_j) \right) \cdot c_i^P + \left( -q_i + q_i(1-p_j) \right) \cdot c_i^F  \nonumber  \\
           &+& \left(  q_{i} p_{i}-\left(1-p_{j}\right)  q_{i}  p_{i} \right) \cdot \mathbb{E} \left[ V_i \mid V_i > \sigma_i^P \right] + \left( q_{i} \left(1-p_{i}\right) p_{j}+\left(1-q_{i}\right)  p_{j} - p_{j}  \right) \cdot \mathbb{E} \left[ V_j \mid V_j > \sigma_i^P \right]   \nonumber \\
           &=& p_{j} \cdot  \left(-c_{i}^{P} - q_{i} c_{i}^{F} + q_{i} p_{i} \cdot  \mathbb{E}\left[V_{i} \mid V_{i}>\sigma_{i}^{P}\right]\right)  - p_i q_i \cdot  \left( -c_{j}^{F}+p_{j} \cdot \mathbb{E}\left[V_{j} \mid V_j>\sigma_{i}^{p}\right] \right)  \nonumber \\
           &=& p_j q_i p_i  \cdot \sigma_i^P  - p_i q_i  \cdot  \left( -c_{j}^{F}+p_{j} \cdot \mathbb{E}\left[V_{j} \mid V_j>\sigma_{i}^{p}\right] \right)  \nonumber \\
           &\geq&  p_j q_i p_i  \cdot \sigma_i^P  - p_i q_i p_j  \cdot \sigma_i^P  \nonumber   \\
           &=& 0.  \nonumber
        \end{eqnarray}
        The inequality holds due to the same reasoning as in the derivation of equation~\eqref{eq:cost_accounting_ineq_1}. 
        Hence, policy $\pi'$ can achieve an expected profit as high as policy $\pi$. 
    
    \item {\em Induction step.} Suppose that the induction hypothesis holds when  $2  \mid \mathcal{C} \mid + \mid \mathcal{P} \mid  = 2, 3, \cdots, n$ with $n \geq 3$. 
    We wish to establish a similar claim when $2  \mid \mathcal{C} \mid + \mid \mathcal{P} \mid  =n + 1$. 
    Suppose an optimal policy $\pi$ does not P-open immediately.  
    We show that we can construct alternative policies that perform at least as well as policy $\pi$.  
    Depending on whether policy $\pi$ F-opens or P-opens a box next, we distinguish between the following two cases. 
\begin{figure}[htbp]
\centering
\begin{subfigure}{.5\textwidth}
  \centering
  \includegraphics[width=0.9\linewidth]{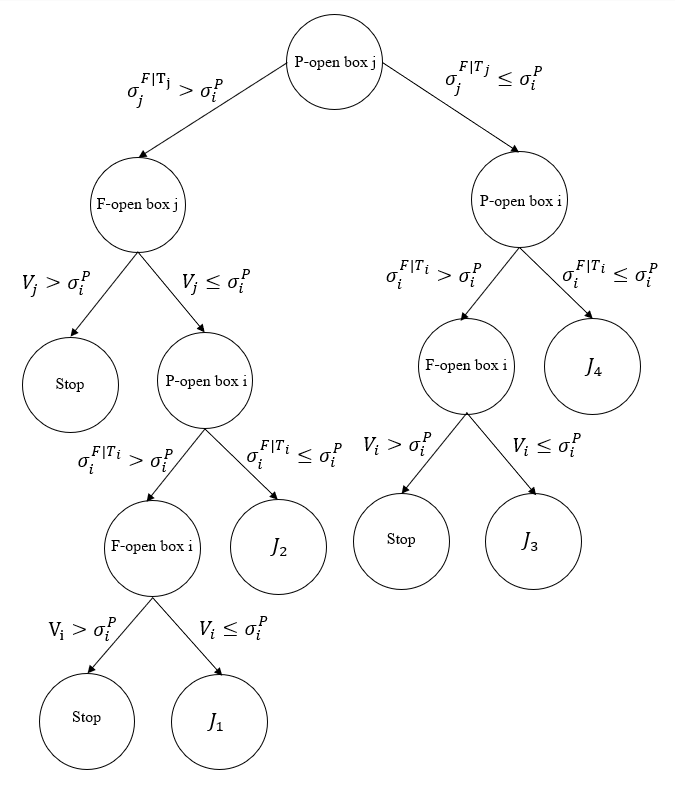}
    \caption{Policy $\pi$}
\end{subfigure}%
\begin{subfigure}{.5\textwidth}
  \centering
  \includegraphics[width=0.9\linewidth]{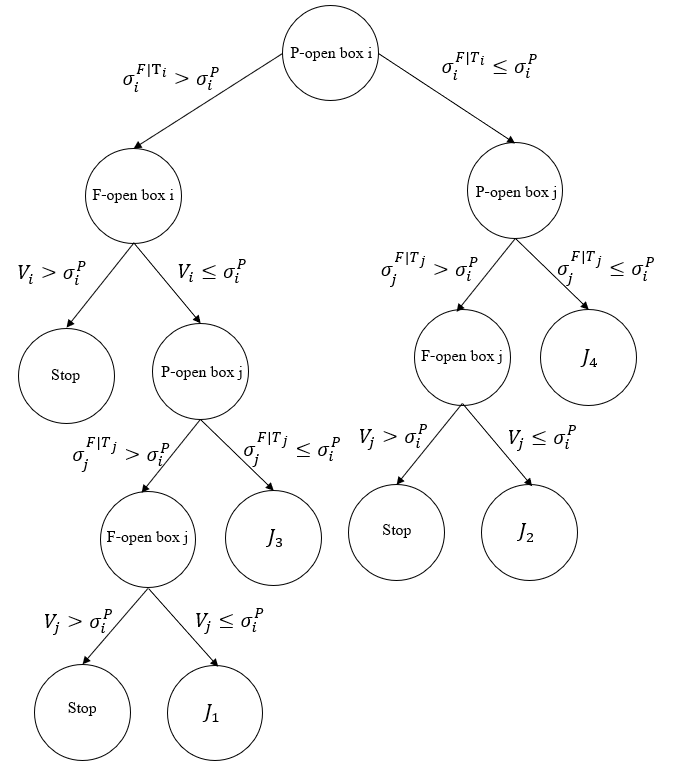}
  \caption{Policy $\pi'$}
\end{subfigure}
\caption{ Graphical illustration of policy $\pi$ and policy $\pi'$, for Case~1 in the induction step, in the proof of Theorem~\ref{thm:just_weak_closed}.   }
\label{fig:just_weak_interchange_ww_policy}
\end{figure}

    \paragraph{Case~1: Policy $\pi$ starts by P-opening.}
           Suppose that policy $\pi$ P-opens a box $j \in \mathcal{C}$ (without loss of generality, we assume has not expired). Consequently, the system  transitions to a state $(\mathcal{C}', \mathcal{P}', y)$ for which $2  \mid \mathcal{C} ' \mid  + \mid \mathcal{P}' \mid  = n-1$. By Theorem~\ref{thm:just_strong}, if the type $T_j$ is found to be such that $\sigma_j^{F \mid T_j} > \sigma_i^P$, policy $\pi$ will further F-open box $j$, and select box $j$ if $V_j > \sigma_i^P$ according to Theorem~\ref{thm:myopic_stopping}. If $V_j\leq \sigma_i^P$, policy $\pi$ will P-open box $i$ by the induction hypothesis, since box $i$ remains well-classified, and $\sigma_i^{P}$ is still the largest threshold. Otherwise, if $\sigma_j^{F \mid T_j} \leq \sigma_i^P$, by the induction hypothesis, policy $\pi$ will P-open box $i$. 
           Let $q_k = \operatorname{Pr} \left[ \sigma_k^{F \mid T_k} > \sigma_i^P  \right] $ and $p_k = \operatorname{Pr} \left[ V_k > \sigma_i^P  \mid \sigma_k^{F \mid t_k} > \sigma_i^P \right]$ for each box $k \in \{i, j\}$. 
           With this definition at hand, the expected profit obtained by policy $\pi$ can be written as
          \begin{eqnarray}
          J^{\pi}( \mathcal{C}, \mathcal{P}, y)  &=& -c_{j}^{P}+q_{j} \cdot \Bigg ( -c_{j}^{F}+p_{j} \cdot E\left[V_{j}  \mid V_{j}>\sigma_{i}^{P} \right]  \nonumber \\
          &+& (1-p_j) \Big( -c_{i}^{P} + q_{i} \left( -c_{i}^{F} + p_{i} \cdot \mathbb{E}\left[V_{i} \mid V_{i} >  \sigma_{i}^{P} \right] + \left( 1 - p_{i} \right) \cdot J_{1} \right) + (1-q_i) \cdot J_2  \Big) \Bigg)   \nonumber \\
          &+& (1-q_j) \Bigg( -c_i^P + q_i \Big( -c_i^F + p_i \cdot \mathbb{E} \left[ V_i \mid V_i > \sigma_i^P  \right] +(1-p_i) \cdot J_3   \Big) + (1-q_i) \cdot J_4   \Bigg)  , \nonumber
        \end{eqnarray}
           where $J_1 = \mathbb{E}\left[J^{\pi}( \mathcal{C} \setminus \{i,j \}, \mathcal{P}, \max\{y, V_i, V_j \}) ~\middle\vert~  \sigma_j^{F \mid T_j} > \sigma_i^P, V_j \leq \sigma_i^P, \sigma_i^{F\mid T_i} > \sigma_i^P, V_i \leq \sigma_i^P \right]$ ,\\
            $J_2 = \mathbb{E}\left[ J^{\pi}( \mathcal{C} \setminus \{i,j \}, \mathcal{P} \cup \{(i, t_i) \}, \max\{y,  V_j \})   ~\middle\vert~  \sigma_j^{F \mid T_j} > \sigma_i^P, V_j \leq \sigma_i^P, \sigma_i^{F\mid T_i} \leq \sigma_i^P \right]$ , \\
           $J_3 = \mathbb{E}\left[ J^{\pi}( \mathcal{C} \setminus \{i,j \}, \mathcal{P} \cup \{(j, T_j) \}, \max\{y, V_i \})  ~\middle\vert~   \sigma_j^{F \mid T_j} \leq \sigma_i^P,  \sigma_i^{F\mid T_i} > \sigma_i^P, V_i \leq \sigma_i^P \right]$ , \\
           $J_4 = \mathbb{E}\left[ J^{\pi}( \mathcal{C} \setminus \{i,j \}, \mathcal{P} \cup \{(j, T_j), (i, t_i) \}, y )  ~\middle\vert~   \sigma_j^{F \mid T_j} \leq \sigma_i^P,  \sigma_i^{F\mid T_i} \leq \sigma_i^P   \right]$.

           We construct an alternative policy $\pi'$ that P-opens box $i$ in state $( \mathcal{C}, \mathcal{P}, y) $, as specified in Figure~\ref{fig:just_weak_interchange_ww_policy}. The expected profit in state $( \mathcal{C}, \mathcal{P}, y)$ under policy $\pi'$ is
           \begin{eqnarray}
            J^{\pi'}( \mathcal{C}, \mathcal{P}, y) &=&   -c_{i}^{P} + q_{i} \Bigg ( -c_{i}^{F}+p_{i}  \mathbb{E} \left[V_{i}  \mid V_{i} > \sigma_{i}^{P} \right]  \nonumber \\
          &+& (1-p_i) \Big( -c_{j}^{P} + q_{j} \left( -c_{j}^{F} + p_{j} \cdot \mathbb{E}\left[V_{j} \mid V_{j} >  \sigma_{i}^{P} \right] + \left( 1 - p_{j} \right) \cdot J_{1} \right) + (1-q_j) \cdot J_3  \Big) \Bigg)   \nonumber \\
          &+& (1-q_i) \Bigg( -c_j^P + q_j \Big(-c_j^F + p_j \cdot \mathbb{E} \left[ V_j \mid V_j > \sigma_i^P  \right] +(1-p_j) \cdot J_2   \Big) + (1-q_j) \cdot J_4   \Bigg) . \nonumber
           \end{eqnarray}
           Consequently, the difference between the expected profits under policies $\pi'$ and $\pi$ is 
           \begin{eqnarray}
              & &  J^{\pi'}( \mathcal{C}, \mathcal{P}, y)  - J^{\pi}( \mathcal{C}, \mathcal{P}, y)  \nonumber  \\
              &=&  \left( -1 + q_j (1-p_j) + (1-q_j)\right) \cdot c_i^P +  \left( -q_i + q_j (1-p_j) q_i + (1-q_j) q_i  \right) \cdot c_i^F     \nonumber  \\
              &+&    \left( -q_i (1-p_i) - (1-q_i) + 1 \right) \cdot c_j^P +    \left( -q_i(1-p_i)q_j - (1-q_i)q_j + q_j \right) \cdot  c_j^F    \nonumber  \\
              &+&  \left( q_{i} p_{i}- q_{j} \left(1-p_{j}\right) q_{i} p_{i} - \left( 1-q_{j} \right) q_{i} p_{i} \right) \cdot \mathbb{E} \left[V_{i}  \mid V_{i} > \sigma_{i}^{P} \right]         \nonumber  \\
              &+&   \left( q_{i} \left(1-p_{i}\right) q_{j} p_{j}+\left(1-q_{i}\right)  q_{j} p_{j} - q_{j} p_{j} \right) \cdot  \mathbb{E} \left[V_{j}  \mid V_{j} > \sigma_{i}^{P} \right]   \nonumber \\
              &=& -q_j p_j c_i^P - q_j p_j q_i c_i^F + q_j p_j q_i p_i  \cdot \mathbb{E} \left[V_{i}  \mid V_{i} > \sigma_{i}^{P} \right]  \nonumber \\
              &+& q_{i} p_{i} c_{j}^{P} + q_{i} p_{i} q_{j} c_{j}^{F} - q_{i} p_{i} q_{j} p_{j} \cdot  \mathbb{E}\left[V_{j} \mid V_j>\sigma_{i}^{p}\right]       \nonumber \\
              &=& q_j p_j  \cdot \left( - c_i^P -  q_i c_i^F + q_i p_i \cdot \mathbb{E} \left[V_{i}  \mid V_{i} > \sigma_{i}^{P} \right]    \right) \nonumber \\
              &-& q_{i} p_{i}  \cdot \left( - c_{j}^{P} - q_{j} c_{j}^{F} + q_{j} p_{j} \cdot \mathbb{E}\left[V_{j} \mid V_j>\sigma_{i}^{p}\right]  \right)        \nonumber \\
              &=& q_j p_j q_i p_i \cdot  \sigma_i^P -q_{i} p_{i}  \cdot \left( - c_{j}^{P} - q_{j} c_{j}^{F} + q_{j} p_{j} \cdot \mathbb{E}\left[V_{j} \mid V_j>\sigma_{i}^{p}\right]  \right)   \nonumber \\
              &\geq& q_j p_j q_i p_i \cdot \sigma_i^P - q_{i} p_{i} q_j p_j \cdot \sigma_i^P \nonumber \\
              &=& 0  . \nonumber
           \end{eqnarray}
           The last inequality is due to the same reasoning as in equation~\eqref{eq:cost_accounting_ineq_2}. 
           Hence, policy $\pi'$ can achieve an expected profit as high as policy $\pi$. 
\begin{figure}[htbp]
\centering
\begin{subfigure}{.5\textwidth}
  \centering
  \includegraphics[width=0.85\linewidth]{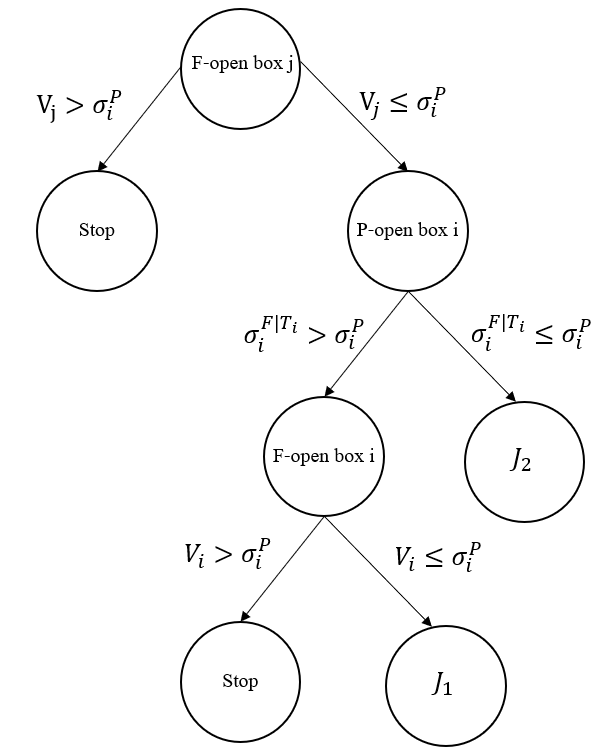}
    \caption{Policy $\pi$}
\end{subfigure}%
\begin{subfigure}{.5\textwidth}
  \centering
  \includegraphics[width=0.85\linewidth]{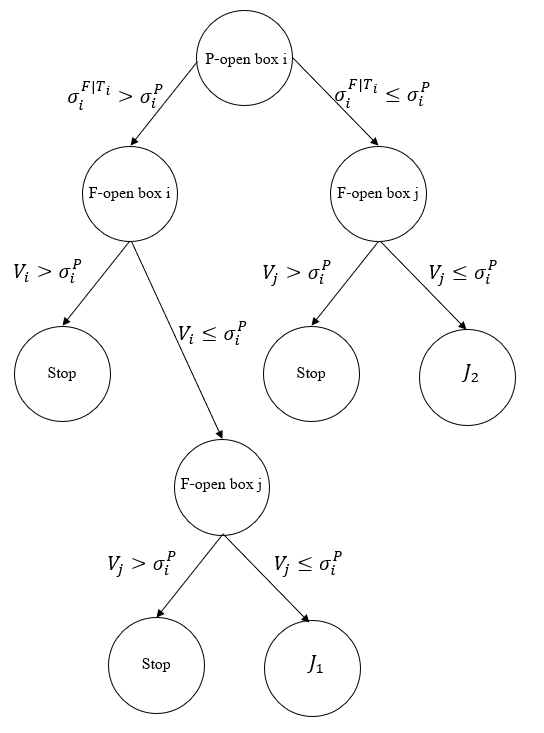}
  \caption{Policy $\pi'$}
\end{subfigure}
\caption{ Graphical illustration of policy $\pi$ and policy $\pi'$, for Case 2 in the induction step, in the proof of Theorem~\ref{thm:just_weak_closed}.   }
\label{fig:just_weak_interchange_sw_policy}
\end{figure}
\paragraph{Case~2: Policy $\pi$ starts by F-opening.}
    Suppose that policy $\pi$ F-opens a box $j \in \mathcal{C}$. 
    Consequently, the system transitions to a state $(\mathcal{C}', \mathcal{P}', y')$ for which $2  \mid \mathcal{C} ' \mid  + \mid \mathcal{P}' \mid  = n-1$. Alternatively, if we suppose that policy $\pi$ F-opens a box $j \in \mathcal{P}$, then it transitions to a state $(\mathcal{C}', \mathcal{P}', y')$ for which $2  \mid \mathcal{C} ' \mid  + \mid \mathcal{P}' \mid  = n$. Observe that, in both settings, box $i$ will remain well-classified. Indeed, after box $j$ is opened, if it is suboptimal to stop, then the P-threshold of box $i$ is still the largest opening threshold. Hence, in view of the induction hypothesis, policy $\pi$ will P-open box $i$. 
    Let $p_j = \operatorname{Pr} \left[ V_j > \sigma_i^{P}\right]$, $q_i = \operatorname{Pr} \left[ \sigma_i^{F \mid T_i}  >  \sigma_i^{P} \right] $ and $p_i = \operatorname{Pr} \left[ V_i > \sigma_i^{P} \mid  \sigma_i^{F\mid T_i}  >  \sigma_i^{P} \right]$. 
    The expected profit in state $( \mathcal{C}, \mathcal{P}, y)$ under policy $\pi$ is
      \begin{eqnarray}
    J^{\pi}( \mathcal{C}, \mathcal{P}, y) &=&  - c_j^F + p_j \cdot \mathbb{E} \left[ V_j \mid V_j > \sigma_i^P \right]  + (1-p_j) \cdot \Bigg( -c_i^P + q_i \left[ -c_i^F + p_i \cdot \mathbb{E} \left[V_i \mid V_i > \sigma_i^P \right] \right]  \nonumber \\
    &+& q_i (1-p_i) \cdot J_1 + (1-q_i) \cdot J_2 \Bigg). \nonumber
    \end{eqnarray}
    
    We construct an alternative policy $\pi'$ that P-opens box $i$ first, as specified in Figure \ref{fig:just_weak_interchange_sw_policy}. The expected profit in state $( \mathcal{C}, \mathcal{P}, y)$ under policy $\pi'$ is
    \begin{eqnarray}
    & & J^{\pi'}( \mathcal{C}, \mathcal{P}, y) \nonumber \\
    &=&  -c_i^P + q_i \cdot \Big( -c_i^F + p_i \cdot  \mathbb{E} \left[ V_i \mid V_i > \sigma_i^P \right] + (1-p_i) \left( -c_j^F + p_j \cdot  \mathbb{E} \left[ V_j \mid V_j > \sigma_i^P \right] + (1-p_j) \cdot J_1  \right) \Big) \nonumber \\
    &+& (1-q_i ) \cdot \Big(  -c_j^F + p_j \cdot  \mathbb{E} \left[ V_j \mid V_j > \sigma_i^P \right]  + (1-p_j) \cdot J_2 \Big).      \nonumber
    \end{eqnarray}
    Hence, the difference in the expected profits under policies $\pi'$ and $\pi$ is
    \begin{eqnarray}
      & &  J^{\pi'}( \mathcal{C}, \mathcal{P}, y)  - J^{\pi}( \mathcal{C}, \mathcal{P}, y)  \nonumber  \\
      &=&  \left( -1 + 1 - p_j \right) \cdot c_i^P +  \left( -q_i + (1-p_j) q_i  \right) \cdot c_i^F +  \left(-q_i (1-p_i) - (1-q_i) + 1 \right) \cdot  c_j^F    \nonumber  \\
      &+&  \left(q_i p_i - (1-p_j) q_i p_i \right) \cdot \mathbb{E} \left[V_{i}  \mid V_{i} > \sigma_{i}^{P} \right]  + \left( q_i (1-p_i) p_j + (1-q_i)p_j - p_j \right) \cdot  \mathbb{E} \left[ V_{j}  \mid V_{j} > \sigma_{i}^{P}  \right]        \nonumber  \\
      &=& p_j \cdot  \left( -c_i^P - q_i c_i^F + q_i p_i \cdot \mathbb{E} \left[V_{i}  \mid V_{i} > \sigma_{i}^{P} \right]   \right)  - q_i p_i \cdot \left( -c_j^F + p_j \cdot \mathbb{E} \left[V_{j}  \mid V_{j} > \sigma_{i}^{P} \right]  \right)   \nonumber \\
      &=& p_j  q_i p_i \cdot \sigma_i^P  - q_i p_i \cdot \left( -c_j^F + p_j \cdot \mathbb{E} \left[V_{j}  \mid V_{j} > \sigma_{i}^{P}   \right]  \right)      \nonumber \\
      &\geq & p_j  q_i p_i  \cdot \sigma_i^P  - p_j  q_i p_i  \cdot \sigma_i^P   = 0 .        \nonumber
    \end{eqnarray}
    The inequality follows from the same reasoning as the derivation of  equation~\eqref{eq:cost_accounting_ineq_1}. 
    Hence, policy $\pi'$ can achieve an as high expected profit as policy $\pi$. This concludes the proof of the induction step.
  \end{itemize}
\end{myproof2}

\subsection{Proof of Theorem~\ref{thm:iid_optimal}}
\label{sec:proof_iid_optimal}

We first introduce notation prior to proceeding to prove each of the five properties. 
Since boxes are identical, for brevity, we drop the subscript of box indices throughout this proof. 
Recall that each box can be of either a good type (G) or a bad type (B), and the rewards are binary with values 0 or 1.
Let $p^G$ and $p^B$ denote the probabilities of a box being of good type and bad type, respectively. Conditional on the box being of good (bad) type, the probability that it contains a reward is $q^G$ ($q^B$).

To encode the state space, it suffices to record the number of closed boxes $n_C$, the number of good-type boxes $n_G$, the number of bad-type boxes $n_B$ and the current best reward: $(n_C, n_G, n_B, y)$. The initial state is then $(n, 0, 0, 0)$. 
Since a good type will be F-opened immediately upon observation and it is optimal to stop once a prize is found, we can ``skip'' states in which good-type boxes were discovered and ``collapse'' states where the reward was found, to obtain a more compact representation using the tuple $(n_C, n_B)$. 
Finally, we denote by $J^F$ ($J^P$) the expected payoff of F-opening (P-opening) a closed box and then following the optimal policy at any state $(n_C, n_B)$ with at least one closed box ($n_C \geq 1$). 

We proceed to prove each stated property.

\textbf{Property: Policy $\pi$ assigns the highest priority to F-opening good-type boxes.} 
Applying Observation~\ref{obs:thresholds_order} to the two-type case, we observe that 
\begin{enumerate}
    \item $\sigma^{F/P} < \sigma^{F} \leq \sigma^{P}$ 
    (the theorem assumes that $\sigma^{F} < \sigma^{P}$), and
    \item $\sigma^{F|B} \leq \sigma^{F / P }$ and $\sigma^{F | G} \geq \sigma^P$.    
\end{enumerate}
Theorem~\ref{thm:just_strong} then implies that an optimal policy directly F-opens a good type upon its discovery. 

\textbf{Property: Policy $\pi$ either discards all bad-type boxes, or  assigns the lowest priority to F-opening them.} 
We consider two cases, depending on whether $\sigma^{F|B}$ is positive. 
For the case when $\sigma^{F|B} \leq 0$, a bad type is not worth opening. 
For the case when $\sigma^{F|B} >0$, the property can be proven by an interchange argument. For brevity, the proof is omitted.

\textbf{Property: Policy $\pi$ is a committing policy.}     
To show that there exists an optimal policy that is a committing policy (see Section~\ref{subsec:apx_policies} for a formal definition and discussion), we first prove that the presence of bad-type boxes does \textit{not} affect the decision to F-open versus P-open the closed boxes. We formalize this in the following lemma, which for ease of reading we defer proving to the end of this section.

\begin{lemma} \label{lemma:iid_diff_not_func_of_nb}
    For a PSI-B2I instance at state $(n_C, n_B)$ where $n_C \geq 1$, 
    the difference $J^F(n_C, n_B) - J^P(n_C, n_B)$ is a function of $n_C$ and is independent of $n_B$.
\end{lemma}

Recall from Theorem~\ref{thm:myopic_stopping} that at state $(1, 0)$, it is optimal to F-open the closed box (see for illustration Figure~\ref{fig:example_1}). Lemma~\ref{lemma:iid_diff_not_func_of_nb} then implies that at any state $(1, n_B)$, it is optimal to F-open the closed box. 
The latter serves as a basis for an inductive argument for the optimal policy being a committing policy. At a state $(n_C, n_B)$, the difference in the expected costs between F-opening versus P-opening depends on actions taken at future states of the form $(n_C-1, n_B)$ and $(n_C-1, n_B+1)$, that are reached depending on the opening mode and whether the opened job is of a bad type or not. Regardless, the policy takes  actions that are independent of the number of bad-type boxes, meaning that whether or not P-opening or F-opening is better at state $(n_C, n_B)$, one can already commit at that state to the opening mode of each of the remaining boxes.

\textbf{Property: There exists a finite threshold $N_C$ such that policy $\pi$ prioritizes P-opening closed boxes over F-opening them iff the number of closed boxes $n_C > N_C$.} 
From Lemma~\ref{lemma:committing}, we know that once an optimal policy commits to an opening mode for each closed box, the optimal opening sequence is prescribed by the opening thresholds (Algorithm~\ref{algo:committing}). Therefore, P-openings will take place prior to F-openings. 

We proceed by induction on $n_C$. At any state $(1, n_B)$, F-opening is optimal. At the state $(n_C, n_B)$ either P-opening or F-opening is optimal. If we add a closed box to create state $(n_C+1, n_B)$, F-opening could be optimal only if P-opening is not optimal at state $(n_C, n_B)$ (because this would imply that a committing policy F-opens prior to P-opening). 

The finiteness of $N_c$ follows from Lemma~\ref{lemma:infinite_iid_boxes} and the asymptotic optimality of the policy that exclusively P-opens closed boxes.

\begin{myproof2}{Lemma~\ref{lemma:iid_diff_not_func_of_nb}}

We prove the claim by induction. 
\begin{itemize}
    \item We start from the base case where $n_C=1$. 
Let the closed box be box $i$. 
Following the reasoning of Lemma~\ref{lemma:committing}, the expected payoff of P-opening box $i$ is given by 
\begin{eqnarray*}
   J^{P} ( 1, n_B ) &=& 
   -c^P + \mathbb{E} \brackets{ 
   \max \braces{
   \min \braces{ V_i, \sigma^{F|T_i}}, 
   \max_{j \in \mathcal{P} } \braces{ \min \braces{ V_j, \sigma^{F|B} } } }  }
\end{eqnarray*}

The expected payoff of F-opening box $i$ is given by 
\begin{eqnarray*}
    J^{F}(1, n_B) &=& \E{  \max \braces{ \min \braces{ V_i, \sigma^F }, 
    \max_{j \in \mathcal{P} } \braces{ \min \braces{ V_j, \sigma^{F|B} } }  } }
\end{eqnarray*}

For brevity let $\mathcal{P}(n_B)$ denote the set of $n_B$ bad boxes. 
We let $X(n_B) \defeq \max_{j \in \mathcal{P} (n_B) } \braces{ \min \braces{ V_j, \sigma^{F|B} } }$. 
We note the difference 
\begin{small}
\begin{eqnarray*}
&&  J^{P} ( 1, n_B ) - J^{F}(1, n_B)   \\ 
    &=&  -c^P + \mathbb{E} \brackets{ 
   \max \braces{
   \min \braces{ V_i, \sigma^{F|T_i}}, 
   X }  }  
   -  \E{  \max \braces{ \min \braces{ V_i, \sigma^F }, 
      X } }                       \\ 
    &\overset{(a)}{=}&  -c^P 
    + \pr{  \min \braces{ V_i, \sigma^{F|T_i}} = 0  } 
     \E{ X }  
     +  \pr{  \min \braces{ V_i, \sigma^{F|T_i}} \neq 0  } 
     \E{ \min \braces{ V_i, \sigma^{F|T_i}}  \mid \min \braces{ V_i, \sigma^{F|T_i}} \neq 0}                     \\ 
     && - \pr{  \min \braces{ V_i, \sigma^{F}} = 0  } 
     \E{ X }  
     -  \pr{  \min \braces{ V_i, \sigma^{F}} \neq 0  } 
     \E{ \min \braces{ V_i, \sigma^{F}}  \mid \min \braces{ V_i, \sigma^{F}} \neq 0} \\
     &\overset{(b)}{=}& - c^P + \pr{  \min \braces{ V_i, \sigma^{F|T_i}} \neq 0  } 
     \E{ \min \braces{ V_i, \sigma^{F|T_i}}  \mid \min \braces{ V_i, \sigma^{F|T_i}} \neq 0} \\
     && -   \pr{  \min \braces{ V_i, \sigma^{F}} \neq 0  } 
     \E{ \min \braces{ V_i, \sigma^{F}}  \mid \min \braces{ V_i, \sigma^{F}} \neq 0}  \ , 
\end{eqnarray*}
\end{small}
which is independent of $X$. 
Step (a) follows by noting that $X$ can only take values 0 or $\sigma^{F|B}$, and $\sigma^{F|T_i} \geq \sigma^{F|B}$ almost surely. Hence, conditioned on the event $\braces{\min \braces{ V_i, \sigma^{F|T_i}} \neq 0}$, we have $X \leq \min \braces{ V_i, \sigma^{F|T_i}}$ almost surely.   
Step (b) follows from the simple observation that $\pr{  \min \braces{ V_i, \sigma^{F|T_i}} = 0  } = \pr{  \min \braces{ V_i, \sigma^{F}} = 0  }$. 

\item 
Suppose the induction hypothesis is true for $n_C \geq 1$. 
At state $(n_C+1, n_B)$ where $n_C \geq 1$, the expected payoff of F-opening a closed box and following the optimal policy afterwards is 
\begin{eqnarray*}
    J^{F} (n_C+1,n_B) &=& - c^F + (p^G q^G + p^B q^B ) \cdot 1 
    + \paren{ p^G (1-q^G) + p^B (1-q^B) } \cdot  J(n_C, n_B)  \ . 
\end{eqnarray*}
The expected payoff of P-opening a closed box and following the optimal policy afterwards is 
\begin{eqnarray*}
    J^{P} (n_C+1,n_B) &=&  -c^P 
    + p^G \cdot \paren{ - c^F + q_G \cdot 1 + (1-q_G) \cdot J( n_C, n_B ) }
    + p^B \cdot \paren{ J( n_C, n_B+1 ) } \ , 
\end{eqnarray*}
where we note that an optimal policy F-opens a good box immediately. 
The difference is 
\begin{scriptsize}
\begin{eqnarray}
    J^{F} (n_C+1,n_B) - J^{P} (n_C+1,n_B) &=&  c^P  
    + p^B \cdot \paren{ q^B   - c^F + (1-q^B) J( n_C, n_B ) - J( n_C, n_B+1 ) } \ .  \label{eq:iid_two_types}
\end{eqnarray}
\end{scriptsize}
To prove the lemma, it suffices to show the right-hand-side of above expression is a constant with respect to $n_B$.

Next, we proceed to show that the term $J(n_C, n_B+1) - (1-q^B) J(n_C, n_B)$ in the above display is indeed independent of $n_B$. 
To this end, we consider how the optimal strategies at state $( n_C, n_B )$ and $( n_C, n_B+1 )$ differ. 
By the induction hypothesis, the decision to whether F-open or P-open a closed box is only a function of $n_C$ and is \textit{independent} of the number of bad type. 
Therefore, the optimal courses of actions starting from state $( n_C, n_B )$ and $( n_C, n_B+1 )$ coincide, up to the moment (if occurs) when the latter has one last bad type and the former has nothing left.

Let $\cal C(n_C)$ be the set of indices of $n_C$ closed boxes.
For brevity, we denote $J(n_C, 0) = \max_{\pi} ~ \E{ \sum_{i\in {\cal C}(n_C) } \left(  \mathbb{S}_i^{\pi} V_i - \mathbb{F}_i^{\pi}  c_i^F - \mathbb{P}_i^{\pi} c_i^P \right)} $. 
Let $\theta(n) = \paren{ p^G (1-q^G) + p^B (1-q^B) }^n$ be the probability that the optimal policy ends up with no rewards at state $(n, 0)$. 
Further, we recall that the optimal policy will postpone opening the bad boxes to the very end. 
Therefore, we can decompose $J(n_C, n_B)$ and $J(n_C, n_B+1)$ in the following way. 
\begin{eqnarray*}
    J(n_C, n_B) = J(n_C, 0) + \theta(n_C) \E{ X(n_B)} \quad \text{and} \quad 
    J(n_C, n_B+1) = J(n_C, 0) + \theta(n_C) \E{ X(n_B+1)}
\end{eqnarray*}
Hence, we have 
\begin{eqnarray*}
     &&  J(n_C, n_B+1) - (1-q^B) J(n_C, n_B) \\ 
     &=& \theta(n_C) \cdot \paren{ \E{ X(n_B+1)} - (1-q^B)  \E{ X(n_B)} } + q_B J(n_C, 0)  \\
     &=& \theta(n_C) \cdot \paren{ \E{ \max_{j \in \mathcal{P} (n_B+1) } \braces{ \min \braces{ V_j, \sigma^{F|B} } }} - (1-q^B)  \E{ \max_{j \in \mathcal{P} (n_B) } \braces{ \min \braces{ V_j, \sigma^{F|B} } } } }  + q_B J(n_C, 0)  \\ 
     &=& \theta(n_C) \cdot \paren{ q^B \sigma^{F|B} + (1-q^B) \E{ \max_{j \in \mathcal{P} (n_B) } \braces{ \min \braces{ V_j, \sigma^{F|B} } }} - (1-q^B)  \E{ \max_{j \in \mathcal{P} (n_B) } \braces{ \min \braces{ V_j, \sigma^{F|B} } } } } \\ 
     && + q_B J(n_C, 0)  \\
     &=& q^B \cdot \paren{ \theta(n_C)  \sigma^{F|B} +  J(n_C, 0) }   
\end{eqnarray*}
which is independent of $n_B$, as desired. 
\end{itemize}

\end{myproof2}

\subsection{ Optimal policy for PSI-B2I  }
\label{sec:algo_psi_b2i}


{  

Algorithm~\ref{algo:psib2i_policy} specifies a class of policies that satisfy the structural properties required by Theorem~\ref{thm:iid_optimal} for PSI-B2I instances. 
It has two parameters $(N_C, \textsc{OpenBad})$, where $N_C$ is the threshold number of closed boxes, above which the policy prioritizes P-opening closed boxes.  
The parameter $\textsc{OpenBad}$ controls whether the policy opens bad-type boxes at all.

An optimal policy can then be obtained by enumerating this finite class and selecting the best-performing configuration, as described in Algorithm~\ref{algo:psib2i_enum_opt}. 

}

\begin{algorithm}[htbp]
\footnotesize
\caption{Policy template $\pi^{N_C,\textsc{OpenBad}}$ for PSI-B2I.  }
\label{algo:psib2i_policy}
\begin{algorithmic}[1]
\State Initialize counts $(n_C,n_G,n_B,y)\gets (N,0,0,0)$;
\While{$y=0$ and $(n_C+n_G+n_B)>0$}
    \If{$n_G>0$}
        \Comment{Highest priority: F-open good-type boxes}
        \State F-open a good-type box; update $y \gets \max\{y,v\}$; $n_G\gets n_G-1$;
    \ElsIf{$n_C>0$ and $n_C>N_C$}
        \Comment{ P-open closed boxes}
        \State P-open a closed box; $n_C\gets n_C-1$;
        \If{revealed type is good}
            \State $n_G\gets n_G+1$;
        \Else
            \State $n_B\gets n_B+1$;
        \EndIf
    \ElsIf{$n_C>0$ and $n_C\le N_C$}
        \State F-open a closed box; update $y \gets \max\{y,v\}$; $n_C\gets n_C-1$;
    \Else
        \Comment{Bad-type boxes are lowest priority}
        \If{$\textsc{OpenBad}=\text{True}$ and $n_B>0$}
            \State F-open a bad-type box; update $y \gets \max\{y,v\}$; $n_B\gets n_B-1$;
        \Else
            \State \textbf{break};
        \EndIf
    \EndIf
\EndWhile
\State Select $y$ and stop;
\end{algorithmic}
\end{algorithm}

\begin{algorithm}[htbp]
\footnotesize
\caption{Computing an optimal PSI-B2I policy by enumerating the threshold class (Theorem~\ref{thm:iid_optimal}).}
\label{algo:psib2i_enum_opt}
\begin{algorithmic}[1]
\State Initialize $V^\star \gets -\infty$; $(N_C^\star,\textsc{OpenBad}^\star)\gets (0,\text{False})$;
\For{$N_C \in \{0,1,\ldots,N\}$}
    \For{$\textsc{OpenBad} \in \{\text{False},\text{True}\}$}
        \State Compute $V(N_C,\textsc{OpenBad})$ = expected payoff of policy $\pi^{N_C,\textsc{OpenBad}}$ prescribed by Algorithm~\ref{algo:psib2i_policy}, using Bellman's recursion;
        \If{$V(N_C,\textsc{OpenBad}) > V^\star$}
            \State $V^\star \gets V(N_C,\textsc{OpenBad})$;
            \State $(N_C^\star,\textsc{OpenBad}^\star)\gets (N_C,\textsc{OpenBad})$;
        \EndIf
    \EndFor
\EndFor
\State \Return the policy $\pi^{N_C^\star,\textsc{OpenBad}^\star}$;
\end{algorithmic}
\end{algorithm}


\section{Proofs of Section~\ref{sec:apx}}

In this section, we present all the proofs for Section~\ref{sec:apx}.
We present the proof of Theorem~\ref{thm:hardness} to the end due to the dependence on other results of which we present the proofs first.

\subsection{Proof of Lemma \ref{thm:upper_bounds}} \label{appendix:thm:upper_bounds}

The proof of the first inequality mostly follows similar lines of reasoning to that of \cite{brown2013optimal}[Proposition 4.2], and for brevity is thus omitted. 
The remainder of the proof is dedicated to showing that $J^W (\mathcal{C}, \mathcal{P}, y)  \leq  J^K (\mathcal{C}, \mathcal{P}, y)$.

Define $X^K$ as follows
    $$
    X^K =   \max \left\{ y,  \underset{i \in \mathcal{C}}{ \max }  \left\{ \max \left\{K_i, \widetilde{K}_{i}  \right\} \right\} ,   \underset{i \in \mathcal{P}}{\max}  \left\{ \widetilde{K}_i^{t_i}  \right\} \right\}. 
    $$ 
    Recall from equation~\eqref{eq:sigma_M_C_P} that $\sigma_M =  \max \left\{  \underset{j \in \mathcal{C} }{\max}  \left\{  \sigma_j^F, \sigma_j^P \right\}, \underset{ j \in \mathcal{P}}{\max }  \left\{ \sigma_j^{F \mid t_j}  \right\}   \right\}$. 
    Since we assume that it is suboptimal to stop in current state $(\mathcal{C}, \mathcal{P}, y)$, we infer that $y<\sigma_M$, and thus, $\sigma_M \geq X^K \geq y \geq 0$. Consequently, we have 
    \begin{eqnarray}
    & & J^K(\mathcal{C}, \mathcal{P}, y) \triangleq \mathbb{E} \left[ X^K \right]  \nonumber \\ 
      &=& \int_{0}^{\sigma_M} \operatorname{Pr} \left[ X^K > s \right] ds   \nonumber \\
       &=&  \sigma_M - \int_{y}^{\sigma_M} \operatorname{Pr} \left[ X^K \leq s \right] ds  \nonumber \\
       &=&  \sigma_M - \int_{y}^{\sigma_M}  \left( \prod_{i \in \mathcal{C}} \operatorname{Pr} \left[ \max \left\{K_i, \widetilde{K}_{i}  \right\} \leq s \right] \right) \cdot \left(  \prod_{i \in \mathcal{P}} \operatorname{Pr} \left[ \widetilde{K}_{i}^{t_i} \leq s \mid T_i=t_i \right]  \right)  dy.  \label{eq:proof_bounds_1}  \\ 
       &\geq& \sigma_M - \int_{y}^{\sigma_M}  \prod_{i=1}^{N} \omega_i(x_i , y) dy  \nonumber \\
       &=&  J^W(\mathcal{C}, \mathcal{P}, y). \nonumber 
    \end{eqnarray}
    Equation~\ref{eq:proof_bounds_1} follows from the independence between boxes. The inequality is justified in the remainder of the proof. 
    Recall that, if  $i \in \mathcal{P}$, we have 
\begin{eqnarray*}
  \omega_i(x_i , y) &=& \operatorname{Pr} \left[ \min\{V_i, \sigma_i^{F \mid t_i} \} \leq y \mid T_i=t_i \right]   =  \operatorname{Pr} \left[ \widetilde{K}_i^{t_i} \leq y \mid T_i=t_i \right] .
\end{eqnarray*}
Now, if $i \in \mathcal{C}$, we have 
\begin{eqnarray*}
  \omega_i(x_i , y) &=&  \BFI\{ y > \sigma_i^{F/P}  \} \operatorname{Pr} \left[ \min \left\{ V_i, \sigma_i^{F \mid T_i}, \sigma_i^P \right\} \leq y \right] + \BFI\{ y \leq \sigma_i^{F/P}  \} \operatorname{Pr} \left[  \min \left\{ V_i,  \sigma_i^F \right\} \leq y \right]    \\
  & =& \BFI\{ y > \sigma_i^{F/P}  \} \operatorname{Pr} \left[ \widetilde{K_i} \leq y \right] + \BFI\{ y \leq \sigma_i^{F/P}  \} \operatorname{Pr} \left[  K_i \leq y \right] \\ 
  &=& \operatorname{Pr} \left[ \max \left\{  \BFI\{ y > \sigma_i^{F/P}  \} \widetilde{K_i}  ,  \BFI\{ y \leq \sigma_i^{F/P}  \} K_i \right\}  \leq  y \right]  \\
  &\geq& \operatorname{Pr} \left[ \max \left\{ \widetilde{K_i}, K_i \right\} \leq y \right].
\end{eqnarray*}

\subsection{Asymptotic optimality} 
\label{appendix:ex_infinite_iid_boxes}

\begin{lemma}[Asymptotic optimality]  \label{lemma:infinite_iid_boxes}
Suppose that there exists $\theta>0$ such that $  N_k > \theta  N $ for all $k\in [K]$.
Let $u=\operatorname{Pr} \left[ V_i > \sigma_i^F \right] $, $q = \operatorname{Pr}\left[ \sigma_i^{F \mid T_i} > \sigma_i^P \right]$ and $p = \operatorname{Pr} \left[ V_i > \sigma_i^P \mid  \sigma_i^{F \mid T_i} > \sigma_i^P\right]$, where $i \in [K]$ is the class of boxes that initially attains the largest opening threshold $\sigma_M([N],\emptyset)$. We distinguish between two cases:
\begin{itemize}
    \item { Case~1: $\sigma_i^F = \sigma_M([N],\emptyset)$.} The policy that always F-opens closed boxes in class $i$ and implements the stopping rule of Theorem~\ref{thm:myopic_stopping} is a factor $(1 - (1-u)^{\theta N})$-close to the optimum.
    \item { Case~2: $\sigma_i^P = \sigma_M([N],\emptyset)$.} The policy that always P-opens closed boxes in class $i$, F-opens partially open boxes when $\sigma_i^{F \mid t_i} > \sigma_i^P$, and implements the stopping rule of Theorem~\ref{thm:myopic_stopping} is a factor $(1 - \left( 1 - p q \right)^{\theta  N })$-close to the optimum.
\end{itemize}
\end{lemma}
The proof is based on the simple observation that the expected profit under the policy described in Cases~1 and~2 asymptotically matches Whittle's integral~\eqref{eq:whittlesintegral}. Intuitively, when we have a sufficiently large number of ``good'' boxes (with associated threshold $\sigma_M({\cal C},{\cal P})$), the actual probability that the effort of opening them is in vain approaches zero, i.e., the gap between Whittle's integral and $\sigma_M({\cal C},{\cal P})$ decreases exponentially in the number of good boxes. Conversely, the policy of Lemma~\ref{lemma:infinite_iid_boxes} has an expected profit approaching $\sigma_M({\cal C},{\cal P})$ from below.

We note in passing, that the Pandora's box model with infinite number of boxes has been commonly used in the literature as a model of consumer search \citep[e.g.,][]{wolinsky1986true}. The fact that our problem admits an asymptotically optimal index policy offers researchers the ability to use the problem as a building block for constructing richer models that capture more realistic behavior.

\begin{myproof}

We let $u=\operatorname{Pr} \left[ V_i > \sigma_i^F \right] $, $q = \operatorname{Pr}\left[ \sigma_i^{F \mid T_i} > \sigma_i^P \right]$ and $p = \operatorname{Pr} \left[ V_i > \sigma_i^P \mid  \sigma_i^{F \mid t_i} > \sigma_i^P\right]$, where $i \in [K]$ is the class of boxes that initially attains the largest opening threshold $\sigma_M([N],\emptyset)$.

We first examine Case~1, where $\sigma_i^F = \sigma_M([N], \emptyset)$. In this setting, our policy achieves an expected profit of at least
\begin{eqnarray*}
    & &  -c_i^F  +  u \mathbb{E} \left[ V_i \mid V_i > \sigma_i^F \right] + \left( 1 - u\right)   \left( -c_i^F + u \mathbb{E} \left[ V_i \mid V_i > \sigma_i^F \right] +   \left( 1 - u\right) (...) \right)  \\
    &\geq& u \sigma^F_i + (1-u)u \sigma^F_i + (1-u)^2 u \sigma^F_i + ... + (1-u)^{ \lfloor  \theta \cdot N \rfloor -1} u \sigma^F_i   \\
    &=&  u \sigma^F_i \frac{1-(1-u)^{ \lfloor  \theta \cdot N \rfloor }}{1-(1-u)} \xrightarrow[]{N \rightarrow \infty} \sigma^F_i,
\end{eqnarray*}
where we use the fact
$$
 u \cdot \sigma_i^F = - c_i^F + u \cdot \mathbb{E} \left[ V_i \mid V_i > \sigma_i^F \right].
$$
By Lemma~\ref{thm:upper_bounds}, we know that $\sigma_i^F$ is an upper bound on the optimal expected profit. Hence, we derive the approximation ratio stated in Lemma~\ref{lemma:infinite_iid_boxes}.

Next, we examine Case~2, where $\sigma_i^P = \sigma_M([N], \emptyset)$. Here, our policy achieves an expected profit of at least
\begin{eqnarray*}
  & &  -c_i^P  + q \left( -c_i^F +  p \mathbb{E}\left[ V_i \mid V_i > \sigma_i^P, \sigma_i^{F \mid T_i} > \sigma_i^P \right]  \right)  \\
   &+&  \left( 1-q + q(1-p) \right) \left( -c_i^P  + q \left( -c_i^F +  p \mathbb{E}\left[ V_i \mid V_i > \sigma_i^P, \sigma_i^{F \mid T_i} > \sigma_i^P \right] + \left( 1-q + q(1-p) \right) \left( ... \right) \right)  \right)  \\
   &\geq & pq \sigma^P_i + (1-pq) pq\sigma^P_i + (1-pq)^2pq \sigma^P_i + ...  + (1-pq)^{\lfloor  \theta \cdot N \rfloor-1} pq \sigma^P_i \\
    &=& pq \sigma^P_i \frac{1-(1-pq)^{\lfloor  \theta \cdot N \rfloor }}{1-(1-pq)} \xrightarrow[]{N \rightarrow \infty} \sigma^P_i  ,
\end{eqnarray*}
where we use the fact 
$$
p_{} q_{} \sigma_i^P = -c_i^P + q_{} \left( -c_i^F +  p_{} \mathbb{E}\left[ V_i  \mid V_i > \sigma_i^P, \sigma_i^{F \mid T_i} > \sigma_i^P \right]  \right).
$$
By Lemma~\ref{thm:upper_bounds}, we know that $\sigma_i^P$ is an upper bound on the optimal expected profit. 
Hence, the approximation ratio of Lemma~\ref{lemma:infinite_iid_boxes} immediately follows. 
\end{myproof}

\subsection{Proof of Lemma \ref{lemma:committing} }
\label{appendix:lemma:committing}

First, we identify necessary and sufficient conditions such that there is no gap between the free-info upper bound and the value function.

\begin{lemma}
\label{lemma:non-exposed_policy}
For every policy $\pi$, we have $J^{\pi,K}(s_0) = J^{\pi}(s_0)$ if and only if policy $\pi$ satisfies the following conditions:
\begin{enumerate}
  \item $\pi$ selects any F-opened box $i$ (which was not P-opened) whose value satisfies $V_i > \sigma^{F}_i$;
  \item $\pi$ fully opens any P-opened box $i$ whose type satisfies $\sigma^{F \mid T_i}_i > \sigma^{P}_i$;
  \item $\pi$ selects any F-opened box $i$ (which was already P-opened) that satisfies $V_i > \min \{\sigma^{P}_i, \sigma^{F \mid T_i}_i \}$.
\end{enumerate}
\end{lemma}

Consistently with the terminology used in previous literature, we call these policies {\em non-exposed}. 
Note that non-exposed policies were introduced by \cite{beyhaghi2019pandora}. Our proof of Lemma~\ref{lemma:non-exposed_policy} in Appendix~\ref{appendix:lemma:non-exposed_policy} requires a slightly more general argument for our sequential inspection processes and the corresponding two-stage (rather than single-stage) call options.

Now, using equation~\eqref{eq:free}, for any committing policy $\pi^{F,P}$, the expected profit in the free-info problem is
\begin{eqnarray}
J^{\pi, K}(s_0) &=& \E{   
\sum_{i=1}^{N} \left(  \mathbb{S}_i^{\pi} K_i (1-\indPi) + \widetilde{\mathbb{S}}_i^{\pi} \widetilde{K}^{t_i}_i \indPi  \right) 
} \nonumber \\
&=& \mathbb{E} \Big[ \sum_{i\in F} \indSi^{ } K_i(1-\indPi^{ } )+\sum_{i\in P}\indSitilde^{ } \widetilde{K}_i \indPi^{ }    \Big]  \nonumber \\ 
&\leq&  \mathbb{E}\left[ \max \left\{ \underset{i \in F}{\max} \{  K_i\}   , \underset{i \in P}{\max}  \{ \widetilde{K}_i \}  \right\}\right] \nonumber, 
\end{eqnarray}
where the first equality holds due to the construction of the committing policy $\pi^{F,P}$, and the inequality follows from the constraints 
$\sum_{i=1}^{N}  \indSi^{\pi} + \sum_{i=1}^{N}\indSitilde^{\pi}  \leq 1 $, and $0 \leq \indPi^{\pi} \leq 1 $.

To conclude the proof, it suffices to show that this upper bound is matched by $\pi^{F,P}$ in the real-world problem. 
To see this, we first remark that the committing policy $\pi^{F,P}$ is non-exposed, and thus, by Lemma~\ref{lemma:non-exposed_policy}, $J^{\pi^{F,P} } (s_0) =J^{\pi^{F,P} , M }(s_0)  $.
Therefore, to prove the lemma, it suffices to show that in the free-info problem, policy $\pi^{F,P}$ achieves the maximal capped realization. 
Letting $\widehat{K_i} = K_i $ if $i \in F$ and $\widehat{K_i} = \widetilde{K_i}$ if $i \in P$ be the capped value of each box $i$, we would like to show that that policy $\pi^{F,P}$ selects $\underset{i \in [N]}{\max} \left\{ \widehat{K}_i \right\}$ under any realization. 

Let $j$ denote the box selected by the policy. 
We note that, at termination, except for $\widehat{K}_j$, there are two groups of boxes: (1) boxes that were F-opened (perhaps after P-opening) ; and (2) boxes that were not F-opened (but possibly were P-opened).
We argue that the capped value of each box in both groups is smaller than or equal to $\widehat{K}_j$. 

Specifically, for any box $i$ in group 1, observe that  
\begin{itemize}
    \item Fact 1: $V_i \leq V_j$. This is due to the fact that policy $\pi^{F,P}$ always selects the best prize so far, and thus $V_i \leq V_j$;
    \item Fact 2: $V_i < \sigma_j$. To see this, note that if $V_j \geq \sigma_j$ the policy immediately stops after F-opening box $j$. In this case, $V_i \leq \sigma_j$ (otherwise box $j$ would not have been opened). Alternatively, if $V_j<\sigma_j$, then it follows from fact 1 that $V_i<\sigma_j$. 
\end{itemize}
Recalling that $\widehat{K_i} \leq V_i$, we have $\widehat{K_j} \geq \widehat{K_i}$ for any box $i$ in group (1). 

Next, for any box $i$ in group (2), we observe that 
\begin{itemize}
    \item Fact 3: $\sigma_j \geq \sigma_i \geq \widehat{K_i}$, since  policy $\pi^{F,P}$ always opens the box with largest threshold;
    \item Fact 4: $V_j \geq \sigma_i \geq \widehat{K_i}$, since  policy $\pi^{F,P}$ obeys the optimal stopping rule. 
\end{itemize}
Hence, $\widehat{K}_j$ is indeed the largest capped value. 
It then follows that policy $\pi^{F,P}$ attains the largest capped realization.

\subsection{Proof of Lemma~\ref{lemma:non-exposed_policy}}
\label{appendix:lemma:non-exposed_policy}

Recall the expressions of $J^\pi(s_0)$ and $J^{\pi, K}(s_0)$
$$
J^\pi(s_0) =  \E{   
\sum_{i=1}^{N} \left(  \mathbb{S}_i^{\pi} V_i - \mathbb{F}_i^{\pi}  c_i^F - \mathbb{P}_i^{\pi} c_i^P \right) + \sum_{i=1}^{N} \left(  \widetilde{\mathbb{S}}_i^{\pi} V_i -  \widetilde{\mathbb{F}}_i^{\pi}  c_i^F \right) 
} \ , 
$$
and 
$$
J^{\pi, K}(s_0) = \E{   
\sum_{i=1}^{N} \left(  \mathbb{S}_i^{\pi} K_i (1-\indPi) + \widetilde{\mathbb{S}}_i^{\pi} \widetilde{K}^{t_i}_i \indPi  \right) 
} \ . 
$$
To prove the lemma\footnote{For brevity, we mute the superscripts of $\mathbb{S}_i^{\pi}, \widetilde{\mathbb{S}}_i^{\pi}, \indFi^{\pi}, \indFitilde^{\pi}, \indPi^{\pi}$ below in this proof. }, it suffices to show that for all $i$,
$$
\mathbb{E}\big[\indSi  V_i + \indSitilde V_i - \indFi c_i^F - \indFitilde c_i^F - \indPi c_i^P \big]  
\leq \mathbb{E}[\indSi K_i(1-\indPi)+\indSitilde \widetilde{K}_i \indPi] \ ,
$$
and that the inequalities hold with equalities if and only if the policy satisfies the three criteria for all boxes $i$. 
We proceed to show that this is indeed true. 

For any box $i \in [N]$, its contribution to the total expected profit is 
\begin{eqnarray*}
\underbrace{\mathbb{E}\big[ \indSi  V_i - \indFi c_i^F \big]}_{\textcircled{1}} + \underbrace{\mathbb{E}\big[ \indSitilde V_i  - \indFitilde c_i^F - \indPi c_i^P \big]}_{\textcircled{2}} \ ,
\end{eqnarray*}
where the term \textcircled{1} corresponds to the expected profit from box $i$ by directly F-opening it, while the term \textcircled{2} corresponds to a P-opening followed by a F-opening. We derive upper bounds on each of two terms, and show these bounds are exactly tight for a non-exposed policy. 

\paragraph{The first condition of non-exposed policies.} For the first term, we note that 
\begin{eqnarray}
  \textcircled{1} &=& \mathbb{E}\big[\indSi  V_i - \indFi c_i^F \mid \indPi=0 \big] \operatorname{Pr}[\indPi=0]  + \mathbb{E}\big[ \underbrace{\indSi  V_i - \indFi c_i^F }_{=0} \mid \indPi=1 \big]  \operatorname{Pr}[\indPi=1]  \label{eq:expiredBox_1_1} \\
  &=& \mathbb{E}\left[ \indSi V_i - \indFi  \mathbb{E}[(V_i-\sigma_i^F)^+ ] \mid \indPi=0\right] \operatorname{Pr}[\indPi=0]  \label{eq:lemma_proof_expired_box_1} \\
   &=& \mathbb{E}\left[ \indSi V_i - \indFi  (V_i-\sigma_i^F)^+  \mid \indPi=0\right]  \operatorname{Pr}[\indPi=0] \label{eq:lemma_proof_expired_box_2} \\
   &\leq&  \mathbb{E}\left[ \indSi V_i - \indSi  (V_i-\sigma_i^F)^+  \mid \indPi=0\right] \operatorname{Pr}[\indPi=0] \label{eq:lemma_proof_expired_box_3} \\
   &=& \mathbb{E}\left[ \indSi K_i  \mid \indPi=0\right] \operatorname{Pr}[\indPi=0]  \label{eq:lemma_proof_expired_box_3_2},
\end{eqnarray}
where equation~\eqref{eq:expiredBox_1_1} is obtained by conditioning on the decision to P-open, and by showing that the underbraced expression is equal to zero. 
The latter fact follows  by noting that $\indSi \leq \indFi = 1 - \indPi=0$ when $\indPi= 1 $.
Equation~\eqref{eq:lemma_proof_expired_box_1} follows from  the definition of the F-threshold $\sigma_i^F$, and equation~\eqref{eq:lemma_proof_expired_box_2} is due to the independence of $\indFi$ and $V_i$, since the decision of directly F-opening or not does not depend on the realization of $V_i$. The inequality again proceeds from constraint $\indSi \leq \indFi$. 
We observe that this inequality is tight if and only if $\mathbb{E}\left[ (\indFi-\indSi) (V_i-\sigma_i^F)^+ \mid \indPi=0 \right] = 0 $. 
In particular, when $V_i \leq \sigma^{F}_i$, this inequality is satisfied with equality.
Moreover, when $V_i$ satisfies $V_i > \sigma^{F}_i$, the first criterion in the definition of non-exposed policies is a sufficient and necessary condition for the box to be selected, in which case $\indSi=\indFi=1$ and the above inequality is satisfied with equality.
Finally, equation~\eqref{eq:lemma_proof_expired_box_3_2} immediately follows from the definition of $K_i$.
We mention in passing that our analysis of term \textcircled{1} is nearly identical to that of \cite{kleinberg2016descending}[Lemma 1].
\paragraph{The second condition of non-exposed policies.}
Now, let us turn to the second term:
\begin{scriptsize}
\begin{eqnarray}
 \textcircled{2}  &=&  \mathbb{E}\left[ \indSitilde V_i  - \indFitilde c_i^F - \indPi c_i^P \mid \indPi=1\right]  \operatorname{Pr}[\indPi=1] +  \mathbb{E}\left[  \underbrace{\indSitilde V_i  - \indFitilde c_i^F - \indPi c_i^P }_{=0} \mid \indPi=0\right]  \operatorname{Pr}[\indPi=0]  \label{eq:expiredBox_2_1} \\
   &=&   \mathbb{E}\left[ \indSitilde V_i  - \indFitilde c_i^F - \indPi \mathbb{E}\big[\max\{0, -c_i^F+\mathbb{E}[ (V_i-\sigma_i^P)^+  \mid T_i]\}\big] \mid  \indPi=1 \right]\operatorname{Pr}[\indPi=1]   \label{eq:lemma_proof_expired_box_4}  \\
   &=&  \mathbb{E}\left[ \indSitilde V_i  - \indFitilde c_i^F - \indPi \max\{0, -c_i^F+\mathbb{E}\big[ (V_i-\sigma_i^P)^+  \mid T_i \big]\} \mid \indPi=1 \right] \operatorname{Pr}[\indPi=1] \label{eq:lemma_proof_expired_box_5} \\
   &\leq&  \mathbb{E}\left[ \indSitilde V_i  - \indFitilde c_i^F - \indFitilde \max\{0, -c_i^F+\mathbb{E}\big[ (V_i-\sigma_i^P)^+  \mid T_i\big]\} \mid \indPi=1 \right]  \operatorname{Pr}[\indPi=1] \label{eq:lemma_proof_expired_box_6},
\end{eqnarray}
\end{scriptsize}
where equation~\eqref{eq:expiredBox_2_1} is obtained by conditioning on the decision to P-open, and the underbraced term is equal to zero when $\indPi=0$ (this fact can be verified from the definition of 
$\indSitilde$ and $\indFitilde$).
Equation~\eqref{eq:lemma_proof_expired_box_4} is due to definition of the P-threshold $\sigma^{P}_i$.
Equation \eqref{eq:lemma_proof_expired_box_5} is due to the independence of $\indPi$ and $T_i$, since the decision to partially open does not depend on the realization of the type $T_i$. 
The inequality \eqref{eq:lemma_proof_expired_box_6} holds since $\indFitilde \leq \indPi$.

 We next examine when the inequality~\ref{eq:lemma_proof_expired_box_6} is tight. To this end, let $ \varphi_i(t_i)=\max \left \{0, -c_i^F+\mathbb{E}[ (V_i-\sigma_i^P)^+  \mid T_i = t_i] \right \} $. We can write the difference between equations \eqref{eq:lemma_proof_expired_box_6} and \eqref{eq:lemma_proof_expired_box_5} as follows:
\begin{eqnarray}
  & & \eqref{eq:lemma_proof_expired_box_6}-\eqref{eq:lemma_proof_expired_box_5} \nonumber\\
  &=& \mathbb{E}\left[ \left(\indPi - \indFitilde\right)\varphi(T_i) \mid \indPi=1 \right] \operatorname{Pr}[\indPi=1]   \label{eq:expiredBox_3_1} \\
   &=& \sum_{t_i \in \Gamma_i} \operatorname{Pr}[T_i=t_i]  \BFI\{ \varphi(t_i)=0 \} \operatorname{Pr}[\indPi=1]  \cdot 0  \label{eq:expiredBox_3_2} \\
   &+&  \sum_{t_i \in \Gamma_i}  \operatorname{Pr}[T_i=t_i] \BFI\{ \varphi(t_i)>0 \}  \operatorname{Pr}[\indPi=1] \operatorname{Pr}\Big[\indFitilde=1\mid \indPi=1, T_i=t_i   \Big]  \cdot 0   \label{eq:expiredBox_3_3} \\
   &+&  \sum_{t_i \in \Gamma_i}  \operatorname{Pr}[T_i=t_i] \BFI\{ \varphi(t_i)>0 \}  \operatorname{Pr}[\indPi=1] \operatorname{Pr}\Big[\indFitilde=0\mid \indPi=1, T_i=t_i  \Big] \cdot \underbrace{  \varphi_i(t_i) }_{>0} \label{eq:expiredBox_3_4},
\end{eqnarray}
In equation \eqref{eq:expiredBox_3_1}, we substitute the function $\varphi_i(t_i)$ into the difference between \eqref{eq:lemma_proof_expired_box_6} and \eqref{eq:lemma_proof_expired_box_5}. 
We then decompose the expectation into three summations based on the type $t_i$.
Specifically, expression \eqref{eq:expiredBox_3_2} corresponds to the ``bad'' types $t_i$, for which F-opening is not worthwhile. 
Expression \eqref{eq:expiredBox_3_3} corresponds to the case where box $i$ is F-opened ($\indFitilde=1$) after being P-opened ($\indPi=1$).
Finally, expression \eqref{eq:expiredBox_3_4} corresponds to the case where box $i$ is P-opened ($\indPi=1$) but not F-opened ($\indFitilde=0$).
Therefore, equations  $\eqref{eq:lemma_proof_expired_box_6}$ and $\eqref{eq:lemma_proof_expired_box_5}$ are equal if and only if $\varphi_i(t_i)>0 $ implies that $\operatorname{Pr} \left[ \indFitilde=0  ~\middle \vert~ \indPi=1, T_i=t_i  \right]=0$. Recall that $c_i^F = \mathbb{E}_{V_i \mid T_i=t_i}\Big[(V_i - \sigma^{F \mid t_i}_i)^+\Big]$ by definition of $\sigma^{F \mid t_i}_i$, and that $ \varphi_i(t_i)=\max \left \{0, -c_i^F+\mathbb{E} \left [ (V_i-\sigma_i^P)^+  \mid T_i = t_i \right] \right\} $.
Therefore, we have 
\begin{eqnarray}
 \varphi_i(t_i)>0  & \Leftrightarrow & -c_i^F+\mathbb{E}[ (V_i-\sigma_i^P)^+  \mid T_i = t_i]>0  \nonumber \\ 
 & \Leftrightarrow & \mathbb{E}[ (V_i-\sigma_i^P)^+  \mid T_i = t_i] > c_i^F \nonumber \\ 
 & \Leftrightarrow & \mathbb{E}_{V_i \mid T_i=t_i}\Big[(V_i - \sigma^{P}_i)^+\Big] > \mathbb{E}_{V_i \mid T_i=t_i}\Big[(V_i - \sigma^{F \mid t_i}_i)^+\Big] \nonumber \\
 & \Leftrightarrow & \sigma_i^{F \mid t_i} > \sigma_i^{P}.\nonumber
\end{eqnarray}
This is exactly the second criterion in the definition of non-exposed policies, which requires that if $\sigma_i^{F \mid t_i} > \sigma_i^{P}$ then F-opening will succeed a P-opening.
\\
\paragraph{The third condition of non-exposed policies.} Now we get back to characterizing an upper bound on term \textcircled{2}:
\begin{scriptsize}
\begin{eqnarray}
   &&  \mathbb{E}\left[ \indSitilde V_i  - \indFitilde \left( c_i^F + \max\{0, -c_i^F+\mathbb{E}[ (V_i-\sigma_i^P)^+  \mid T_i]\} \right) \mid \indPi=1 \right]  \operatorname{Pr}[\indPi=1] \nonumber   \\
   &=&  \mathbb{E}\left[ \indSitilde V_i  - \indFitilde \left( \max\{c_i^F, \mathbb{E}[ (V_i-\sigma_i^P)^+  \mid T_i]\} \right) \mid \indPi = 1\right]   \operatorname{Pr}[\indPi=1]     \nonumber  \\
   &=&  \sum_{t_i \in \Gamma_i} \operatorname{Pr}[T_i = t_i]  \BFI\{ c_i^F \geq \mathbb{E}\left[(V_i-\sigma_i^P)^+ \mid T_i=t_i\right] \}  \mathbb{E}  \left[ \indSitilde V_i  -  \indFitilde c_i^F  \mid T_i=t_i, \indPi = 1 \right]\operatorname{Pr}[\indPi=1]      \nonumber     \\
   &+&  \sum_{t_i \in \Gamma_i}   \operatorname{Pr}[T_i = t_i]  \BFI\{  c_i^F < \mathbb{E}\left[(V_i-\sigma_i^P)^+ \mid T_i=t_i\right] \}   \mathbb{E} \left[ \indSitilde V_i  -  \indFitilde \mathbb{E}\left[(V_i-\sigma_i^P)^+ \mid T_i=t_i\right] \mid T_i=t_i, \indPi = 1 \right]\operatorname{Pr}[\indPi=1]     \label{eq:expiredBox_4_1}     \\
   &=& \sum_{t_i \in \Gamma_i}  \operatorname{Pr}[T_i = t_i]  \BFI\{ c_i^F \geq \mathbb{E}\left[(V_i-\sigma_i^P)^+ \mid T_i=t_i\right] \}  \mathbb{E}  \left[ \indSitilde V_i  -  \indFitilde \mathbb{E}\left[(V_i - \sigma_i^{F \mid t_i})^+ \mid T_i=t_i \right]  \mid T_i=t_i, \indPi = 1 \right] \operatorname{Pr}[\indPi=1]   \nonumber     \\
   &+&  \sum_{t_i \in \Gamma_i}  \operatorname{Pr}[T_i = t_i]  \BFI\{  c_i^F < \mathbb{E}\left[(V_i-\sigma_i^P)^+ \mid T_i=t_i \right] \}  \mathbb{E} \left[ \indSitilde V_i  -  \indFitilde \mathbb{E}\left[(V_i-\sigma_i^P)^+ \mid T_i=t_i\right] \mid T_i=t_i, \indPi = 1 \right] \operatorname{Pr}[\indPi=1]    \label{eq:expiredBox_4_2} \\
   &=&     \sum_{t_i \in \Gamma_i} \operatorname{Pr}[T_i = t_i]  \BFI\{ c_i^F \geq \mathbb{E}\left[(V_i-\sigma_i^P)^+ \mid T_i= t_i\right] \}  \mathbb{E}  \left[ \indSitilde V_i  -  \indFitilde (V_i - \sigma_i^{F \mid t_i})^+  \mid T_i = t_i , \indPi=1 \right]  \operatorname{Pr}[\indPi=1]   \nonumber    \\
   &+&  \sum_{t_i \in \Gamma_i} \operatorname{Pr}[T_i = t_i]  \BFI\{  c_i^F < \mathbb{E}\left[(V_i-\sigma_i^P)^+  \mid T_i =  t_i \right] \} \mathbb{E} \left[ \indSitilde V_i  -  \indFitilde (V_i-\sigma_i^P)^+   \mid T_i =  t_i, \indPi=1 \right]  \operatorname{Pr}[\indPi=1]   \label{eq:lemma_proof_expired_box_7a}   
\end{eqnarray}
\end{scriptsize}
The first equation is identical to equation~\eqref{eq:lemma_proof_expired_box_6}.
Equation~\eqref{eq:expiredBox_4_1} is obtained by conditioning on the type $t_i$ revealed after P-opening box $i$. 
Equation~\eqref{eq:expiredBox_4_2} is derived by substituting the definition of $\sigma^{F \mid t_i}_i$ (cf. Definition~\ref{def:testing_threshold}). 
We obtain equation\eqref{eq:lemma_proof_expired_box_7a} by consolidating the expectation operator around
$(V_i-\sigma_i^P)^+$ and $(V_i - \sigma_i^{F \mid t_i})^+$ and by observing the following sequence of equalities: 
\begin{eqnarray}
  &&\mathbb{E}\left[ \indFitilde \mathbb{E}\left[(V_i-\sigma_i^{F \mid t_i})^+ \mid T_i=t_i \right] \mid T_i=t_i, \indPi=1 \right] \nonumber\\
  &=& 
   \mathbb{E}\left[ \indFitilde  \mid T_i=t_i, \indPi=1 \right] \mathbb{E}\left[(V_i-\sigma_i^{F \mid t_i})^+  \mid T_i=t_i, \indPi=1  \right] \label{eq:independence_1_2}  \nonumber \\
   &=& \mathbb{E}\left[ \indFitilde (V_i-\sigma_i^{F \mid t_i})^+  \mid T_i=t_i, \indPi=1 \right] \label{eq:independence_1_3}. \nonumber 
\end{eqnarray}
The first equality follows from the fact that $\mathbb{E}\left[(V_i-\sigma_i^{F \mid t_i})^+ ~\middle \vert ~ T_i=t_i \right] $ is a constant, independent of $\indFitilde$. The second equality follows from the independence between $\indFitilde$ and $V_i$ conditional on $T_i=t_i$.
Moving forward, we have:
\begin{scriptsize}
\begin{eqnarray}
\eqref{eq:lemma_proof_expired_box_7a} 
&=& \sum_{t_i \in \Gamma_i} \operatorname{Pr}[T_i = t_i]  \BFI\{ c_i^F \geq \mathbb{E}\left[(V_i-\sigma_i^P)^+ \mid T_i= t_i\right] \}  \mathbb{E}  \left[ \indSitilde V_i  -  \indFitilde (V_i - \sigma_i^{F \mid t_i})^+  \mid T_i = t_i , \indPi=1 \right]  \operatorname{Pr}[\indPi=1]   \nonumber    \\
   &+&  \sum_{t_i \in \Gamma_i} \operatorname{Pr}[T_i = t_i]  \BFI\{  c_i^F < \mathbb{E}\left[(V_i-\sigma_i^P)^+  \mid T_i =  t_i \right] \} \mathbb{E} \left[ \indSitilde V_i  -  \indFitilde (V_i-\sigma_i^P)^+   \mid T_i =  t_i, \indPi=1 \right]  \operatorname{Pr}[\indPi=1]   \label{eq:lemma_proof_expired_box_7}   \\
   &\leq&\sum_{t_i \in \Gamma_i} \operatorname{Pr}[T_i = t_i]  \BFI\{ c_i^F \geq \mathbb{E}\left[(V_i-\sigma_i^P)^+  \mid T_i= t_i\right] \}  \mathbb{E}  \left[ \indSitilde V_i  -  \indSitilde (V_i - \sigma_i^{F \mid t_i})^+  \mid T_i = t_i, \indPi=1 \right] \operatorname{Pr}[\indPi=1]    \nonumber     \\
   &+& \sum_{t_i \in \Gamma_i} \operatorname{Pr}[T_i = t_i]  \BFI\{  c_i^F < \mathbb{E}\left[(V_i-\sigma_i^P)^+  \mid T_i= t_i \right] \} \mathbb{E} \left[ \indSitilde V_i  -  \indSitilde (V_i-\sigma_i^P)^+   \mid T_i =  t_i , \indPi=1 \right] \operatorname{Pr}[\indPi=1]  \label{eq:lemma_proof_expired_box_8}     \\
   &=&\sum_{t_i \in \Gamma_i} \operatorname{Pr}[T_i = t_i]  \BFI\{ \sigma_i^{F \mid t_i} \leq \sigma_i^P \}  \mathbb{E}  \left[ \indSitilde V_i  -  \indSitilde (V_i - \sigma_i^{F \mid t_i})^+  \mid T_i = t_i, \indPi=1\right] \operatorname{Pr}[\indPi=1]    \nonumber     \\
   &+& \sum_{t_i \in \Gamma_i} \operatorname{Pr}[T_i = t_i]  \BFI\{  \sigma_i^{F \mid t_i} > \sigma_i^P \} \mathbb{E} \left[ \indSitilde V_i  -  \indSitilde (V_i-\sigma_i^P)^+  \mid T_i = t_i , \indPi=1\right] \operatorname{Pr}[\indPi=1]  \label{eq:lemma_proof_expired_box_9}  \\
   &=& \sum_{t_i \in \Gamma_i} \operatorname{Pr}[T_i = t_i]  \BFI\{ \sigma_i^{F \mid t_i} \leq \sigma_i^P \}  \mathbb{E}  \left[ \indSitilde \min\{ V_i, \sigma_i^{F \mid t_i}\}  \mid T_i = t_i, \indPi=1\right] \operatorname{Pr}[\indPi=1]    \nonumber     \\
   &+& \sum_{t_i \in \Gamma_i} \operatorname{Pr}[T_i = t_i]  \BFI\{  \sigma_i^{F \mid t_i} > \sigma_i^P \} \mathbb{E} \left[ \indSitilde \min\{ V_i, \sigma_i^P\}  \mid T_i = t_i, \indPi=1 \right] \operatorname{Pr}[\indPi=1]   \label{eq:lemma_proof_expired_box_10}   \\
   &=& \mathbb{E}\Big[\indSitilde \min\{V_i, \sigma_i^P, \sigma^{F \mid T_i}_i \} \mid \indPi=1 \Big] \operatorname{Pr}[\indPi=1]  \label{eq:lemma_proof_expired_box_11}    \\
   &=&  \mathbb{E}[\indSitilde \widetilde{K}_i \mid \indPi=1] \operatorname{Pr}[\indPi=1] \label{eq:lemma_proof_expired_box_12}.
\end{eqnarray}
\end{scriptsize}
The first equation is identical to equation~\ref{eq:lemma_proof_expired_box_7a}. The inequality~\ref{eq:lemma_proof_expired_box_8} follows from the inequality $ \indSitilde  \leq \indFitilde $. 
For equation\eqref{eq:lemma_proof_expired_box_9}, note that by definition of $\sigma^{F \mid t_i}_i$, 
\begin{eqnarray}
 c_i^F \geq \mathbb{E}\left[(V_i-\sigma_i^P)^+ \mid T_i=t_i\right] &\Leftrightarrow& \mathbb{E}\left[(V_i-\sigma_i^{F \mid t_i})^+  \mid T_i=t_i\right] \geq \mathbb{E}\left[(V_i-\sigma_i^P)^+ \mid T_i=t_i\right] \nonumber\\
 &\Leftrightarrow& \sigma_i^{F \mid t_i} \leq \sigma_i^P,\nonumber
\end{eqnarray}
where the last equivalence follows from the monotonicity of the mapping $x \rightarrow  \E{ (V_i-x)^+ \mid T_i=t_i }  $. 
Equation~\eqref{eq:lemma_proof_expired_box_10} follows from the identity $x-(x-y)^+=\min\{x,y \}$. 
Equation~\eqref{eq:lemma_proof_expired_box_11} can be justified  by conditioning on $T_i$, and finally, equation\eqref{eq:lemma_proof_expired_box_12} follows from the definition of $\widetilde{K}_i$. 

To identify under which condition we have $\eqref{eq:lemma_proof_expired_box_7} = \eqref{eq:lemma_proof_expired_box_8}$, we consider the difference between these terms:
\begin{eqnarray}
  \eqref{eq:lemma_proof_expired_box_8}-\eqref{eq:lemma_proof_expired_box_7} &=&  \sum_{t_i \in \Gamma_i} \operatorname{Pr}[T_i = t_i]  \BFI\{ \sigma_i^{F \mid t_i} \leq \sigma_i^P \}  \mathbb{E}  \left[(\indFitilde-\indSitilde) (V_i - \sigma_i^{F \mid t_i})^+  \mid T_i = t_i\right] \operatorname{Pr}[\indPi=1]    \nonumber     \\
   &+& \sum_{t_i \in \Gamma_i} \operatorname{Pr}[T_i = t_i]  \BFI\{  \sigma_i^{F \mid t_i} > \sigma_i^P \} \mathbb{E} \left[ (\indFitilde-\indSitilde) (V_i-\sigma_i^P)^+  \mid T_i = t_i \right] \operatorname{Pr}[\indPi=1].\nonumber
\end{eqnarray}
We note that the latter sum-expressions are equal to zero if and only if the policy $\pi$ is designed such that 
\begin{itemize}
    \item   $  \left\{ V_i > \sigma_i^{F \mid t_i} \right\}  \cap \left\{  \sigma_i^{F \mid t_i} \leq \sigma_i^P \right\}  \cap \left\{ \indFitilde=1 \right\} \Rightarrow \left\{  \indSitilde=1 \right\}$ , and
    \item   $  \left \{ V_i > \sigma_i^P \right\} \cap  \left \{  \sigma_i^{F \mid t_i} > \sigma_i^P \right\}  \cap  \left\{  \indFitilde=1 \right\} \Rightarrow \left\{ \indSitilde=1 \right\} $.
\end{itemize}
The latter conditions can be reformulated as $ \left\{ V_i > \min \{\sigma_i^P, \sigma_i^{F \mid t_i} \} \right\} \cap  \left\{ \indFitilde=1 \right\} \Rightarrow \left\{ \indSitilde=1 \right\} $. 
This is precisely the third criterion in the definition of non-exposed policies.

Finally, combining inequalities \eqref{eq:lemma_proof_expired_box_3_2} and \eqref{eq:lemma_proof_expired_box_12} yields that the expected profit from box $i$ is bounded above by:
\begin{eqnarray}
   \mathbb{E}\big[\indSi  V_i + \indSitilde V_i - \indFi c_i^F - \indFitilde c_i^F - \indPi c_i^P \big] &\leq &   \mathbb{E}\left[ \indSi K_i \mid \indPi=0 \right] \operatorname{Pr}[\indPi=0] + \mathbb{E}[\indSitilde \widetilde{K}_i \mid \indPi=1] \operatorname{Pr}[\indPi=1] \nonumber \\
   &=& \mathbb{E}[\indSi K_i(1-\indPi)+\indSitilde \widetilde{K}_i \indPi] \ .  \nonumber
\end{eqnarray}
Following from the discussion above, we know that the inequalities hold as equalities for all $i$ if and only if the policy is non-exposed.

\subsection{Proof of Theorem~\ref{thm:1-1/e}} \label{appendix:thm:1-1/e}
In the spirit of \cite{beyhaghi2019pandora}, we cast our free-info problem into a respective stochastic submodular optimization problem. 
The adaptivity gap result by \cite{asadpour2015maximizing}  guarantees that the performance of some committing policy is comparable to a fully adaptive policy, up to a constant-factor approximation.

The proof is organized as follows.
We begin by summarizing results from \cite{asadpour2015maximizing} on stochastic monotone submodular optimization, in section~\ref{subsec:summary_submodular}. In section~\ref{subsec:submodular_problems}, we then formulate specific submodular optimization problems related to our Pandora's box with sequential inspection. In section~\ref{subsec:submodular_transformation}, we continue to define the transformation of policies between the free-info problem and the submodular problem and prove relations between these transformations. Finally, in section~\ref{subsec:proof_of_thm:1-1/e}, we combine earlier results to prove the main theorem.

\subsubsection{Submodular monotone nonnegative maximization subject to a matroid constraint.} \label{subsec:summary_submodular}
We consider the class of stochastic optimization problems, represented by a three-tuple $(f, \mathcal{A}, \mathcal{F})$, where the goal is to maximize the expectation of a stochastic monotone submodular function $f$, subject to a matroid constraint $\mathcal{M}=(\mathcal{A}, \mathcal{F})$. That is, $f: \mathbb{R}_+^n \rightarrow \mathbb{R}_+$ satisfies
$$
\forall~x, y \in \mathbb{R}_+^n: f(x \vee y ) + f(x \wedge y) \leq f(x) + f(y),
$$
where we denote by $x \vee y$ the element-wise maximum and $x \wedge y$ the element-wise minimum of vectors $x$ and $y$. 
The matroid $\mathcal{M}$ captures the feasibility constraint, where  $\mathcal{A}=\{X_1, X_2, ..., X_n \}$ is a ground set of $n$ \textit{independent} random variables, and $\mathcal{F}$ is a collection of independent sets.\footnote{Definition of a matroid $\mathcal{M}=(\mathcal{A}, \mathcal{F})$: all elements of $\mathcal{F}$ are subsets of $\mathcal{A}$ satisfying the following properties: (i) $\emptyset \in \mathcal{F}$; (ii) If $A \in \mathcal{F}$ and $B \subseteq A$, then $B \in \mathcal{F}$; (iii) If $A \in \mathcal{F}$, $B \in \mathcal{F}$ and $ \mid B \mid  >  \mid A \mid $, then there exists $b \in B \setminus A$ with $A \cup \{b\} \in \mathcal{F}$. }

A policy $\pi$ generates a sequence of \textit{selections} of elements in the ground set. 
More precisely, it is a mapping from the set of all possible states to a distribution over elements in the ground set. 
After each selection, the realization of the corresponding element is revealed and the policy may adapt to this new information accordingly. 
The state of this problem $\Theta$ can be described by an $n$-tuple $(\theta_1, \theta_2, ..., \theta_n)$, where $\theta_i$ denotes the realization of $X_i$, if $X_i \in \mathcal{A}$ is selected by the policy, and is equal to $\circ$ otherwise, which means this element has not been selected yet. 
We say a policy is adaptive if the selection decisions are based on the outcomes of the previous actions; otherwise, we say that a policy is nonadaptive. 
Denote by $\Theta^\pi$ the random $n$-tuple corresponding to the last state reached by the policy $\pi$. The value of this policy is defined to be $\mathbb{E}\Big[f(\Theta^\pi)\Big]$. 
We denote by $\mathfrak{F}(\mathcal{A}, \mathcal{F})$ the set of feasible policies, in which the set of elements chosen by $\pi$ always belong to $\mathcal{F}$.

The goal of this optimization problem is to find a feasible policy with respect to $\mathcal{M}$ that obtains the maximum expected value of its final state, i.e.,
\begin{equation}\label{eq:submodular_form}
  \max_{\pi \in \mathfrak{F}(\mathcal{A}, \mathcal{F})}  \mathbb{E} \Big[f(\Theta^\pi)\Big].    \nonumber
\end{equation}

In what follows, we restate the main result of \cite{asadpour2015maximizing} as a theorem, which we will use as the main machinery for our analysis.
\begin{theorem} \label{lemma:borrowed_machinery} (\cite{asadpour2015maximizing}, Theorem 1).
There exists a nonadaptive policy that achieves $1-\frac{1}{e} \approx 0.63$ fraction of the optimal policy in maximizing a stochastic monotone submodular function with respect to a matroid constraint $(f, \mathcal{A}, \mathcal{F})$.
\end{theorem}

\subsubsection{Modeling Pandora's box problem with sequential inspection as a submodular optimization problem.} \label{subsec:submodular_problems}
The Pandora's box problem can be essentially viewed as a selection problem. 
Following Lemma \ref{lemma:committing} and the discussion above, we define two associated problems that capture exactly the tradeoff of inspection modes. 

\begin{definition} \label{def:submodular_optimization}
(The associated stochastic submodular optimization problems). Given an instance of a real-world problem $\mathcal{I}$, which has $N$ boxes with costs and distributions $\left\{c_i^F,c_i^P, \mathcal{D}_i \right\}_{i=1}^{N}$, we define two stochastic submodular optimization problems $\mathcal{J}=\left(f, \mathcal{A}, \mathcal{F}\right)$ and $\mathcal{J}'=\left(f, \mathcal{A}', \mathcal{F} \right)$ associated with the free-info problem:
\begin{itemize}
    \item {\em Ground sets}. We define the ground set $\mathcal{A}$ is
        $$
        \mathcal{A} = \{ K_1, \widetilde{K}_1, K_2, \widetilde{K}_2, ..., K_N, \widetilde{K}_N  \},
        $$
        where $K_i$ and $\widetilde{K}_i$ are correlated and induced from $\mathcal{D}_i$. Thus, $\mathcal{I}$ and $\mathcal{J}$ are defined over the same probability space. However, the ground set $\mathcal{A}'$ is
        $$
        \mathcal{A}' = \{ K_1', \widetilde{K}_1', K_2', \widetilde{K}_2', ..., K_N', \widetilde{K}_N'  \},
        $$
        where $K_i'$ and $\widetilde{K}_i'$ are drawn independently from the same distribution as $K_i$ and $\widetilde{K}_i$.
        
    \item { \em Matroid constraint}. The matroid constraint $\mathcal{M}$ is defined by the partition matroid whose independent sets are all the subsets of $\mathcal{\mathcal{A}}$ (resp., $\mathcal{\mathcal{A}}'$) that contain at most one element of each pair $\{K_i, \widetilde{K}_i \}_{i=1}^{N}$ (resp., $\{K_i', \widetilde{K}_i' \}_{i=1}^{N}$).
    
    \item { \em Objective function}. The submodular objective function $f: \mathbb{R}^{2N}_{+} \rightarrow \mathbb{R}_{+} $ is defined as 
    $$
    f(\Theta) = \max\{ \Theta_i: 1 \leq i \leq 2N , \theta_i \neq \text{ NULL}\}. 
    $$
    We denote the expected profit of policy $\pi$ in $\mathcal{J}$ by $J^{\pi, M}(s_0)$ 
    and in $\mathcal{J}'$ by $J^{\pi, M'}(s_0)$. 
\end{itemize}

\end{definition}

\paragraph{Remark.}  Clearly, any feasible policy $\pi$ for problem $\mathcal{J}$ is also feasible for problem $\mathcal{J}'$ (that is, $\pi \in \mathfrak{F}(\mathcal{A}, \mathcal{F}) \Leftrightarrow \pi \in \mathfrak{F}(\mathcal{A}', \mathcal{F})$).
A perhaps simple, yet crucial observation is that for any feasible policy $\pi$, we have $J^{\pi, M}(s_0)=J^{\pi, M'}(s_0)$. 
The equality holds by noting that only one single variable from each pair $\{ K_i, \widetilde{K}_i\}$ or $\{ K_i', \widetilde{K}_i'\}$, is picked by any feasible policy.

\begin{table}[htbp]
  \centering
    \begin{tabular}{cl}
    \toprule
    \toprule
    \multicolumn{2}{c}{Summary of Notations for Theorem \ref{thm:1-1/e}} \\
    \midrule
    $\mathcal{I}$ & A real-world instance \\ 
    $\mathcal{J}$ & The corresponding stochastic submodular problem \\
    $\rho$ & A nonadaptive policy in $\mathcal{J}$ \\
    $\Psi(\rho)$ & A committing policy in $\mathcal{I}$ induced by $\rho$ \\
    $\pi$ & Any policy in $\mathcal{I}$ \\
    $\Phi(\pi)$ & The transformed policy in $\mathcal{J}$ from policy $\pi$ \\
    \bottomrule
    \end{tabular}%
  \label{tab:addlabel}%
\end{table}%

\subsubsection{Transformation of policies.} \label{subsec:submodular_transformation} 

Let $\rho$ be a \textit{nonadaptive} policy in $\mathcal{J}$ and $A(\rho) \subseteq \mathcal{A}$ be the set of random variables that $\rho$ selects. Let $F(\rho)$ denote the (deterministic) set of boxes $\{ i \in [N] \mid K_i \in A(\rho) \}$ and let $P(\rho)$ denote the (deterministic) set of boxes $\{ i \in [N] \mid \widetilde{K_i} \in A(\rho) \} $. Clearly, the expected profit of policy $\rho$ in $\mathcal{J}$ is
\begin{equation}\label{eq:u_III}
  J^{\rho, M}(s_0)  = \mathbb{E} \Big[  \max \big\{  \max_{i \in F(\rho) }\{ K_i \},    \max_{i \in P(\rho)}\{ \widetilde{K}_i  \}  \big\}  \Big].  \nonumber
\end{equation}
Moreover, $\rho$ naturally induces a corresponding committing policy $\Psi(\rho)$ in $\mathcal{I}$ by specifying $F=F(\rho)$ and $P=P(\rho)$. 
We mention in passing that $F \cup P$ might not be equal to $[N]$, but as long as $F \cap P = \emptyset$, Lemma \ref{lemma:committing} still holds. Thus, we have
\begin{equation}
    J^{\Psi(\rho)}(s_0)  =  \mathbb{E} \Big[  \max \big\{  \max_{i \in F(\rho) }\{ K_i \},    \max_{i \in P(\rho)}\{ \widetilde{K}_i  \}  \big\}  \Big].  \nonumber
\end{equation}
Since $\mathcal{I}$ and $\mathcal{J}$ share the same probability space, we have the following lemma:

\begin{lemma} \label{lemma:committing_nonadaptive}
Let $\rho$ be a nonadaptive policy in $\mathcal{J}$, and $\Psi(\rho)$ be the corresponding committing policy in $\mathcal{I}$. We have 
$J^{\Psi(\rho)}(s_0) = J^{\rho, M}(s_0) = J^{\rho, M'}(s_0)$.
\end{lemma}

Conversely, let $\pi$ be any policy in $\mathcal{I}$. 
Since problem $\mathcal{I}$ and problem $\mathcal{J}$ share the same probability space, we transform $\pi$ into policy $\Phi(\pi)$ in $\mathcal{J}$ in the following way: 
(i) if policy $\pi$ directly F-opens box $i$ without a P-opening in $\mathcal{I}$, then policy $\Phi(\pi)$ selects $K_i$ in $\mathcal{J}$;
(ii) if policy $\pi$ P-opens box $i$ in $\mathcal{I}$, then policy $\Phi(\pi)$ selects $\widetilde{K}_i$ in $\mathcal{J}$;
(iii) if if policy stops in $\mathcal{I}$, then policy $\Phi(\pi)$ does not select any other box as well.

\begin{lemma} \label{lemma:adaptivity_gap_bridge}
Let $\pi$ be a policy in $\mathcal{I}$, and $\Phi(\pi)$ be the corresponding transformed policy in $\mathcal{J}$. We have 
$J^{\Phi \left(\pi \right), M'}(s_0) = J^{\Phi \left(\pi \right), M}(s_0) \geq J^{\pi}(s_0)$.
\end{lemma}

\begin{myproof}
By construction, given any policy $\pi$ in $\mathcal{I}$,  the \textit{random} set of random variables selected by the transformed policy $\Phi(\pi)$ in $\mathcal{J}$  is $\{K_i \mid \indFi^{\pi}=1 \} \cup \{ \widetilde{K}_i \mid \indPi^{\pi} = 1 \}$.  
Let $F(\pi) = \{ i \in [N] \mid \indFi^{\pi}=1 \}$ be the random set of boxes that is directly F-opened by policy $\pi$, and similarly, let $P(\pi) = \{ i \in [N] \mid \indPi^{\pi}=1 \}$ be the set of boxes that will be P-opened first by policy $\pi$.
Thus, the expected profit of $\Phi(\pi)$ in $\mathcal{J}$ is 
\begin{equation}
 J^{\Phi \left(\pi \right), M}(s_0) =   \mathbb{E} \Big[  \max \big\{  \max_{i \in F(\pi) }\{ K_i \},    \max_{i \in P(\pi)}\{ \widetilde{K}_i  \}  \big\}  \Big].    \label{eq:lemma_gap_1}
\end{equation}
We emphasize that the expectation is taken over $F(\pi)$ and $P(\pi)$ in addition to $K_i$ and $\widetilde{K}_i $, since the policy can adaptively adjust which inspection mode to apply on boxes based on the observed information.
On the other hand, we note that 
\begin{eqnarray}
J^{\pi}(s_0)  &\leq& \mathbb{E} \Big[ \sum_{i=1}^{N} \indSi^{\pi} K_i(1-\indPi^{\pi} )+\indSitilde^{\pi} \widetilde{K}_i \indPi^{\pi}    \Big]  \nonumber \\ 
   &=& \mathbb{E} \Big[ \sum_{i \in F(\pi) } \indSi^{\pi} K_i + \sum_{i \in P(\pi) } \indSitilde^{\pi} \widetilde{K}_i \Big] \nonumber  \\
   &\leq&  \mathbb{E} \Big[  \max \big\{  \max_{i \in F(\pi) }\{ K_i \},    \max_{i \in P(\pi)}\{ \widetilde{K}_i  \}  \big\}  \Big],   \label{eq:lemma_gap_2}  \\
   &=& J^{\Phi(\pi), M}(s_0)  ,   \label{eq:lemma_gap_3}
\end{eqnarray}
where inequality~\ref{eq:lemma_gap_2} proceeds from the constraint that only one box can be selected. 
Equation~\ref{eq:lemma_gap_3} follows by noting equation~\ref{eq:lemma_gap_1}, and that $\mathcal{I}$ and $\mathcal{J}$ share the same probability space.
\end{myproof}

Armed with the preceding results,  we are ready to establish our main theorem. 

\subsubsection{Proof of Theorem~\ref{thm:1-1/e}.}  \label{subsec:proof_of_thm:1-1/e}

Let $\pi$ be the optimal policy in $\mathcal{I}$. By Lemma \ref{lemma:adaptivity_gap_bridge}, we have
\begin{equation}
 J^{\Phi(\pi), M'}(s_0)  = J^{\Phi(\pi), M}(s_0)  \geq J^{\pi}(s_0) \triangleq  J^{}(s_0) \ . \label{eq:thm:1-1/e_1}  
\end{equation}
By Theorem~\ref{lemma:borrowed_machinery}, there exists a nonadaptive policy $\rho$ in $\mathcal{J}'$ such that
\begin{equation}
  J^{\rho, M}(s_0)  \geq \left( 1- \frac{1}{e} \right)   J^{\Phi(\pi), M'}(s_0) \ .  \label{eq:thm:1-1/e_2}
\end{equation}
By Lemma~\ref{lemma:committing_nonadaptive}, we know that the transformed policy $\Psi(\rho)$ in $\mathcal{I}$ satisfies
\begin{equation}
  J^{\Psi(\rho)}(s_0)  = J^{\rho, M}(s_0) = J^{\rho, M'}(s_0) \ .  \label{eq:thm:1-1/e_3}
\end{equation}
Combining \ref{eq:thm:1-1/e_1}, \ref{eq:thm:1-1/e_2} and \ref{eq:thm:1-1/e_3} together yields
\begin{equation}
  J^{\Psi(\rho)}(s_0) \geq \left( 1- \frac{1}{e} \right) J(s_0)  \ .  \nonumber
\end{equation}

Finally, we remark here that a good committing policy can be found efficiently, by invoking another technical result established in \cite{asadpour2015maximizing}, which we restate as follows. 
\begin{theorem} \label{lemma:borrowed_machinery_2} (\cite{asadpour2015maximizing}, Theorem 2).
For any $\epsilon > 0$ and any instance $(f, \mathcal{A}, \mathcal{F})$, a nonadaptive policy that obtains a $(1 - 1/e - \epsilon)$-fraction of the value of the optimal adaptive policy can be found in polynomial time with respect to $\mid \mathcal{A} \mid$ and $1 / \epsilon$.
\end{theorem}

\subsection{Proof of Theorem~\ref{thm:hardness}}
\label{sec:proof_thm_hardness}

In this section, we show that computing optimal policies for \textit{Pandora box problem with sequential inspection} (PSI) is NP-hard. 
We devise the reduction from a family of NP-hard subset sum problems.

\paragraph{\bf The hard instances.} We devise a reduction from the subset sum problem (SSP). We are given positive integers $a_1,\ldots,a_{n}$ and the goal is to find $W \subseteq [n]$ such that $\sum_{i\in W} a_i = T$. Without loss of generality, we assume that $T \leq \frac{1}{2} \sum_{i=1}^{n} a_i $ as otherwise, we consider the complementary target $\sum_{i=1}^{n} a_i - T$, corresponding to selecting the complement in $[n]$. 
For brevity, we use the notation $L \defeq \sum_{i=1}^{n} a_i $ throughout.  
Moreover, we focus on a family of SSP which we call the $\varepsilon$-SSP, where we have $ \frac{a_i}{\sum_{i=1}^{n} a_i} \leq \varepsilon $ for every $i \in [n]$ for some $\varepsilon \in (0,1)$.  
This family is still NP-hard. 
To see this, suppose there exists a polynomial-time algorithm that solves the $\varepsilon_0$-SSP for some fixed $\varepsilon_0$. 
Now given a general SSP instance $\{ \braces{a_i}_{i=1}^{n}, T \}$, we can transform it to a $\varepsilon_0$-SSP instance by adding ``fake items" $a_0 > L$. 
Since $a_0 > L$, we know for sure we will not include them in a feasible solution. 
In addition, we can ensure for every $i \in [n]$, we have $ \frac{a_i}{ N \cdot a_0 + \sum_{i=1}^{n} a_i} < \frac{a_0}{ N \cdot a_0 + \sum_{i=1}^{n} a_i } < \varepsilon_0$ for large enough $N = O(\frac{1}{\varepsilon_0})$ (independent of $n$). 
The following lemma ensues.

\begin{lemma}
    The $\varepsilon$-subset sum problem is NP-hard for any fixed $\varepsilon \in (0,1)$.
\end{lemma}

Our reduction makes use of the following class of PSI instances.

\begin{definition}[The 2T2L-PSI Instance]
    \label{def:2T2L-PSI_instance}
    Given positive integers $\braces{a_i}_{i=1}^{n}$, an instance of PSI with $n$ boxes is a two-type two-support low-cost-low-return (2T2L) instance  with parameter $(\nu, \alpha)$ where $\nu \in (0,1), \alpha > 0$ if the following conditions hold: 
    \begin{enumerate}
        \item all rewards are binary random variables $V_i \in \{ 0, 1 \}$. Specifically, for box $i$, we set its success probability as 
        $$
    p_i \defeq \pr{V_i = 1} = 1-  \exp\left(- \alpha a_i \right) \ , 
        $$
        for some constant $\alpha>0$. We define $p_i'$ as 
        $$
    p_i ' \defeq  1-  \exp\left(- \frac{1}{2} \alpha a_i \right) \ ; 
        $$
        \item each box $i$ has 2 types: good and bad, $\Gamma_i = \{ G, B\}$, and moreover $\pr{T_i=G}=(1-\nu) p_i $, $\pr{T_i=B}= 1 - (1-\nu) p_i  $, $\pr{V_i=1 \mid T_i=G} = \frac{p_i'}{p_i (1-\nu)} + p_i - p_i' $, $ \pr{V_i=1 \mid T_i=B} = p_i - p_i' $; 
        \item box $i$ has F-opening cost $c_i^F = \frac{3}{5} p_i$ and P-opening cost  $c_i^P = (1-\nu)  p_i \paren{ \frac{1}{3} ( \frac{p_i'}{ p_i (1-\nu) } + p_i - p_i') - \frac{3}{5} p_i } $. 
    \end{enumerate}
\end{definition}

\paragraph{\bf The reduction.}
Given an instance of the $\varepsilon$-Subset Sum Problem ($\varepsilon$-SSP) $\{\{a_i\}_{i=1}^n, T\}$, we construct an associated 2T2L-PSI instance by setting the parameters $\alpha = \frac{\ln(\frac{3}{2})}{L-T}$ and $\nu = \frac{1}{2}$.

We will show that there exists a universal constant $c \in (0,1)$ such that, whenever $\varepsilon < c$, there exists an optimal policy of the associated PSI instance that admits the form of a committing policy (see Section~\ref{subsec:apx_policies} for a formal definition and discussion). 
Specifically, in this associated PSI instance, the good type and the bad type are highly distinguishable. Learning the type of a box directly informs the decision to either F-open it or discard it. If a box is discarded, it has no further impact on any future decisions.
Additionally, as the rewards are binary, it is optimal to stop and select a box immediately upon finding one with a reward of 1. Consequently, there exists an optimal policy for this instance that is a committing policy.
If a PSI instance admits the above-mentioned properties, we call it a \textit{well-separated} instance. 
Suppose that there is an oracle which returns an optimal policy for such a well-separated instance. 
Then one can read off a partition of boxes, of which the policy is committed to the opening mode.

Moreover, the parameter $\alpha = \frac{\ln(\frac{3}{2})}{L-T}$ of the associated 2T2L-PSI instance is carefully set such that the optimal partition $(F^*, P^*)$ is exactly achieved at the condition $\sum_{i \in P^*} a_i = T$, if such a subset $P^*$ exists.

One caveat, however, is that in the associated PSI instance, the inputs $p_i$ and $p_i'$ are irrational, potentially requiring exponential bit complexity.
We note that for the purpose of a polynomial-sized reduction,\footnote{The reduction must run in polynomial time in $n \log(\max_{i \in [n]} a_i)$, which is the input size of the $\varepsilon$-SSP instance.} it might be infeasible to feed the \emph{true} associated low-cost-low-return PSI instance to the Pandora oracle. Instead, we approximate the inputs and feed the Pandora oracle an \emph{approximated version}, using $\hat{p}_i$ and $\hat{p}_i'$ as substitutes for $p_i$ and $p_i'$, respectively.

We use 
$$
\delta \defeq \max_{i \in [n]} ~\max \braces{ \frac{\abs{ p_i - \hat{p}_i }}{p_i} ,  \frac{\abs{ p_i' - \hat{p}_i' }}{p_i'} }  
$$ 
to measure the encoding precision. 
In the following lemma, we make it precise the conditions needed for the \textit{true} associated 2T2L-PSI instance to be well-separated.

Moreover, we characterize the degree of precision needed in the approximated instance to ensure the same optimal policy structure, in terms of a requirement on $\delta$.

\begin{lemma}  \label{lemma:hard_psi_is_committing}
    Consider a 2T2L-PSI instance, with parameter $(\nu, \alpha)$ set to $\alpha=\frac{\ln(\frac{3}{2})}{ L - T}$ where $ 0 < T \leq \frac{1}{2} L$.
    There exists a universal constant $c \in (0,1)$ such that if $\frac{a_i}{L} \leq c $ for every $i$, then an optimal policy F-opens a good box immediately upon observation and discards a bad box (never opens it afterwards) upon discovery. 
    Consequently, there exists an optimal policy which is a committing policy. 
    
    Moreover, $\delta = \Theta(\frac{1}{L})$ ensures that the approximated PSI instance admits an optimal committing policy as well. 

    In addition, an optimal policy must be a committing policy in both cases. 
\end{lemma}

\begin{myproof}

\textbf{First half: the true instance. }

We will show that by construction, there exists a universal constant $c \in (0,1)$ such that if $\frac{a_i}{L} \leq c $ for every $i \in [n]$, then
(1) $\sigma_{i}^{F} = \frac{2}{5} $, 
(2) $\sigma_{i}^{F|G} > \frac{2}{3} $, 
(3) $\sigma_{i}^{F|B} < 0$,
(4) $\sigma_{i}^{P} = \frac{2}{3}$,
(5) $\sigma_{i}^{F/P} > 0$. 
Therefore, by virtue of Theorem~\ref{thm:just_strong}, an optimal policy shall immediately F-open a good box upon finding it. 
By Theorem~\ref{thm:myopic_stopping}, an optimal policy shall discard any bad boxes.  Not only will one never open such a bad box, but also it does not affect any of the subsequent decisions.
Consequently, there exists an optimal policy that is a committing policy.  
It is worth noting that our constructed instance does \textit{not} satisfy the conditions of Corollary~\ref{cor:support_2}, where we have identified the optimal committing policy.

In what follows, we compute each of the above-mentioned thresholds. 
Observe that in order to ensure that $p_i < \theta$ for some $\theta \in (0,1)$, it suffices to require 
\begin{equation} \label{eq:p_i_to_a_L_ratio}
    \frac{a_i}{L} < \frac{ \ln(\frac{3}{2})}{2  \ln(\frac{1}{1-\theta})} \ .  
\end{equation}

\begin{enumerate}
    \item \textbf{The F-threshold $\sigma_{i}^{F} = \frac{2}{5} $.} 
        We recall that $c_i^F = \frac{3}{5} p_i$, $\pr{V_i=1}=p_i$ and
        $$
            \E{ \paren{V_i - \sigma_i^F}^+ } = \pr{V_i=1} \paren{1 - \sigma_i^F}  \ . 
        $$
        Hence, $\sigma_i^F = \frac{2}{5} $ exactly. 
    \item \textbf{The conditional F-threshold $\sigma_{i}^{F|G} > \frac{2}{3} $ .}         
        To see this, we note that by definition 
        $$
        c_i^F = \E{ \paren{V_i - \sigma_i^{F|G} }^+ \mid T_i = G } = \pr{V_i=1 \mid T_i=G} \paren{1 - \sigma_i^{F|G} } \ . 
        $$ 
        Plugging $\pr{V_i=1 \mid T_i=G} = \frac{p_i'}{p_i (1-\nu)} + p_i - p_i'$ into the above equation, we have that 
        \begin{equation} \label{eq:sigma_F_G_true_hard_instance}
            \sigma_i^{F|G} = 1 - \frac{3}{5} \frac{p_i^2}{ \frac{1}{1-\nu} p_i' + p_i (p_i - p_i') } \ . 
        \end{equation}
        In order for $\sigma_i^{F|G} > \frac{2}{3}$, it suffices to set $\frac{p_i^2}{ \frac{1}{1-\nu} p_i' + p_i (p_i - p_i') } < \frac{5}{9}$. 
        By Jensen's inequality, we know 
        $
        \exp \paren{- \frac{1}{2} \alpha a_i  } < \frac{1}{2} \paren{ \exp \paren{- \alpha a_i } + 1  } 
        $
        which implies that 
        \begin{equation} \label{eq:consequence_of_jensen_p_p_prime}
            p_i >  p_i' > \frac{1}{2} p_i \ .     
        \end{equation}
        Hence, $\frac{p_i^2}{ \frac{1}{1-\nu} p_i' + p_i (p_i - p_i') } 
        < \frac{p_i^2}{ \frac{1}{1-\nu} \frac{1}{2} p_i + p_i (p_i - p_i') } 
        < 2 p_i $. It suffices to impose $2p_i < \frac{5}{9}$, for which to happen we can require 
        \begin{equation}
            \frac{a_i}{L} \leq \frac{1}{2} \frac{\ln(18/13)}{\ln(3/2)} = O(1) \ . 
        \end{equation}

    \item \textbf{The conditional F-threshold $\sigma_{i}^{F|B} < 0$ .}
        By straightforward calculation, we see that $
        \sigma_i^{F|B} = 1 - \frac{c_i^F}{\pr{V_i=1 \mid T_i = B}} = 1 - \frac{\frac{3}{5}p_i}{p_i - p_i'} < 1 - \frac{\frac{3}{5}p_i}{p_i - \frac{1}{2}p_i} = - \frac{1}{5}  
        $.

    \item \textbf{The P-threshold $\sigma_{i}^{P} = \frac{2}{3}$.} 
        Recall the definition
        \begin{small}
        $$
        c_i^P = \pr{T_i=G} \paren{ \underbrace{  -c_i^F + \pr{V_i=1 \mid T_i = G} \paren{ 1 - \sigma_i^{P} } }_{(a)} }^+ 
        +  \pr{T_i=B} \paren{ \underbrace{ -c_i^F + \pr{V_i=1 \mid T_i = B} \paren{ 1 - \sigma_i^{P} } }_{(b)} }^+ \ . 
        $$
        \end{small}
        If $(a) >0$ and $(b)<0$, it is easy to verify that $ \sigma_i^{P} = \frac{2}{3} $ is indeed implied from the above equation. 
        In what follows, we show that term $(a) >0$ and $(b)<0$ when setting  $ \sigma_i^{P} = \frac{2}{3} $.
        To see this, for $(b)$, we observe that 
        $
        -c_i^F + \pr{V_i=1 \mid T_i = B} \paren{ 1 - \frac{2}{3}} = - \frac{3}{5} p_i + \frac{1}{3} (p_i - p_i') < 0 \ . 
        $
        Next, for term (a): we have 
        \begin{small}
        $$
        -c_i^F + \pr{V_i=1 \mid T_i = G} \paren{ 1 - \frac{2}{3}} 
        = - \frac{3}{5} p_i +  \frac{1}{3} \paren{ \frac{p_i'}{p_i (1-\nu)} + p_i - p_i' }
        = \frac{1}{3} \frac{p_i'}{p_i (1-\nu)} - \frac{4}{15} p_i' - \frac{1}{3} p_i 
        > \frac{1}{3} \frac{1}{2}  - \frac{4}{15} p_i' - \frac{1}{3} p_i \ , 
        $$ 
        \end{small}
        as $p_i' > \frac{1}{2} p_i$ and $ \frac{1}{1-\nu} > 1$. 
        For the right-hand-side to be positive, it suffices to have 
        \begin{equation}
            \frac{4}{15} p_i + \frac{1}{3} p_i' < \frac{4}{15} p_i + \frac{1}{3} p_i < \frac{1}{6}
        \end{equation}
        which translates to $\frac{a_i}{L} = O(1)$, by Equation~\eqref{eq:p_i_to_a_L_ratio}. 
    
    \item \textbf{The F/P threshold $\sigma_{i}^{F/P} > 0$.}
        To prove $\sigma_i^{F/P} > 0$, in view of Theorem~\ref{thm:myopic_stopping}, it suffices to show that the expected payoff of F-opening is greater than the expected payoff of P-opening in the single-box case. 
        The difference of the expected payoff of P-opening and the expected payoff of F-opening is
        \begin{eqnarray}
          && -c_i^P + \pr{T_i = G} \paren{-c_i^F + \pr{V_i=1 \mid T_i=G} \cdot 1 } - \paren{ - c_i^F + \pr{V_i=1} \cdot 1} \nonumber \\ 
            &=& \frac{2}{3} p_i' +  \frac{2}{3}  (1-\nu) p_i (p_i - p_i')  - \frac{2}{5} p_i \label{eq:instance_condition_} \\
            &=& \frac{2}{15} e^{-2  \alpha a_i   } \paren{e^{\frac{1}{2} \alpha a_i   } - 1 } 
            \paren{ \underbrace{ 2 e^{\frac{3}{2}  \alpha a_i  } + e^{ \alpha a_i  } (2-5\nu) - 5(1-\nu) }_{ (*) } } \nonumber  \ . 
        \end{eqnarray}
        Hence, the sign is determined by the $(*)$ term in the last parenthesis. 
        By using $e^x \leq 1+x+x^2$ for $\forall~x\leq 1.79$, we know that whenever
        \begin{equation}  \label{eq:instance_condition_3}
            \frac{a_i}{L} < \frac{1.79}{3} \frac{1}{\ln(\frac{3}{2})} 
        \end{equation}
        for every $i$, we have 
        $
        (*) = 2 e^{\frac{3}{2}  \alpha a_i  } + e^{ \alpha a_i  } (2-5\nu) - 5(1-\nu)
        \leq -1 + 5 (1-\nu) \alpha a_i + ( \frac{13}{2} - 5 \nu ) (\alpha a_i)^2
        $. 
        For the latter to be less than $0$, it suffices to impose 
        \begin{equation} \label{eq:instance_condition_4}
            \alpha a_i < \frac{1}{2} \min \braces{ \frac{1 }{5 (1-\nu)} ,\sqrt{\frac{1}{\frac{13}{2} - 5 \nu }}}  \ ,
        \end{equation}
        which again translates to $\frac{a_i}{L} = O(1)$ for every $i$. 
        
\end{enumerate}

\textbf{Second half: the approximated instance. }
In the approximated instance, we use the notation $\hat{\cdot}$ to denote the approximated version of the threshold, obtained by replacing $p_i, p_i'$ with $\hat{p}_i, \hat{p}_i'$ in the calculation. 
We proceed to show that by construction, for the same constant $c \in (0,1)$ prescribed in the first half of the proof, as long as $\delta = \Theta(\frac{1}{L})$, we have 
(1) $\hat{\sigma}_{i}^{F} = \frac{2}{5} $, 
(2) $\hat{\sigma}_{i}^{F|G} > \frac{2}{3} $, 
(3) $\hat{\sigma}_{i}^{F|B} < 0$,
(4) $\hat{\sigma}_{i}^{P} = \frac{2}{3}$,
(5) $\hat{\sigma}_{i}^{F/P} > 0$. 
As a result, there exists an optimal policy that is a committing policy as well in the approximated instance. 

Before calculating the thresholds exhaustively, we make some algebraic observations. 
We recall $\delta \defeq \max_{i \in [n]} ~\max \braces{ \frac{\abs{ p_i - \hat{p}_i }}{p_i} ,  \frac{\abs{ p_i' - \hat{p}_i' }}{p_i'} }  $.
A direct consequence is that 
\begin{equation}
    \abs{p_i - \hat{p}_i} \leq p_i \delta \leq \delta \ . 
\end{equation}
In view of Equation~\ref{eq:consequence_of_jensen_p_p_prime}, we have 
\begin{equation} \label{eq:lower_bound_p_hat_prime_to_p_prime}
    \frac{\hat{p}_i'}{\hat{p}_i} > \frac{(1-\delta) p_i' }{ (1+\delta) p_i } > \frac{1}{2} \frac{1-\delta}{1+\delta}
\end{equation}
as well.

Moreover, we have  $ \hat{p}_i > \hat{p}_i'$ as long as $\delta = \Theta (\frac{1}{L})$.
To see this, we observe $ \hat{p}_i -  \hat{p}_i' \geq p_i - p_i' - 2 \delta $, and by the Mean Value Theorem, we have 
$
p_i - p_i' = \exp \paren{ - \frac{1}{2} \alpha a_i } - \exp \paren{ - \alpha a_i } = \exp \paren{ \xi} \frac{1}{2} \alpha a_i 
$
for some $\xi$  between $- \frac{1}{2} \alpha a_i$ and $-  \alpha a_i$.
Hence, $\exp \paren{ \xi } \frac{1}{2} \alpha a_i \geq  
\exp \paren{ - \alpha a_i } \frac{1}{2} \alpha a_i 
= \exp \paren{ - \ln(\frac{3}{2}) \frac{a_i}{L-T} } \frac{1}{2} \ln(\frac{3}{2}) \frac{a_i}{L-T}
\geq \exp \paren{ - 2 \ln(\frac{3}{2})  } \frac{1}{2} \ln(\frac{3}{2}) \frac{1}{L}
$, as $\frac{a_i}{L-T} \leq \frac{2a_i}{L} \leq 2$. 
Therefore, in order for 
$$
p_i - p_i' = \exp \paren{ \xi} \frac{1}{2} \alpha a_i  
\geq \exp \paren{ - 2 \ln\frac{3}{2}  } \frac{1}{2} \ln \frac{3}{2} \frac{1}{L} 
> 2 \delta \ , 
$$
it suffices to require 
$$
\delta = \exp \paren{ - 2 \ln\frac{3}{2}  } \frac{1}{4} \ln(\frac{3}{2}) \frac{1}{L} = \Theta \left( \frac{1}{L} \right)  \ .
$$

Now, we turn to the computation of thresholds. 
\begin{enumerate}
    \item \textbf{The F-threshold $\hat{\sigma}_{i}^{F} = \frac{2}{5} $.} 
    Recall the definition $ c_i^F = \E{ \paren{V_i - \sigma_i^F}^+ } = \pr{V_i=1} \paren{1 - \sigma_i^F} $. In the approximated version, $\hat{\sigma}_{i}^{F}$ is the solution to $ \frac{3}{5} \hat{p}_i = \hat{p}_i \cdot \paren{1 - \hat{\sigma}_{i}^{F}}  $, which can be calculated exactly. 
    \item \textbf{The F-threshold $\hat{\sigma}_{i}^{F|G} > \frac{2}{3} $.}  
    Following the same reasoning as Equation~\ref{eq:sigma_F_G_true_hard_instance}, we have 
    $$
        \hat{\sigma}_{i}^{F} = 1 - \frac{3}{5} \frac{\hat{p}_i^2}{ \frac{1}{1-\nu} \hat{p}_i' + \hat{p}_i ( \hat{p}_i - \hat{p}_i') } \ . 
    $$
      Since we ensure $ \hat{p}_i > \hat{p}_i'$ and Inequality~\ref{eq:lower_bound_p_hat_prime_to_p_prime}, we can upper bound
      $$
\frac{\hat{p}_i^2}{ \frac{1}{1-\nu} \hat{p}_i' + \hat{p}_i ( \hat{p}_i - \hat{p}_i') }
< \frac{\hat{p}_i^2}{ \frac{1}{1-\nu} \hat{p}_i' }
< (1-\nu) 2 \frac{1+\delta}{1-\delta} \hat{p}_i
\leq 2 \frac{1+\delta}{1-\delta} (p_i + \delta) \ . 
      $$
      Moreover, $ \frac{1+\delta}{1-\delta} (p_i + \delta) < 2  (p_i + \delta)$ if $\delta < \frac{1}{3}$. 
      Hence, it suffices to impose $p_i < \frac{5}{36} - 0.0001$ and $\delta < 0.0001$ to make sure that $\hat{\sigma}_{i}^{F|G} > \frac{2}{3} $. 

    \item \textbf{The F-threshold $\hat{\sigma}_{i}^{F|B} < 0 $.}   
            For the approximated threshold  $
        \hat{\sigma}_i^{F|B} =  1 - \frac{\frac{3}{5} \hat{p}_i}{\hat{p}_i - \hat{p}_i'} 
        < 1 - \frac{\frac{3}{5} \hat{p}_i}{ \hat{p}_i - \frac{1}{2} \frac{1-\delta}{1+\delta} \hat{p}_i} = 1 - \frac{3}{5} \frac{2+2 \delta}{1+3\delta}
        $. It is easy to verify that $\hat{\sigma}_i^{F|B} < 0$ as long as $\delta < \frac{1}{9}$. 
    \item \textbf{The P-threshold $\hat{\sigma}_{i}^{P} = \frac{2}{3} $.} 
    Recall the definition
        \begin{small}
        $$
        c_i^P = \pr{T_i=G} \paren{ \underbrace{  -c_i^F + \pr{V_i=1 \mid T_i = G} \paren{ 1 - \sigma_i^{P} } }_{(a)} }^+ 
        +  \pr{T_i=B} \paren{ \underbrace{ -c_i^F + \pr{V_i=1 \mid T_i = B} \paren{ 1 - \sigma_i^{P} } }_{(b)} }^+ \ . 
        $$
        \end{small}
        In the approximated instance, when setting $\sigma_i^P = \frac{2}{3}$, it is easy to see $(b) = - \frac{3}{5} \hat{p}_i + \frac{1}{3} (\hat{p}_i - \hat{p}_i') < 0$ as long as $\delta = O(1)$. 
        Likewise, for term (a), we observe that  
    $
\frac{1}{3} \frac{\hat{p}_i'}{\hat{p}_i (1-\nu)} - \frac{4}{15} \hat{p}_i' - \frac{1}{3} \hat{p}_i 
> \frac{1}{3} \frac{1}{2} \frac{1-\delta}{1+\delta} \frac{1}{1-\nu} - \frac{4}{15} \hat{p}_i' - \frac{1}{3} \hat{p}_i
> \frac{1}{3} \frac{1}{2} \frac{1-\delta}{1+\delta}  - \frac{4}{15} (p_i'+\delta) - \frac{1}{3} ( p_i+\delta) . 
    $
    One can verify that the latter is strictly positive as long as $\delta = O(1)$.
    \item \textbf{The F/P-threshold $\hat{\sigma}_{i}^{F/P} > 0 $.} 
    In the approximated version, the difference of the expected payoff of P-opening and the expected payoff of F-opening is
\begin{eqnarray}
  && -c_i^P + \pr{T_i = G} \paren{-c_i^F + \pr{V_i=1 \mid T_i=G} \cdot 1 } - \paren{ - c_i^F + \pr{V_i=1} \cdot 1} \nonumber \\ 
    &=& \frac{2}{3} \hat{p}_i' + \frac{2}{3}  (1-\nu) \hat{p}_i (\hat{p}_i - \hat{p}_i')  - \frac{2}{5} \hat{p}_i \nonumber \label{eq:instance_condition_}  \\ 
    &\leq& \frac{2}{3} (1+\delta) p_i' +  \frac{2}{3}  (1-\nu)  (1+\delta)  p_i ( (1+\delta) p_i - (1-\delta) p_i')  - \frac{2}{5} (1-\delta) p_i \nonumber \\ 
    &\defeq& E(\delta, \nu, \alpha)  \ . 
\end{eqnarray}

Recall the fact that $ \frac{x}{1+x} < 1 - e^{-x} < \frac{x+x^2}{1+x+x^2}$ for $-1 < x < 1.79$. 
Moreover, $\frac{x+x^2}{1+x+x^2} = x - x^3 + O(x^4)$ and $\frac{x}{1+x} = x + O(x^2) $.  
We hence have 
$$
  \frac{\alpha a_i}{1+\alpha a_i}   < p_i <   \frac{\alpha a_i + \alpha^2 a_i^2}{1 + \alpha a_i + \alpha^2 a_i^2}
\text{ and }  
\frac{ \frac{1}{2} \alpha a_i}{1+ \frac{1}{2} \alpha a_i}  
< p_i' < 
\frac{ \frac{1}{2} \alpha a_i + \frac{1}{4} \alpha^2 a_i^2}{1 + \alpha a_i + \frac{1}{4} \alpha^2 a_i^2} 
$$
as long as $ \alpha a_i < 1.79$, which translates to the condition $ \frac{a_i}{L} < \frac{1.79}{ \ln(3/2) } = O(1) $.
Therefore, 
\begin{eqnarray*}
    & & E(\delta, \nu, \alpha)  \\ 
    &<& \frac{2}{3} (1+\delta) p_i' +  \frac{2}{3}  (1-\nu)  (1+\delta)  p_i ( (1+\delta) p_i - (1-\delta) p_i')  - \frac{2}{5} (1-\delta) p_i  \\ 
    &<& \frac{2}{3} (1+\delta) \frac{ \frac{1}{2} \alpha a_i + \frac{1}{4} \alpha^2 a_i^2}{1 + \alpha a_i + \frac{1}{4} \alpha^2 a_i^2} \\
    & & +  \frac{2}{3}  (1-\nu)  (1+\delta)  \frac{\alpha a_i + \alpha^2 a_i^2}{1 + \alpha a_i + \alpha^2 a_i^2}
    \paren{ (1+\delta)  \frac{\alpha a_i + \alpha^2 a_i^2}{1 + \alpha a_i + \alpha^2 a_i^2} - (1-\delta) \frac{ \frac{1}{2} \alpha a_i}{1+ \frac{1}{2} \alpha a_i}   }  \\ 
    & & - \frac{2}{5} (1-\delta) \frac{\alpha a_i}{1+\alpha a_i}  \\ 
    &=& \left( - \frac{1}{15} + \frac{11}{15} \delta \right) \alpha a_i + O( \alpha^2 a_i^2) \ . 
\end{eqnarray*}
In order for the latter to be negative, it suffices to impose $\delta < \frac{1}{11}$ and $\alpha a_i = O(1)$.

\textbf{An optimal policy must be a committing policy.}
We proceed by induction on the number of closed boxes.
\begin{itemize}
    \item Base case: For $n=1$, the claim is immediate.
    \item Assume the statement holds for all well-separated instances of \(n \ge 1\) closed boxes. 
Consider an instance with \(n+1\) closed boxes and let \(\pi^*\) be an optimal policy. 
\(\pi^*\) must either F-open or P-open a closed box:
    \begin{itemize}
      \item \emph{F-open}: If \(\pi^*\) finds a reward, it stops immediately. 
      Otherwise, we transition to a state with \(n\) boxes, where by the inductive hypothesis, a committing policy is optimal.
      \item \emph{P-open}: If the box is good, it would be suboptimal not to F-open it immediately. 
      If it is bad, the box is discarded and does not affect subsequent decisions.
      Either way, we again reach a state with \(n\) boxes, which by the inductive hypothesis admits a committing policy.
    \end{itemize}
In both cases, the policy transitions to a scenario where a committing policy is optimal, so \(\pi^*\) itself must be committing.
\end{itemize}

\end{enumerate}

\end{myproof}

The next Lemma shows the parameter $\alpha$ is carefully set so that we can use a Pandora oracle to solve a $\varepsilon$-SSP instance.

\begin{lemma}
There exists a universal constant $c \in (0,1)$, such that given any instance of $\varepsilon$-SSP where $\varepsilon < c$, the answer to it is YES if and only if the optimal policy for the associated well-separated PSI P-opens a subset $W^*$ that satisfies $T = \sum_{i \in W^*} a_i$.
\end{lemma}
\begin{myproof}
    The ``if" part is obvious, as such a subset $W^*$ is an solution to the SSP. 

    For the ``only if" part, we first note that by construction of the PSI instance and in light of Lemma~\ref{lemma:hard_psi_is_committing},  finding an optimal policy for this PSI instance reduces to identifying an optimal committing policy. 
    Recall $K_i =  \min \braces{V_i, \sigma^F } $, $\tilde{K}_i = \min \braces{ V_i, \sigma^P, \sigma_{i}^{ F | T_i}  }$. 
    For a committing policy $(F,P)$, we define  $ x(P) \defeq \frac{ \sum_{j \in P}^{} a_j }{\sum_{i=1}^{n} a_i} $. 
    Recall $\alpha = \frac{\ln(\frac{3}{2})}{L-T}$. 
    Lemma~\ref{lemma:committing} shows that the expected payoff of a committing policy $(F,P)$ is exactly 
\begin{eqnarray}
 \Psi( x(P)   ) &=&   \E{ 
    \max \braces{ 
    \max_{i \in F} \braces{K_i} ,  
    \max_{j \in P } \braces{ \tilde{K}_j }
    } } \nonumber \\ 
    &=& \sigma^P \cdot 
    \pr{  \max_{j \in P } \braces{ \tilde{K}_j } = \sigma^P  } \nonumber   \\
    && + \sigma^F \cdot 
    \pr{ \max_{i \in F} \braces{K_i} = \sigma^F } 
    \paren{ 1 -  \pr{  \max_{j \in P } \braces{ \tilde{K}_j } = \sigma^P }  }  \nonumber  \\ 
    && +  0 \cdot 
    \pr{ \max_{i \in F} \braces{K_i} = 0 } 
    \pr{  \max_{j \in P } \braces{ \tilde{K}_j } \leq 0}  \nonumber  \\
    &=& \sigma^P  \cdot 
    \paren{ 1 - \prod_{j \in P} \paren{ 1 - p_j' } } 
    + \sigma^F \cdot 
    \paren{ 1 -  \prod_{i \in F } \paren{1-p_i}   }  
    \paren{ \prod_{j \in P} \paren{1- p_j'  }  }  \\
    &=&  \frac{2}{3} \paren{ 1 - \exp \paren{ - \alpha \sum_{ i \in P} \frac{1}{2} a_i  } } 
    + \frac{2}{5}  \paren{ 1 - \exp \paren{  - \alpha \sum_{i \in F} a_i  }  } 
    \exp \paren{ - \alpha \sum_{ i \in P} \frac{1}{2} a_i  }  \nonumber  \\ 
    &=&  \frac{2}{3} \paren{ 1 - \exp \paren{ - \frac{1}{2} \frac{ \ln(\frac{3}{2}) }{ 1 - \frac{T}{\sum_{i=1}^{n} a_i}  } x } } 
    + \frac{2}{5}  \paren{ 1 - \exp \paren{  - \frac{ \ln(\frac{3}{2}) }{ 1 - \frac{T}{\sum_{i=1}^{n} a_i}  } (1-x)  }  } 
    \exp \paren{ - \frac{1}{2}  \frac{ \ln(\frac{3}{2}) }{ 1 - \frac{T}{\sum_{i=1}^{n} a_i}  } x  }  \nonumber \ .  
\end{eqnarray}
When viewing $\Psi$ as a function defined on $[0,1]$, its derivative is 
$$
\frac{d \Psi}{d x}(x)  = \frac{2}{15} \left(5-e^{-\frac{L x \ln (3)}{2 L-2 T}} \left(3 e^{\frac{L (x-1) \ln (3)}{L-T}}+2\right)\right)  \ . 
$$
It is easy to verify that this function is unimodal on $[0,1]$, and hence the first-order-condition  $\frac{d \Psi}{d x} = 0$ leads to the optimal solution $x^* = \frac{T}{L}$. 

Suppose the answer to the SSP instance is YES. 
Namely, there exists a partition $W^*$ such that $\sum_{i \in W^*} a_i = T$. 
We can use the PSI oracle to solve the associated PSI instance. 
As there is an optimal policy that is a committing policy, we can read off the partition $(F^*,P^*)$. From the discussion above, we must have $\sum_{i \in P^*} a_i = T = \sum_{i \in W^*} a_i $, as desired. 
\end{myproof}

\paragraph{\bf Polynomial-sized encoding.} 
At this point, the remaining challenge is to prove that the computational effort required to compute the approximated associated PSI instance, such that it retains the same optimal committing policy as the true associated instance, is still polynomial in $n \log( \max_{i \in [n]} a_i )$, which is number of bits required to encode the  original SSP instance.

Given $P \subseteq [n]$, the true expected payoff of a committing policy $(F,P)$ is 
\begin{small}    
$$ 
\Psi(x(P)) = \E{ 
    \max \braces{ 
    \max_{i \in F} \braces{K_i} ,  
    \max_{j \in P } \braces{ \tilde{K}_j }
    } } 
    = \sigma^P  \cdot 
    \paren{ 1 - \prod_{j \in P} \paren{ 1 - p_j' } } 
    + \sigma^F \cdot 
    \paren{ 1 -  \prod_{i \in F } \paren{1-p_i}   }  
    \paren{ \prod_{j \in P} \paren{1- p_j'  }  } \ . 
$$
\end{small}

In the reduction, instead of computing exactly $p_i, p_i'$'s, we can only afford to compute the approximated inputs $\hat{p}_i, \hat{p}_i'$'s and feed them to the PSI oracle. Hence, the approximated expected payoff function is 
\begin{equation}
    \widehat{\Psi} ( P ) = \sigma^P  \cdot 
    \paren{ 1 - \prod_{j \in P} \paren{ 1 - \hat{p}_j' } } 
    + \sigma^F \cdot 
    \paren{ 1 -  \prod_{i \in F } \paren{1- \hat{p}_i}   }  
    \paren{ \prod_{j \in P} \paren{1- \hat{p}_j'  }  }  \ . 
\end{equation}
Hence, we are able to characterize the approximation error $\abs{ \Psi(x(P)) - \widehat{\Psi}(P) }$ in terms of the encoding precision $\delta$. 
\begin{claim}  \label{lemma:obj_truncation_error}
    For all $P \subseteq [n]$, $ \abs{ \Psi(x(P)) - \widehat{\Psi}(P) } \leq (\sigma^P + \sigma^F) \cdot \paren{ n \delta }$.   
\end{claim}

\begin{myproof}
    
Given fixed set $P \in 2^{[n]}$, consider the following function $f: [0,1]^n \rightarrow \mathbb{R}$
$$
f( \bfx ) = \sigma^P \cdot \paren{ 1 - \prod_{j \in P}  (1-x_j) } + \sigma^F \cdot \paren{  1 - \prod_{j \in [n] \setminus P } (1-x_j)  } \paren{  \prod_{j \in P} (1-x_j) } \ , 
$$
where $\bfx$ denotes a $n$-dimensional vector and $x_j$ is the $j$-th element of it. 
Its partial derivative with respect the $i$th element is 
\begin{eqnarray*}
    \frac{\partial f}{ \partial x_i } &=& \one{ i \in P  }  \brackets{
    \sigma^P \cdot \prod_{ j \in P \setminus \{ i \} } (1-x_j) 
    + (- \sigma^F ) \cdot \paren{ 1 - \prod_{j \in [n] \setminus P} (1-x_j) } \paren{ \prod_{ j \in P \setminus  \{ i \} } (1-x_j) } 
    }  \\ 
    && + \one{ i \in [n] \setminus P } \brackets{ \sigma^F \cdot \paren{ \prod_{ j \in P} (1-x_j) } \paren{ \prod_{ j \in [n] \setminus ( P \cup \{i\} )    }  (1-x_j) }   }    \ . 
\end{eqnarray*}
It is easy to see that $\abs{ \frac{\partial f}{ \partial x_i }(\bfx) } \leq \sigma^P + \sigma^F $ for $ \bfx \in [0,1]^n$. 
When we calculate $\hat{p}_i, \hat{p}_i$'s, we can easily keep them in the range of $[0,1]$. 
Therefore, applying Mean Value Theorem yields that 
$$
\abs{ \Psi(x(P)) - \widehat{\Psi}(P) } \leq 
(\sigma^P + \sigma^F) \cdot \abs{ \sum_{ j \in P  } ( p_j' - \hat{p}_j' )   +  \sum_{ j \in [n] \setminus P } ( p_j - \hat{p}_j )  } 
\leq (\sigma^P + \sigma^F) \paren{n \delta }  \ . 
$$
We note that the above argument works for any fixed $P$. The proof is hence complete. 

\end{myproof}

Moreover, the function $\Psi$ has enough curvature at the point $\frac{T}{L}$. 
\begin{claim} \label{lemma:curvature}
For any $0 \leq T \leq \frac{L}{2}$, where $L \geq 2$, 
    $
     \Psi(\frac{T}{L}) - \Psi( \frac{T}{L}  + \frac{1}{L} )  \geq  (\frac{\ln(3)}{2})^2 3^{-3} \frac{1}{L^2}
    $
    and 
    $
    \Psi(\frac{T}{L}) - \Psi( \frac{T}{L} - \frac{1}{L} )  \geq (\frac{\ln(3)}{2})^2 3^{-3} \frac{1}{L^2} \ . 
    $
\end{claim}

\begin{myproof}
We only prove the first inequality. The second can be handled similarly. 
When viewing $\Psi$ as a function defined on $[0,1]$,
\begin{small}
\begin{eqnarray*}
  \Psi(x)  &=&  \frac{2}{3} \paren{ 1 - \exp \paren{ - \frac{1}{2} \ln(3)  \frac{L}{ L-T }  x   } }  + \frac{1}{2}  \paren{ 1 - \exp \paren{  - \ln(3) \frac{L}{L-T} \paren{1-x}     }  } 
    \exp \paren{ - \frac{1}{2} \ln(3)  \frac{L}{L-T} x  }  \ .  \nonumber 
\end{eqnarray*}
\end{small}
As it is smooth enough, by Taylor's expansion, we have 
$$
\Psi \paren{ \frac{T}{L}   + \frac{1}{L} } 
= \Psi \paren{ \frac{T}{L} } + \underbrace{\Psi' \paren{ \frac{T}{L} }}_{=0}   \frac{1}{L} + \frac{1}{2} \Psi'' \paren{ \xi } \paren{\frac{ 1 }{ L }}^2  \ ,
$$
for some $\xi$ on the line segment of $ \frac{T  \pm 1}{L} $ and $\frac{T}{L}$. 
In what follows, we proceed to lower bound $ \abs{\Psi'' \paren{ \xi }} $. To straightforward calculation, we have 
$$
\Psi''(x) = -  \frac{ \ln(3)^2 }{8} \cdot (\frac{L}{L-T})^2 \cdot  3^{\frac{L (x-4)+2 T}{2 (L-T)}} \cdot \paren{ 3^{\frac{L-L x}{L-T}}+3 } \ . 
$$
By noting that for any $x \in [0,1]$ and $0 \leq T \leq \frac{L}{2}$: 
\begin{itemize}
    \item $3^{\frac{L-L x}{L-T}} \geq 3^0 = 1$,
    \item $3^{\frac{L (x-4)+2 T}{2 (L-T)}} \geq 3^{\frac{-4 L +2 T}{2 (L-T)}} = 3^{\frac{-2 L +T}{L-T}} \geq 3^{-3}$,
    \item $\frac{L}{L-T} \geq 1$,
\end{itemize}
we conclude that $
\abs{ \Psi''(x) } \geq (\frac{L}{L-T})^2 \frac{4 \ln(3)^2}{8} 3^{-3}  
\geq \frac{ \ln(3)^2}{2} 3^{-3} 
$
for any $x \in [0,1]$ and $0 \leq T \leq \frac{L}{2}$.

\end{myproof}

Combining Lemma~\ref{lemma:obj_truncation_error} and Lemma~\ref{lemma:curvature}, we have a chain of inequalities
$$
\hat{\Psi} \paren{ \frac{T}{L} } \geq \Psi \paren{ \frac{T}{L} } - O \paren{ n \delta } 
\geq  \Psi \paren{ \frac{T  \pm 1}{L} } +  \bigOmega{ \frac{1}{L^2} } - O \paren{ n \delta }  
\geq \hat{\Psi} \paren{ \frac{T  \pm 1}{L} }   +  \bigOmega{ \frac{1}{L^2} } - O \paren{ n \delta }   \ . 
$$
Therefore, it suffices to make $\delta$ small enough so that $\bigOmega{ \frac{1}{L^2} } \geq O \paren{ n \delta }  $ to make $ \hat{\Psi} \paren{ \frac{T}{L} } \geq \hat{\Psi} \paren{ \frac{T  \pm 1}{L} }$. 
In addition,  we note that it is easy to show  $\hat{\Psi}$ is unimodal as a function defined on $[0,1]$. 
Hence, when viewed as functions on subsets, both $\hat{\Psi}$ and $\Psi$ will attain the global maximum at the same subset.

Lastly, we conclude that the polynomial-sized encoding is feasible, by virtue of the following classic result (stated in our context). 
\begin{lemma}[\cite{brent1976fast}]
    Assuming access to multitape Turing machine, let $M(d)$ be the number of bit-operations required to multiply $d$-bit numbers with $d$-bit accuracy.  
    It takes $O(\log(d) M(d))$  bit operations to ensure that $\max \braces{ \frac{\abs{p_i - \hat{p}_i}}{p_i} ,  \frac{\abs{p_i'- \hat{p}_i'}}{p_i'}}  \leq 2^{-d}$
    for each $i$, for the parameter $\alpha = \frac{\ln(\frac{3}{2})}{L-T}$. 
\end{lemma}

\begin{remark}[Relation to \cite{fu2022pandora}]
\label{remark:connections_to_fu}
    At their core, both \cite{fu2022pandora} and our work study variations of the Pandora's box problem: one problem with non-obligatory inspections, and another where there are two types of inspections and where one type must be followed by the other type prior to selection. 
    Both hardness proofs are based on restricted problem instances for which different notion of ``committing policies" are optimal. 
    Both proofs use reductions to the Subset Sum problem \citep[][use the Partition problem, a special case of the Subset Sum problem]{fu2022pandora}.

    However, the proofs also differ along multiple dimensions. 
\begin{enumerate}
    \item 
    The combinatorial hardness in the two problems stems from different sources: ordering $n!$ potential permutations of boxes \citep{fu2022pandora} versus choosing one of $2^n$ inspection modes for the boxes.

    To clarify, in \cite{fu2022pandora},  the hardness comes \textit{solely} from deciding the ordering $\sigma$ of the boxes. 
    The support set of the value of boxes is $\{ 0, \frac{1}{2}, 1\}$ (the three-point support is essential, as the problem admits a polynomial-time computable optimal policy if the distributions are only supported on $\{ 0, 1 \}$).
    The optimal policy inspects boxes according to the order $\sigma$. Whenever it sees the value $1$, it selects the box immediately. 
    Whenever it sees the value $\frac{1}{2}$, it runs Weitzman's index policy on the remaining boxes, without using the option to select a box without inspection. 
    If it never sees the value $\frac{1}{2}$ and reaches the last box, then it exercises the option to select without inspection. 

    For a given instance of Partition with $n$ items, they construct the hard instance by introducing two  auxiliary boxes $B_{n+1}$ and $B_{n+2}$. 
    The construction ensures that $B_{n+2}$ optimally appears in the last position of the ordering, and is therefore the only box that may be selected without inspection. 
    The computational difficulty lies in determining the position of $B_{n+1}$ in the permutation: the optimal placement induces a partition of the remaining boxes into two segments with equal total weights, whenever such a partition exists. 

    By contrast, in our hard instance, the support set of the value of boxes is $\{0,1\}$. 
    Our instance is carefully crafted so that a committing policy is optimal. 
    The combinatorial decision is therefore not an ordering problem, but rather the choice between full (F) versus partial (P) inspection for each box. Once the inspection modes are fixed, the remaining ordering decision is straightforward. 
    
    \item In both cases, constructing the hard instance relies on a deep structural characterization of optimal policies, which makes it possible to explicitly derive the analytical form of the expected payoff as a function of the underlying combinatorial decision. 
    Namely, the inspection ordering in \cite{fu2022pandora} and the choice between full versus partial inspection in our setting.
    Building on insights from \cite{guha2008information}, \cite{fu2022pandora} strengthen the characterization of optimal policies and exploit this structure in their hardness construction. 
    In contrast, our reduction builds on the structural results established in our paper, e.g., Theorems~\ref{thm:myopic_stopping} and~\ref{thm:just_strong}, which characterize optimal behavior under obligatory inspection and sequential opening modes.
    
    \item There are also several technical differences between the two hardness constructions. For instance, \cite{fu2022pandora} design their instances so that all primitives lie on a finite grid of rational numbers. In contrast, to facilitate multiplicative computations, we parameterize the Bernoulli probabilities in exponential form, which results in primitives that are no longer confined to a finite rational grid. Accordingly, we provide a careful discussion of the encoding length required to represent our problem instances.  
    Moreover, in their construction, boxes have heterogeneous opening thresholds. In our instance, by contrast, the full-inspection (F) and partial-inspection (P) thresholds are identical across boxes (at least in the true instance of the reduction).

\end{enumerate}
\end{remark}

\section{Additional Tables}
\label{apx:tables}

\begin{table}[htbp]
 \footnotesize
  \centering
  \caption{Runtime (in seconds) of different policies. }
    \begin{tabular}{cccrccccccccccc}
    \toprule
    \multirow{2}[3]{*}{$N$} & \multicolumn{2}{c}{$\pi^\textrm{OPT}$} &    & \multicolumn{2}{c}{$\pi^{\textrm{index}}$} &    & \multicolumn{2}{c}{$\pi^{\textrm{W}}$} &    & \multicolumn{2}{c}{$\pi^{F^*,P^*}$} &    & \multicolumn{2}{c}{$\pi^{STP}$} \\
\cmidrule{2-3}\cmidrule{5-6}\cmidrule{8-9}\cmidrule{11-12}\cmidrule{14-15}       & mean  & std &    & mean  & std &    & mean  & std &    & mean  & std &    & mean & std \\
\midrule
    2  & 0.002 & 0.002 &    & 0.001 & 0.001 &    & 0.005 & 0.004 &    & 0.001 & 0.001 &    & 0.000 & 0.000 \\
    3  & 0.014 & 0.013 &    & 0.002 & 0.002 &    & 0.021 & 0.024 &    & 0.004 & 0.004 &    & 0.001 & 0.001 \\
    4  & 0.101 & 0.078 &    & 0.004 & 0.005 &    & 0.084 & 0.121 &    & 0.013 & 0.014 &    & 0.002 & 0.002 \\
    5  & 0.738 & 0.577 &    & 0.007 & 0.010 &    & 0.276 & 0.466 &    & 0.032 & 0.049 &    & 0.004 & 0.003 \\
    6  & 5.698 & 5.128 &    & 0.012 & 0.017 &    & 0.884 & 1.729 &    & 0.081 & 0.144 &    & 0.007 & 0.008 \\
    7  & 41.049 & 39.646 &    & 0.025 & 0.050 &    & 2.933 & 6.081 &    & 0.248 & 0.563 &    & 0.015 & 0.020 \\
    8  & 316.648 & 311.921 &    & 0.047 & 0.114 &    & 10.831 & 30.604 &    & 0.707 & 1.902 &    & 0.028 & 0.031 \\
    9  & 2472.509 & 2585.495 &    & 0.088 & 0.323 &    & 29.127 & 72.949 &    & 1.662 & 3.831 &    & 0.052 & 0.060 \\
\cmidrule{1-1}      
    10 & -  & -  &    & 0.128 & 0.316 &    & 75.226 & 228.431 &    & 3.292 & 7.010 &    & 0.072 & 0.104 \\
    11 & -  & -  &    & 0.261 & 1.018 &    & 152.678 & 1817.640 &    & 9.082 & 25.463 &    & 0.141 & 0.220 \\
    12 & -  & -  &    & 0.372 & 0.778 &    & 718.722 & 3458.725 &    & 20.250 & 40.365 &    & 0.227 & 0.382 \\
    13 & -  & -  &    & 0.823 & 1.851 &    & 1613.774 & 4178.379 &    & 79.878 & 266.576 &    & 0.438 & 0.594 \\
    14 & -  & -  &    & 1.556 & 6.139 &    & 3668.967 & 10860.112 &    & 245.827 & 1494.645 &    & 0.793 & 1.365 \\
    15 & -  & -  &    & 2.977 & 3.895 &    & 10325.838 & 13620.728 &    & 810.565 & 1658.033 &    & 1.885 & 2.392 \\
    16 & -  & -  &    & 5.475 & 11.173 &    & 10168.432 & 15753.618 &    & 869.394 & 5183.688 &    & 2.025 & 3.310 \\
    \bottomrule
    \end{tabular}%
 \label{fig:optimality_boxplot}
\end{table}%

\newpage
\section{When is P-opening worthwhile?}

\label{sec:numerical-apx}

We conduct an additional numerical study with the goal of examining when P-opening is worthwhile. 
We apply a similar process for generating instances (Section~\ref{sec:numerical_instances}) with the exception choosing 100 as the number prototypical boxes, which are illustrated in Figure~\ref{fig:scatter_boxes}.

\begin{figure}[!h]
  \centering
  \includegraphics[width=7cm]{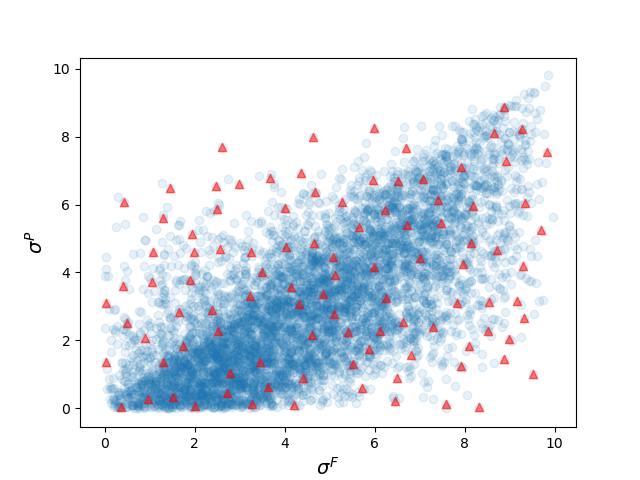}
  \caption{An example of how prototypical boxes are generated. Each point is a box sampled from the distribution. Red triangles are those which are selected, while blue circles are those which are not. }
  \label{fig:scatter_boxes}
\end{figure}

To get a sense for when P-opening is particularly useful under an optimal policy, we define the \textit{P-ratio}. This is the ratio between the expected number of P-openings and the total expected number of P-openings and F-openings (of closed-boxes), both computed for an optimal policy at the initial state of a problem instance.

\begin{figure}[htbp]
\centering
\begin{subfigure}{.5\textwidth}
  \centering
  \includegraphics[width=0.8\textwidth]{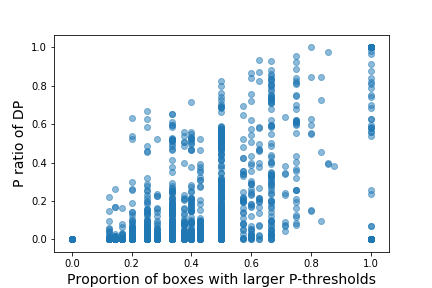}
\end{subfigure}%
\hfill
\begin{subfigure}{.5\textwidth}
  \centering
  \includegraphics[width=0.8\textwidth]{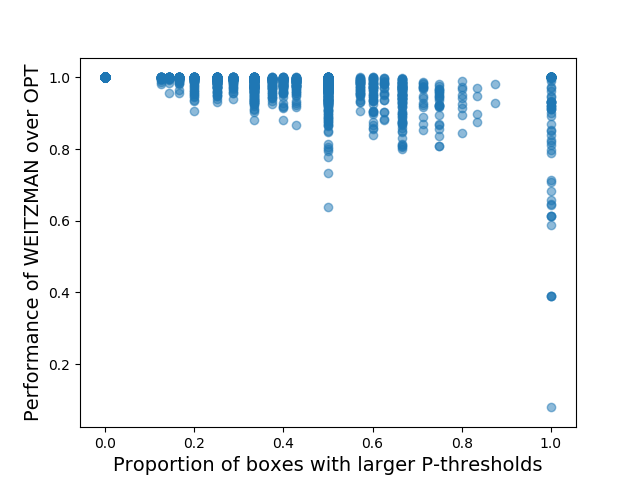}
  \label{diag:Weitzman_performance}
\end{subfigure}
\caption{ When is P-opening important? \\
Left figure: P-ratio of $\pi^{\textrm{OPT}}$ versus proportion  of boxes with $\sigma_i^P > \sigma_i^F$. \\
Right figure:   Suboptimality (ratio of the value of $\pi^{\textrm{Weitzman}}$ to that of the optimal policy)) of $\pi^{[N], \emptyset}$ versus proportion of boxes with $\sigma_i^P > \sigma_i^F$. }
\label{fig:when_P_important}
\end{figure}

Figure \ref{fig:when_P_important} (left) plots the P-ratio of $\pi^{\textrm{OPT}}$ versus the proportion of boxes with $\sigma_i^P > \sigma_i^F$. We observe a positive correlation between these two quantities, indicating that when there are more boxes with $\sigma_i^P > \sigma_i^F$, the optimal policy tends to take advantage of P-opening more frequently.
We also calculate the performance of the policy $\pi^{\textrm{Weitzman}}$, which simply ignores all the P-opening options, i.e., the committing policy $\pi^{[N],\emptyset}$. Figure \ref{fig:when_P_important} (right) shows its suboptimality versus the proportion of boxes with $\sigma_i^P > \sigma_i^F$. We similarly observe that it tends to perform worse when there are more boxes with larger P-thresholds.
These observations further suggest that the interaction between the P-thresholds and F-thresholds impacts optimal decisions, and the relative magnitude of P-thresholds compared to F-thresholds determine the extant to which P-opening should be utilized. 
This may also explain why the simple index policy $\pi^{\textrm{index}}$ performs so well.

Next, we consider the heterogeneity of boxes. 
For each problem instance, we define \textit{dispersion} as the sum of standard deviations of the P-thresholds and F-thresholds that are associated with closed boxes. 
Intuitively, when the dispersion is higher, boxes are more heterogeneous. The upper panels of Figure~\ref{fig:P_F_dispersion} illustrate dispersion. Each panel shows a problem instance, where every point on the scatter plot represents a box (e.g., the left panel shows an instance with 9 boxes, while the right panel shows a problem instance with 10 boxes). The coordinates of each box correspond to its F-opening and P-opening thresholds. 
Intuitively, the instance on the left has boxes that are more homogeneous while the instance on the right has boxes that are more heterogeneous, which are reflected by the dispersion metric (the dispersion of the left instance is smaller than 5, while the dispersion of the right instance is greater than 20).

\begin{figure}[htbp]
        \centering
        \begin{subfigure}[b]{0.475\textwidth}
            \centering
            \includegraphics[width=0.8\textwidth]{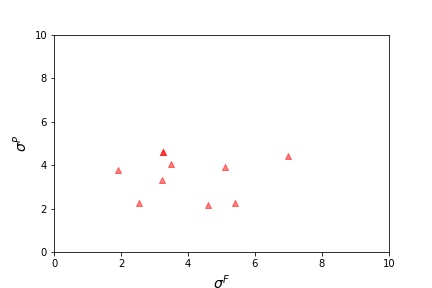}
        \end{subfigure}
        \hfill
        \begin{subfigure}[b]{0.475\textwidth}
            \centering
            \includegraphics[width=0.8\textwidth]{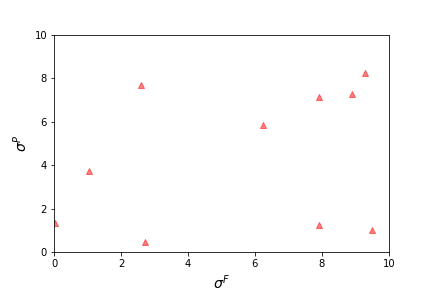}
        \end{subfigure}
        \vskip \baselineskip
        \begin{subfigure}[b]{0.475\textwidth}
            \centering
            \includegraphics[width=0.8\textwidth]{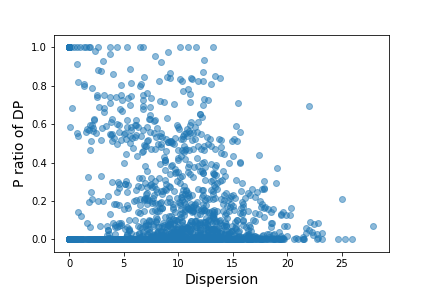}
        \end{subfigure}
        \hfill
        \begin{subfigure}[b]{0.475\textwidth}
            \centering
            \includegraphics[width=0.8\textwidth]{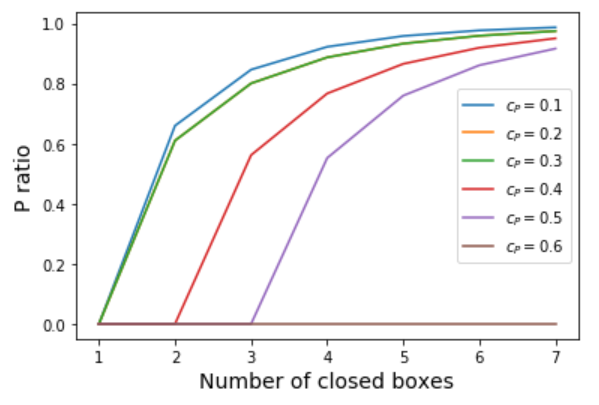}
            \label{fig:more_boxes}
        \end{subfigure}
        \caption[]
        {\small Heterogeneity of boxes. 
        Figures from left to right, top to bottom correspond to:  example instance where dispersion is relatively small ($<5$),
        example instance where dispersion is relatively large ($>20$),
        P-ratio of the optimal policy versus dispersion, and
        P-ratio of the optimal policy versus the number of boxes. 
        }
        \label{fig:P_F_dispersion}
\end{figure}

The lower-left panel of Figure~\ref{fig:P_F_dispersion} plots the P-ratio versus dispersion. We observe that, when the dispersion is large (i.e., above 15), the P-ratio tends to be small, meaning that there are relatively fewer P-openings in the optimal policy. Intuitively, when boxes are very different in terms of the potential information gain which is proxied by the opening thresholds, boxes are more likely to dominate each other. In such settings, an optimal policy tends to directly F-open the boxes according to a predefined order.

Finally, the lower-right panel of Figure \ref{fig:P_F_dispersion} shows the P-ratio versus the number of boxes of the problem instance. To generate this figure, we randomly sample several box instances, vary their P-opening costs, and investigate how an optimal policy behaves when there are $N$ i.i.d. such boxes. We observe that the optimal policy exploits P-opening more often when there are sufficiently many boxes. The panel complements panel (c), showing that when the dispersion is low (here instances are homogeneous), there is interaction affect with the number of boxes.

\end{APPENDICES}

\end{document}